\documentclass[namedreferences]{solarphysics}
\usepackage[optionalrh,solaromanenum]{spr-sola-addons} 
\usepackage{graphicx}                    
\usepackage{geometry}
\usepackage{rotating}
\usepackage{color}                       
\usepackage{url}                         
\usepackage{lscape}
\usepackage{longtable}
\usepackage{caption}
\usepackage{afterpage}
\usepackage[normalem]{ulem}

\newcommand{\apj}{    {\it Astrophys. J.}}
\newcommand{\apjs}{   {\it Astrophys. J. Supp.}}
\newcommand{\apjl}{   {\it Astrophys. J. Lett.}}

\newcommand{\grl}{    {\it Geophys. Res. Lett.}}

\newcommand{\jgr}{    {\it J. Geophys. Res.}}

\newcommand{\solphys}{{\it Solar Phys.}}
 
\newcommand{\ssr}{    {\it Space Sci. Rev.}}

\begin{document}

\begin{article}

\begin{opening}

\title{Ensemble modeling of CMEs using the WSA-ENLIL+Cone model}

%
\author{M.L.~\surname{Mays}$^{1,2}$\sep
        A.~\surname{Taktakishvili}$^{1,2}$\sep
        A.~\surname{Pulkkinen}$^{2}$\sep
        P.J.~\surname{MacNeice}$^{2}$\sep
        L.~\surname{Rast\"{a}tter}$^{2}$\sep
        D.~\surname{Odstrcil}$^{3,2}$\sep
        L.K.~\surname{Jian}$^{2,4}$\sep
        I.G.~\surname{Richardson}$^{4,5}$\sep
        J.A.~\surname{LaSota}$^{6}$\sep
        Y.~\surname{Zheng}$^{2}$\sep
        M.M.~\surname{Kuznetsova}$^{2}$ 
       }

%
\runningauthor{Mays et al.}
\runningtitle{Ensemble modeling of CMEs using the WSA-ENLIL+Cone model}

%
\institute{$^{1}$ Catholic University of America, Washington, DC, USA
                     email: \url{m.leila.mays@nasa.gov}\\ 
             $^{2}$ Heliophysics Science Division, NASA Goddard Space Flight Center, Greenbelt, MD, USA\\
             $^{3}$ George Mason University, Fairfax, VA, USA\\
             $^{4}$ Department of Astronomy, University of Maryland, College Park, MD, USA\\
             $^{5}$ CRESST\\
             $^{6}$ University of Illinois at Urbana-Champaign, Champaign, IL, USA\\
             }

\begin{abstract}
Ensemble modeling of coronal mass ejections (CMEs) provides a probabilistic forecast of CME arrival time which includes an estimation of arrival time uncertainty from the spread and distribution of predictions and forecast confidence in the likelihood of CME arrival. The real-time ensemble modeling of CME propagation uses the Wang-Sheeley-Arge (WSA)-ENLIL+Cone model installed at the {\it Community Coordinated Modeling Center} (CCMC) and executed in real-time at the CCMC/{\it Space Weather Research Center}. The current implementation of this ensemble modeling method evaluates the sensitivity of WSA-ENLIL+Cone model simulations of CME propagation to initial CME parameters. We discuss the results of real-time ensemble simulations for a total of 35 CME events which occurred between January 2013 - July 2014.  For the 17 events where the CME was predicted to arrive at Earth, the mean absolute arrival time prediction error was 12.3 hours, which is comparable to the errors reported in other studies. For predictions of CME arrival at Earth the correct rejection rate is 62\%, the false-alarm rate is 38\%, the correct alarm ratio is 77\%, and false alarm ratio is 23\%.  The arrival time was within the range of the ensemble arrival predictions for 8 out of 17 events. The Brier Score for CME arrival predictions is 0.15 (where a score of 0 on a range of 0 to 1 is a perfect forecast), which indicates that on average, the predicted probability, or likelihood, of CME arrival is fairly accurate.   The reliability of ensemble CME arrival predictions is heavily dependent on the initial distribution of CME input parameters (e.g. speed, direction, and width), particularly the median and spread.  Preliminary analysis of the probabilistic forecasts suggests undervariability, indicating that these ensembles do not sample a wide enough spread in CME input parameters.  Prediction errors can also arise from ambient model parameters, the accuracy of the solar wind background derived from coronal maps, or other model limitations. Finally, predictions of the $K_P$ geomagnetic index differ from observed values by less than one for 11 out of 17 of the ensembles and $K_P$ prediction errors computed from the mean predicted $K_P$ show a mean absolute error of 1.3.
\end{abstract}

%

\end{opening}

%
\section{Introduction}\label{intro} 
Ensemble modeling has been employed in weather forecasting in order to quantify prediction uncertainties and determine forecast confidence \cite{sivillo1997}.  Individual forecasts which constitute an ensemble forecast represent possible scenarios which approximate a probability distribution that reflects forecasting uncertainties.  Such uncertainties which when considered as a group include those associated with initial conditions (such as observational uncertainties), techniques and models. Different forecasts in the ensemble can start from different initial conditions and/or be based on different forecasting models/procedures.  In the simplest application, the ensemble mean or a weighted mean can be taken as a single forecast.  The ensemble mean should perform better than individual ensemble members by emphasizing systematic features found in all members. However, an ensemble also contains additional information about possible scenarios and their probabilities and thus provides a probabilistic forecast.  For example, ensemble modeling provides a quantitative description of the forecast probability that an event will occur by giving event occurrence predictions as a percentage of ensemble size.  This conveys the level of uncertainty in a given forecast in contrast to a categorical yes/no forecast.  Additionally, all ensemble forecast members can be plotted together to allow visualization of the uncertainty among ensemble members, and their clustering distribution. An example of such a visualization is hurricane track ``plume'' maps in weather forecasting.  Regions where members tend to coincide/cluster can be taken to have a higher forecast confidence. 

To understand the uncertainties in space weather forecasting, ensemble coronal mass ejection (CME) forecasting efforts have now begun in space weather models of the heliosphere. \inlinecite{fry2003}, \inlinecite{mckenna2006}, and \inlinecite{smith2009} compared the performance of real-time shock arrival time forecasts following solar events (since 1997) from the three ``Fearless Forecast'' models: Shock Time of Arrival (STOA)\cite{dryer1974}, Interplanetary Shock Propagation Model (ISPM)\cite{smith1990}, and Hakamada-Akasofu-Fry (HAFv.2)\cite{dryer2001}.  
While there are many models predicting the evolution of CMEs (see \cite{zhao2014} and references therein), only the Wang-Sheeley-Arge (WSA) coronal model \cite{arge2000,arge2004} coupled with the global heliospheric ENLIL solar wind model \cite{odstrcil2003} has been used extensively in space weather operations world-wide. 
The first effort in utilizing this model for ensemble forecasting of CME propagation was reported by \inlinecite{pulkkinen2011}.  \inlinecite{emmons2013} performed WSA-ENLIL ensemble CME modeling using 100 ensemble members for 15 historical events with automatically determined cone model CME parameters \cite{pulkkinen2009}.  They found that the observed CME arrival was within the ensemble prediction spread for 8 out of the 15 events.  \inlinecite{lee2013} discuss ensemble modeling of CME propagation with WSA-ENLIL for an event study using eight ensemble members and various synoptic background maps.  Differences found in the predicted arrival time of each individual simulation were mostly due to CME initial speed and the time at which the CME was inserted at the WSA-ENLIL inner boundary, resulting in propagation through a different background solar wind.  They used National Solar Observatory Global Oscillation Network Group (GONG) \cite{harvey1996} synoptic magnetograms and Air Force Data Assimilative Photospheric flux Transport (ADAPT) maps \cite{arge2010,henney2012}. For their CME event they show that when using ADAPT maps, the WSA-ENLIL model values were in better agreement with in-situ observations, and the arrival time predictions were improved due to the more accurate background solar wind representation. However, the overall spread in CME arrival times did not change significantly.

This paper describes the WSA-ENLIL+Cone ensemble modeling system installed at the {\it Community Coordinated Modeling Center} (CCMC) and results from the past 1.5 years of real-time execution at the CCMC/{\it Space Weather Research Center}.  This is the first ensemble space weather prediction system for CME propagation of its kind employed in a real-time environment.  The current version of the system evaluates the sensitivity of CME arrival time predictions from the WSA-ENLIL+Cone model to initial CME parameters.  The CCMC, located at NASA Goddard Space Flight Center, is an interagency partnership to facilitate community research and accelerate implementation of progress in research into space weather operations. The SWRC is a CCMC sub-team which provides space weather services to NASA robotic mission operators and science campaigns, and prototypes new models, forecasting techniques and procedures.
The CCMC also serves the {\it CME Scoreboard} website\footnote{\url{http://kauai.ccmc.gsfc.nasa.gov/CMEscoreboard}} to the research community who may submit CME arrival time predictions in real-time for a variety of forecasting methods.  The website facilitates model validation under real-time conditions and enables collaboration.  For every CME event table on the site, the average of all submitted forecasts is automatically computed, thus itself providing a world-wide ensemble mean CME arrival time forecast from a variety of models/methods.

In Section \ref{model} a brief description of the WSA-ENLIL+Cone model is given.  The triangulation algorithm for determining CME parameters for the ENLIL model is described in Section \ref{params}.   The real-time ensemble modeling methodology is explained in Section \ref{method} followed by an example of an ensemble simulation given in Section \ref{event}.  Results and the evaluation of the first 1.5 years of simulations are described in Section \ref{results}.   In Section \ref{case} we discuss a parametric event case study of the sensitivity of the CME arrival time prediction to model free parameters for the CME and ambient solar wind.  Finally, a summary and discussion are presented in Section \ref{disc}.

\section{WSA-ENLIL+Cone Model Description}\label{model} 
The global 3D MHD WSA-ENLIL model provides a time-dependent description of the background solar wind plasma and magnetic field into which a CME can be inserted \cite{ods1996,odstrcil1999_1,odstrcil1999_2,odstrcil2003,odstrcil2004}. This modeling system does not simulate CME initiation but uses kinematic properties of CMEs inferred from coronagraphs to launch a CME-like hydrodynamic structure into the solar wind and interplanetary magnetic field computed from the WSA coronal model \cite{arge2000,arge2004}. A common method to estimate the 3D CME kinematic and geometric parameters is to assume that the geometrical CME properties are approximated by the Cone model \cite{zhao2002,xie2004} which assumes isotropic expansion, radial propagation, and constant CME cone angular width.  Generally, a CME disturbance is inserted in the WSA-ENLIL model as slices of a homogeneous spherical plasma cloud with uniform velocity, density, and temperature as a time-dependent inner boundary condition at 21.5 solar radii ($R_{\odot}$) with an unchanged background magnetic field. While the simplest geometrical case is employed in this work, the WSA-ENLIL+Cone model can also support an elliptical geometry including tilt, an elongated spheroid or ellipsoid, and leading and trailing edge velocities. Measurements derived from coronagraphs (described in Section \ref{stereocat}) determine the cloud velocity, location, and width. The CME cloud density (\url{dcld}) is a free parameter which by default is 4 times larger than typical mean values in the ambient fast wind providing a pressure of four times larger than that in the ambient fast wind.  The cloud temperature is taken to be equal to the ambient fast wind temperature. Another ENLIL free CME parameter is the cavity ratio which allows the CME to be represented by a spherical shell of plasma and is based on coronagraph observations of CME cavities.  The cavity ratio \url{radcav} is defined as the ratio of the radial CME cavity width to the CME width, with the default being no cavity (\url{radcav}=0).

WSA-ENLIL+Cone runs performed for research and operations have shown that accurate descriptions of the heliosphere and transients are achieved only when the background solar wind is well-reproduced and if coronagraph observations from multiple views, for example from the {\it SOlar and Heliospheric Observatory} (SOHO) spacecraft near the Earth \cite{soho} and the Solar TErrestrial RElations Observatory (STEREO) spacecraft \cite{stereo}, are used to derive CME parameters \cite{lee2013,millward2013}. WSA coronal maps provide the magnetic field and solar wind speed at the boundary between the coronal and heliospheric models, usually at 21.5 $R_{\odot}$, and they are generated from synoptic magnetograms. Small latitudinal shifts in the magnetogram-derived coronal maps caused by inaccuracies in solar magnetic field observations, particularly in the polar regions, can cause large longitudinal shifts in the solar wind structure, for example in characterizing high speed stream arrival times (e.g. \inlinecite{macneice2009}; \inlinecite{jian2011}).  Other coronal models, such as MAS (MHD around a Sphere) \cite{riley2001} or heliospheric tomography from interplanetary scintillation (IPS) \cite{jackson2011} can also provide the background solar wind and have been coupled with ENLIL heliospheric simulations.

CCMC/SWRC has been carrying out routine WSA-ENLIL+Cone simulations for several years using solar magnetic synoptic maps and CME geometric and kinematic properties inferred from coronagraph observations \cite{zheng2013}. Each ENLIL run uses a WSA model synoptic map computed from the single GONG daily-updated synoptic magnetogram (see e.g. \inlinecite{arge2000}) closest to the time the simulation is executed.  These low 4$^{\circ}$ resolution real-time simulations complete in $\sim$20 minutes running on 2 nodes with 16 processors/node on a spherical grid size of 256$\times$30$\times$90 ($r,\theta,\phi$) with 5-10 minute output cadence at locations of interest.  The simulation range is 0.1 to 2 AU in radius $r$, -60$^{\circ}$ to +60$^{\circ}$ in latitude $\theta$, and 0$^{\circ}$ to 360$^{\circ}$ in longitude $\phi$.  CME parameters are derived using real-time coronagraph observations from spacecraft and a geometric triangulation algorithm. The measurements are an approximation of the true 3D speed and width of the CME at 21.5$R_{\odot}$ (ENLIL inner boundary). However, often the coronagraph derived measurements are inferred from just a few data points, and some CMEs may be missed due to real-time data gaps. CME parameters derived in real-time and simulation graphical outputs are publicly available from the CCMC {\it Space Weather Database Of Notifications, Knowledge, Information} {\it (DONKI)} database\footnote{\url{http://kauai.ccmc.gsfc.nasa.gov/DONKI}}. 

\section{Ensemble CME Parameters}\label{params}
\subsection{STEREOCAT TRIANGULATION ALGORITHM FOR DETERMINING CME PARAMETERS}\label{stereocat}
CME parameters are determined using the Stereoscopic CME Analysis Tool (StereoCAT), developed by the CCMC for real-time CME analysis carried out by the CCMC/SWRC forecasting team. The goal was to develop a tool that can be used quickly, yet reliably in a real-time environment with any possible combination of spacecraft available for analysis. It was also required that the tool was intuitive and simple enough to be employed by a wide variety of users such as space weather forecasters, scientists, students, and citizen scientists.  The basic methodology of the tool, i.e., tracking of CME kinematic properties from two different fields-of-view, is similar to that of the {\it NOAA Space Weather Prediction Center} CME Analysis Tool (CAT) by \inlinecite{millward2013} and the geometric localization developed by \inlinecite{pizzo2004}. However, StereoCAT does not attempt to capture the volumetric structure of CMEs but is based on tracking specific CME features. The algorithm is most similar to the CME geometric triangulation method of \inlinecite{liu2010a}. For a more detailed discussion of different CME analysis techniques in the context of cone model-based CME simulations, see \cite{pulkkinen2009,millward2013}. Other stereoscopic methods for determining the kinematic properties of CMEs include those by \inlinecite{thernisien2006}, \inlinecite{lugaz2010}, and \inlinecite{davies2013}.  

StereoCAT is based on triangulation of transient CME features from two different coronagraph fields-of-view or planes-of-sky. We will call these planes-of-sky $A$ and $B$, which may designate, for example, fields-of-view of the SECCHI/COR2 instruments onboard the STEREO A and STEREO B spacecraft \cite{secchi}.  The tool is used to manually identify the same CME features in two consecutive images which are then used to calculate the plane-of-sky velocities for $A$ and $B$, ${\bf v}'_A$ and ${\bf v}''_B$, respectively. Note that these velocities are in local plane-of-sky coordinates indicated by $'$ and $''$. These data need to be brought into the same coordinate system (Heliospheric Earth Equatorial (HEEQ) coordinates in this case), which can be accomplished by rotations:

\begin{eqnarray}
{\bf v}_A = {\bf R}_A \cdot {\bf v}'_A \label{eq:va} \\
{\bf v}_B = {\bf R}_B \cdot {\bf v}''_B \label{eq:vb}
\end{eqnarray}
where operators ${\bf R}_A$ and ${\bf R}_B$ carry out transformations from planes-of-sky $A$ and $B$ coordinates into a common base such as HEEQ, respectively.

We then define two projection matrices as

\begin{eqnarray}
{\bf P}_A = {\bf 1} - {\bf e}_A {\bf e}_A^T \label{eq:Pa} \\
{\bf P}_B = {\bf 1} - {\bf e}_B {\bf e}_B^T  \label{eq:Pb}
\end{eqnarray}
 where ${\bf 1}$ is a 3 $\times$ 3 identity matrix. The unit vectors normal to the planes-of-sky of coronagraphs $A$ and $B$ are defined as ${\bf e}_A$ and ${\bf e}_B$, where ${\bf e}_A^T$ is the transpose of matrix ${\bf e}_A$. The matrices ${\bf P}_A$ and ${\bf P}_B$ project any vector to plane-of-skies $A$ and $B$, respectively. Consequently, plane-of-sky speeds can be expressed as

\begin{eqnarray}
{\bf v}_A = {\bf P}_A \cdot {\bf v} \label{eq:vaPa} \\
{\bf v}_B = {\bf P}_B \cdot {\bf v} \label{eq:vbPb}
\end{eqnarray}
where ${\bf v}$ is the three dimensional vector pointing toward the propagation direction of the CME. While individual projection matrices are not invertible, we can combine Eqs. (\ref{eq:vaPa}) and (\ref{eq:vbPb}) to obtain

\begin{equation}
 ({\bf P}_A + {\bf P}_B) \cdot {\bf v} = {\bf v}_A + {\bf v}_B
\end{equation}
from which we can solve

\begin{equation}
{\bf v} =  ({\bf P}_A + {\bf P}_B)^{-1} \cdot ( {\bf v}_A + {\bf v}_B )
\end{equation}
Importantly, $({\bf P}_A + {\bf P}_B)^{-1}$ exists as long as planes-of-sky $A$ and $B$ are different, i.e. when $e_A$ and $e_B$ are not co-linear (parallel to each other).   Therefore large triangulation errors occur when the spacecraft separation angle is very small or around 180$^{\circ}$.

A similar approach can be used to track the three-dimensional location ${\bf r}$ of a feature from plane-of-sky measurements ${\bf r}_A$ and ${\bf r}_B$ as

\begin{equation}
{\bf r} =  ({\bf P}_A + {\bf P}_B)^{-1} \cdot ( {\bf r}_A + {\bf r}_B )
\label{eq:invr}
\end{equation}

Often the time stamps of coronagraph imagery from spacecraft $A$ and $B$ do not match exactly. This is handled in StereoCAT by propagating the tracked feature in $A$ with speed ${\bf v}_A$ to a new ${\bf r}_A$ that matches the $B$ time stamp. Consequently, matching time stamps are used for ${\bf r}_A$ and ${\bf r}_B$ in Eq. (\ref{eq:invr}).

 The angular size of a CME is estimated in StereoCAT simply by manually selecting the two outer edges of the CME. These two lines that connect through the center of the Sun are then used to compute the opening angle of the CME. It is noted that this process does not take into account projection of the outer CME edges to the spacecraft plane-of-sky, and is therefore a measurement of the projected CME width. While this is not an issue if the CME propagation direction is not too far away from the plane-of-sky of the spacecraft which is used to measure the opening angle, one needs to be very careful with events with propagation directions substantially away from the plane-of-sky, as in such cases the opening angle can be overestimated. This issue will be addressed in the future versions of StereoCAT.

Other limitations of StereoCAT arise from the user's ability to reliably identify the same structures in images from both spacecraft due to ambiguities from the different viewing angles. It may at times be difficult or impossible to track the same structure since different sections of the CME contribute most strongly to images in different planes-of-sky \cite{howard2012}.  Consequently, StereoCAT is not suitable to use with coronagraph data in which the CME appears as a halo, since the CME leading edge is not visible.


\begin{figure}
\centerline{\includegraphics[width=0.7\textwidth,origin=c]{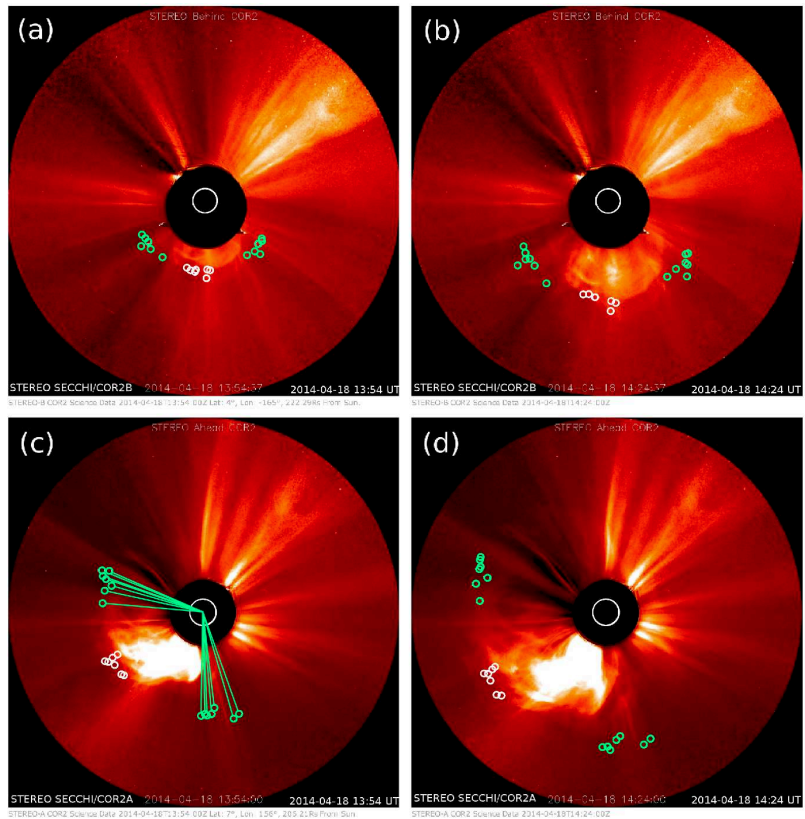}}
\caption{Example screenshot of $m$=6 ``two-timepoint'' measurements performed for the 18 April 2014 CME using StereoCAT in ``ensemble'' mode.   Two image pairs are shown from the SECCHI/COR2 instruments from STEREO B (a-b; top row) and STEREO A (c-d; bottom row), for two different time steps,  2014-04-18 13:54 UT (a, c; left column) and 2014-04-18 14:24 UT (b, d; right column).  The white circles indicate the 6 individual ``two-timepoint'' plane-of-sky leading edge height measurements (near the center of the CME front) and the width measurements are marked by the green circles (near the CME edges).  The green lines in panel c illustrate the CME opening angle measurements for one of the coronagraph images. The plane-of-sky leading edge measurements (central white circles) are later combined together using the triangulation algorithm discussed in Sections \ref{stereocat}-\ref{stereocatmodes} to generate $6^2=36$ ensemble members. The distribution of the resulting CME parameters which are used as initial conditions for 36 WSA-ENLIL+Cone simulations is shown in Figure \ref{fig:catmeashist1}.}
\label{fig:catmeas}
\end{figure}

\subsection{PERFORMING CME MEASUREMENTS WITH STEREOCAT}\label{stereocatmodes}
StereoCAT has three modes: ``two-timepoint'', ``ensemble'', and ``frame series'', and is available online via a web interface\footnote{\url{http://ccmc.gsfc.nasa.gov/analysis/stereo}} \cite{lasota2013}.  Available coronagraphs include the LASCO C2 and C3 instruments on board the SOHO spacecraft \cite{lasco}, and the SECCHI/COR2 instruments on the STEREO A and B spacecraft.  All three modes are based on the same triangulation algorithm, described in section \ref{stereocat}.  In the basic  ``two-timepoint'' mode the user manually measures the CME leading edge height for two different times in each coronagraph image for two different coronagraph viewpoints. The plane-of-sky sky speed for each viewpoint is calculated, from which the triangulated speed and direction is computed using the algorithm described in section \ref{stereocat}.  The user also manually measures the CME opening angle in each coronagraph view. Because this is a projected width measurement, both widths and their average are displayed for the user.

 In ``ensemble'' mode the user manually repeats the same procedure as for the ``two-timepoint'' mode, by measuring the same feature for the same pair of coronagraphs at two different times.  Between each ``two-timepoint'' measurement, the display is fully reset such that the user is forced to carefully remeasure the CME leading edge height and opening angle.  This series of repeated measurements leads to a range of CME parameters which can be used to initialize an ensemble simulation.  For every $m$ ``two-timepoint'' measurements made, $n=m^2$ ensemble CME parameter members are automatically generated by combining different spacecraft measurement pairs.  For example, for $m=2$ ``two-timepoint'' measurements, there are $n=2^2=4$ ways to combine the first and second time step height measurements in viewpoints $A$ and $B$ to triangulate the CME.  Since the two projected width measurements made for each measurement $m$ are not triangulated, they are randomly assigned to each ensemble member. An example screenshot of $m$=6 ``two-timepoint'' measurements performed in ``ensemble'' mode using StereoCAT is shown in Figure \ref{fig:catmeas}.   Two image pairs are shown from the SECCHI/COR2 instruments from STEREO B (top row) and STEREO A (bottom row), for two different times separated by 30 minutes in the left and right columns.  The white circles indicate the 6 individual ``two-timepoint'' plane-of-sky leading edge height measurements (near the center of the CME front) and the width measurements are marked by the green circles (near the CME edges).  The green lines in panel c of Figure \ref{fig:catmeas} illustrate the CME opening angle measurements for one of the coronograph images. In this example the 6 individual ``two-timepoint'' measurements were combined by the algorithm to create $6^2=36$ ensemble members.  

 After completing the measurements the user may inspect histograms of their CME parameters.  The web interface allows the user to remove any ensemble members, and add any ``custom''  members. Generally, members are removed when they have nearly identical parameters, or members for which the triangulation appears unreliable.  Custom members can be measurements from different image time pairs, from plane of sky estimates which incorporate the source location, or from any other CME measurement technique.  The same procedure can be applied to create $n$ individual ensemble measurements for $x$ CMEs for a series of events which are then combined one-to-one to be simulated together such that there are $n$ ensemble members containing $x$ CMEs each.

In ``frame series'' mode the user can measure a series of different frames (times) for each spacecraft, which are then triangulated to create a CME height-time profile.   The user selects a range of time, and steps through the images available from each instrument, measuring the CME in as many images as they choose. The software chooses time pairs of measurements for triangulation based on a user-specified maximum allowed time difference. From these measurements, plane-of-sky and triangulated height-time, velocity, latitude, and longitude profiles of the CME are generated.   Triangulations made with different spacecraft pairs are shown as separate height-time profiles.  Several methods are used to calculate the CME speed, acceleration, the time the CME passes 21.5 ${\rm R}_{\rm S}$ (ENLIL inner boundary), and the time it erupts from the Sun. These include least-squares linear and quadratic fits, averages over selected data points, and averages from only the first and last data points.  Results for each method are reported separately, allowing the user to choose the most appropriate fitting technique depending on the acceleration profile of the CME. Plane-of-sky values are also reported, which can be used when coronagraph projection effects make this triangulation method unreliable. This can occur if the CME is very wide, appears as a halo, or is heavily projected in the coronagraph data.  In these cases the user will not be able to identify the same CME leading edge feature in the data from two coronagraphs.  The user can inspect the triangulated height values directly on the height-time plot to evaluate triangulation accuracy in these cases. 

\section{Ensemble Modeling with WSA-ENLIL+Cone}\label{method}
The current implementation of this ensemble modeling method evaluates the sensitivity of WSA-ENLIL+Cone model simulations of CME propagation to initial CME parameters. As described in Section \ref{stereocat}, StereoCAT is used to create an ensemble of $n$ CME parameters which are used as input to $n$ WSA-ENLIL+Cone simulations.   We have observed that $n\sim$36 to 48 provides an adequate spread of input parameters, but this number can be increased if necessary. For $n$=48 a typical run takes ~130 minutes to complete on 24 nodes with 4 processors/node on the initial development system. We estimate that the same run will take $\sim$80 minutes on the CCMC production system that has 16 processors/node. 

The simulations provide $n$ profiles of MHD quantities (density, velocity, temperature, and magnetic field components) and a distribution of $n$ predicted arrival times at locations of interest within the computational domain.  Currently, ensemble modeling is performed for spacecraft at the following locations: Mercury (MESSENGER), Venus (VEX), Earth (ACE, Wind, SOHO, and orbiting spacecraft), Mars (MSL, MAVEN, MEX), Spitzer Space Telescope, STEREO-A and B.   The CME-associated disturbance/shock arrival time is then automatically computed in post-processing from any sharp increases in the modeled solar wind dynamic pressure at a given location.   In this work, we focus on the ensemble results of the Earth-directed events.

For Earth-directed CMEs, the CCMC/SWRC also computes $n$ estimates of the geomagnetic $K_P$ index using the WSA-ENLIL+Cone model plasma parameters at Earth.   The geomagnetic three-hour planetary $K$ index, $K_P$, is a measure of general planetary wide geomagnetic disturbances at mid-latitudes based on ground-based magnetic observations \cite{bartels1939,rostoker1972,menvielle1991}. The $K_P$ index is created from standardized $K$ indices from individual stations, which measure the magnitude of horizontal geomagnetic field disturbances (not including daily variations).  $K_P$ is a quasi-logarithmic index ranging from 0 to 9. Real-time estimated planetary $K_P$ indices are available from NOAA using real-time data from a limited number of geomagnetic observatories, and the final definitive $K_P$ is from the Helmholtz Center Potsdam GFZ German Research Centre for Geosciences.

The predicted $K_P$ estimate is made by using the \inlinecite{newell2007} coupling function arising from their correlation of 20 candidate coupling functions with geomagnetic indices.  The function which represents the rate of magnetic flux $d \Phi_{\rm MP}/dt$ opening at the magnetopause and correlated best with 9 out of 10 indices is given as
\begin{equation}
d \Phi_{\rm MP}/dt = {v_{\rm bulk}}^{4/3} B_{\rm T}^{2/3}{\sin}^{8/3}(\frac{\theta_{\rm C}}{2}),\label{eq:new}
\end{equation} 
where $v_{\rm bulk}$ is the bulk solar wind speed, the interplanetary magnetic field (IMF) clock angle $\theta_{\rm C}$ is given by $\tan^{-1}(B_y/B_z)$, and the perpendicular component of the magnetic field is given by $B_{\rm T}=(B_y^2+B_z^2)^{1/2}$ (in GSM coordinates). An exponential fit to the correlation of this coupling function with the  $K{\rm p}$ index yields the following relation used for the estimate
\begin{equation}
K_P = 9.5 - e^{2.17676-5.2001 (d \Phi_{\rm MP}/dt)}. \label{eq:kp}
\end{equation}
\inlinecite{emmons2013} showed for their sample of 15 events, that $K_P$ predictions using Eq. \ref{eq:kp} computed directly from in-situ solar wind observations had a mean absolute error of 0.5.  Because ENLIL modeled CMEs do not contain an internal magnetic field and the magnetic field amplification is caused mostly by plasma compression, only the magnetic field magnitude is used and three magnetic field clock angles scenarios of 90$^{\circ}$ (westward), 135$^{\circ}$(south-westward), 180$^{\circ}$ (southward) are assumed.   This provides a simplistic estimate of three possible maximum values which the $K_P$ index might reach following arrival of the predicted CME shock/sheath. For the forecast, the $K_P$ estimates are rounded to the nearest whole number. 

Another commonly used activity index is the $Dst$ (disturbance storm time) index, which is a measure of magnetosphere storm activity primarily from the strength of the ring current.  The index is obtained from the measurement of the perturbations in the horizontal component of the Earth's magnetic field from ground-based observatories that are sufficiently distant from the auroral and equatorial electrojets, are located at approximately $\pm20^{\circ}$ geomagnetic latitude, and are evenly
distributed in longitude \cite{sugiuraDst}.   Although the ring current makes the largest contribution to the $Dst$, all magnetospheric current systems contribute, such as the Chapman-Ferraro magnetopause current which is strengthened during sudden storm commencement (SSC) and increases the Earth's surface field and gives a sudden positive jump in $Dst$.  Currently, ENLIL model results are not used to predict the $Dst$, however in principle this can be computed in a similar manner to the $K_P$ index by using the \inlinecite{newell2007} $Dst$ relation.

\section{Example Ensemble: 18 April 2014 CME}\label{event}
\begin{figure}  
\centerline{\includegraphics[width=0.335\textwidth,angle=0,origin=c]{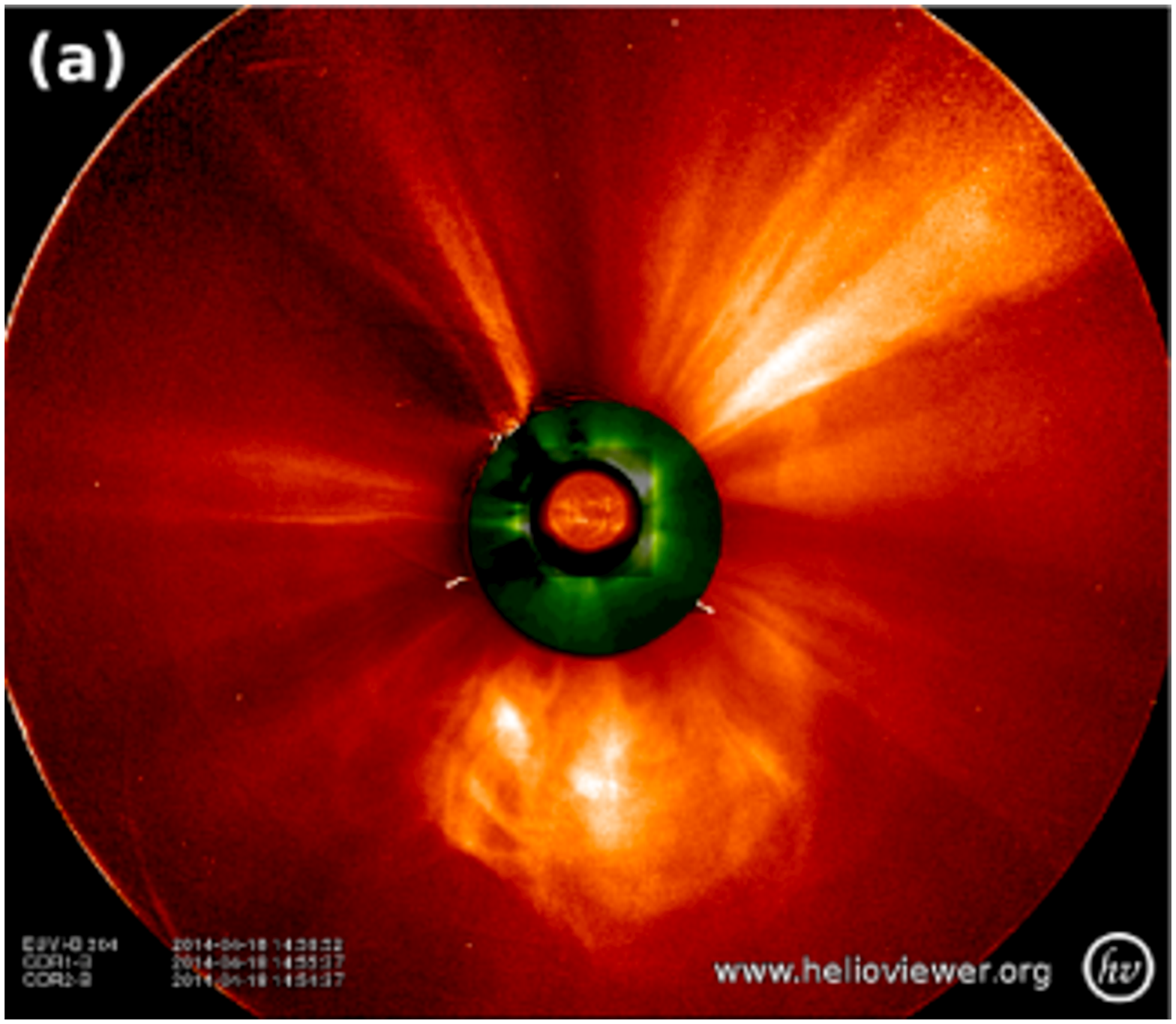}\includegraphics[width=0.335\textwidth,angle=0,origin=c]{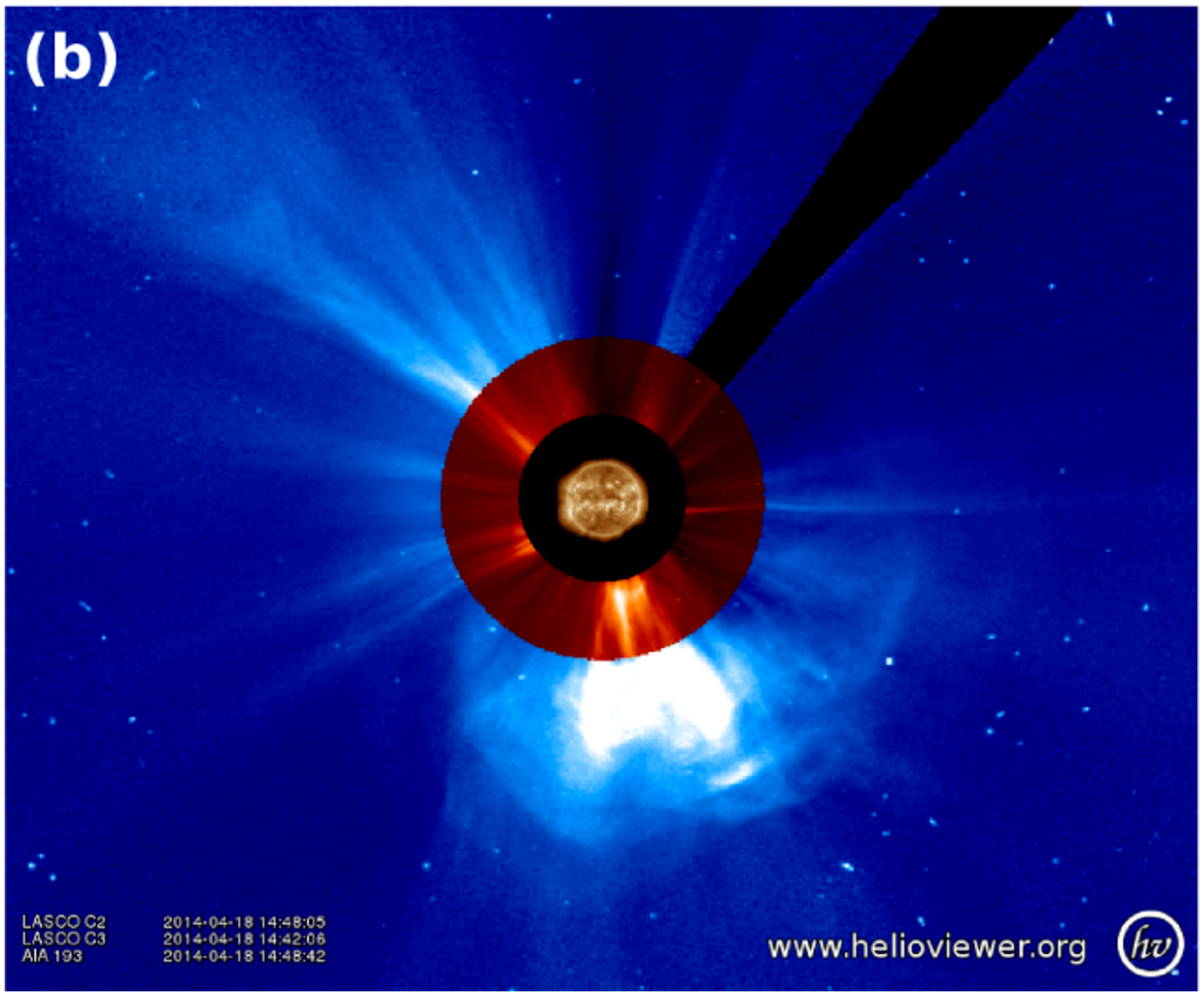}\includegraphics[width=0.335\textwidth,angle=0,origin=c]{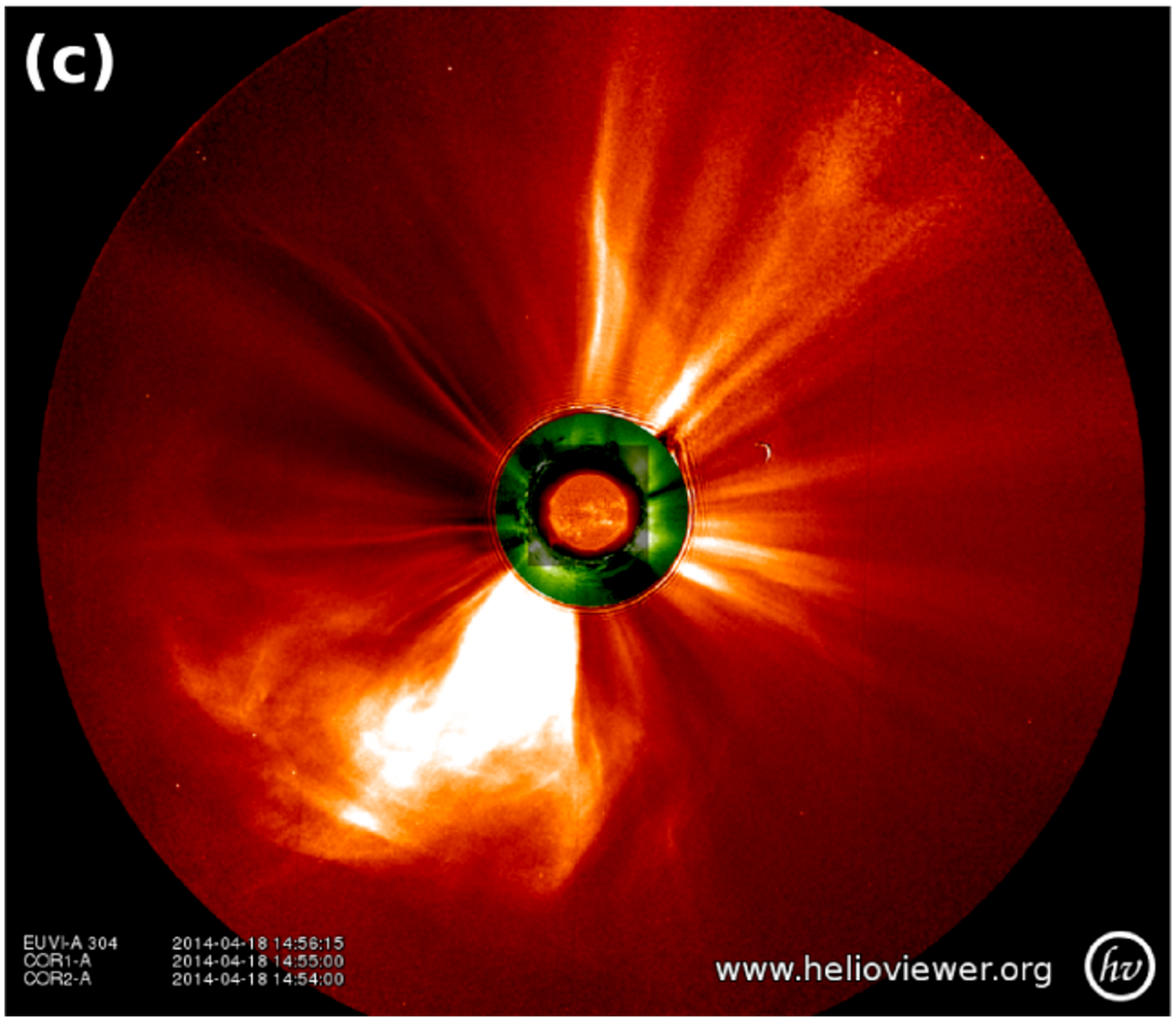}}
\caption{Coronagraph observations of the 18 April 2014 CME with an onset time at 13:09 UT as viewed from (b) SOHO LASCO/C2 and C3, (a) STEREO SECCHI/COR2 B and (c) A near the time of 14:50 UT. The fields-of-view of LASCO/C2, C3, and SECCHI/COR2 are  2.2-6$R_{\odot}$, 2.8-32$R_{\odot}$ (shown here cropped to 17$R_{\odot}$), and 2.5-15$R_{\odot}$ respectively. Images from helioviewer (\protect\url{http://www.helioviewer.org}) \protect\cite{muller2009}.}\label{fig:apr2014cme}
 \end{figure}

In this section we describe the real-time ensemble modeling of an Earth-directed partial halo CME that was first observed at 13:09 UT on 18 April 2014 by by SECCHI/COR2A. Figure \ref{fig:apr2014cme} shows this CME as viewed from SOHO LASCO/C2 and C3, STEREO SECCHI/COR2 A, and B near 14:50 UT.  This CME was associated with an M7.3 class solar flare from Active Region (AR) 12036 located at S18$^{\circ}$W29$^{\circ}$ with peak at 13:03 UT.  The eruption and a coronal wave were visible south of the active region in SDO/AIA 193{\AA} and a nearby filament eruption was visible in AIA 304{\AA}. Subsequently starting at 13:35 UT, an increase in solar energetic particle proton flux above 0.1 pfu/MeV (1 pfu = 1 particle cm$^{-2}$ sr$^{-1}$ s$^{-1}$) was observed by the GOES-13 EPEAD instrument in Earth orbit.  

\begin{figure}  
\centerline{\includegraphics[width=0.9\textwidth,angle=0,origin=c]{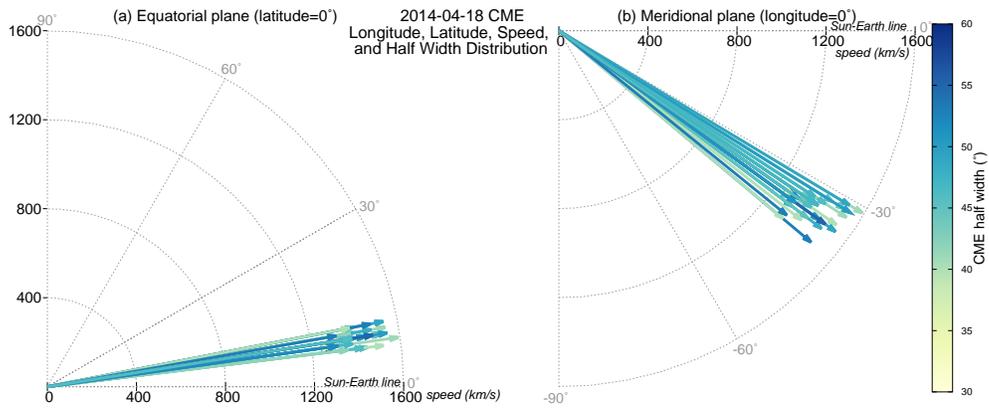}}
\caption{Distribution of the 18 April 2014 CME input parameters shown in the (a) equatorial plane (latitude=0$^{\circ}$) and (b) meridional plane (longitude=0$^{\circ}$).  The plots show the CME speed vectors in spherical HEEQ coordinates with the grids showing the degrees longitude (a) and latitude (b), and the radial coordinate showing the speed in km/s. The Sun-Earth line is along 0$^{\circ}$ longitude and latitude. The arrow directions on the grid indicate the CME central longitude and latitude respectively, with CME half width indicated by the color of the vector. The arrow lengths correspond to the CME speed. CME propagation directions are clustered between -30 to -40$^{\circ}$ latitude, and around 10$^{\circ}$ west of the Sun-Earth line in longitude, while CME speeds range from $\sim$1300 to 1600 km/s.  Median CME parameters are: speed of 1394 km/s, direction of 9$^{\circ}$ longitude, -35$^{\circ}$ latitude, and a half-width of 46$^{\circ}$.}
\label{fig:catmeashist1}
\end{figure}

Figure \ref{fig:catmeas} shows StereoCAT measurements for the 18 April 2014 CME.  As discussed above, the central white circles indicate the individual leading edge measurements and the  green outer circles near the CME edges are the projected width measurements.  The six leading edge measurements are combined together using the triangulation algorithm discussed in Sections \ref{stereocat}-\ref{stereocatmodes}  to generate $6^2=36$ ensemble members. The distribution of the resulting CME parameters which are used as initial conditions for $n$=36 WSA-ENLIL+Cone simulations is shown in Figure \ref{fig:catmeashist1} in the (a) equatorial plane (latitude=0$^{\circ}$) and (b) meridional plane (longitude=0$^{\circ}$).  The plots show the CME velocity vectors in spherical HEEQ coordinates with the grids showing the degrees longitude (a) and latitude (b), and the radial coordinate showing the speed in km/s. The Sun-Earth line is along 0$^{\circ}$ longitude and latitude. The arrow directions on the grid indicate the CME central longitude and latitude respectively, with CME half width indicated by the color of the vector. The arrow lengths correspond to the CME speed.  CME propagation directions are clustered between -30 to -40$^{\circ}$ latitude, and around 10$^{\circ}$ west of the Sun-Earth line in longitude, while CME speeds range from $\sim$1300 to 1600 km/s.  Median CME parameters are: speed of 1394 km/s, direction of 9$^{\circ}$ longitude, -35$^{\circ}$ latitude, and a half-width of 46$^{\circ}$. 

\begin{figure}  \centerline{\includegraphics[width=0.75\textwidth,angle=0,origin=c]{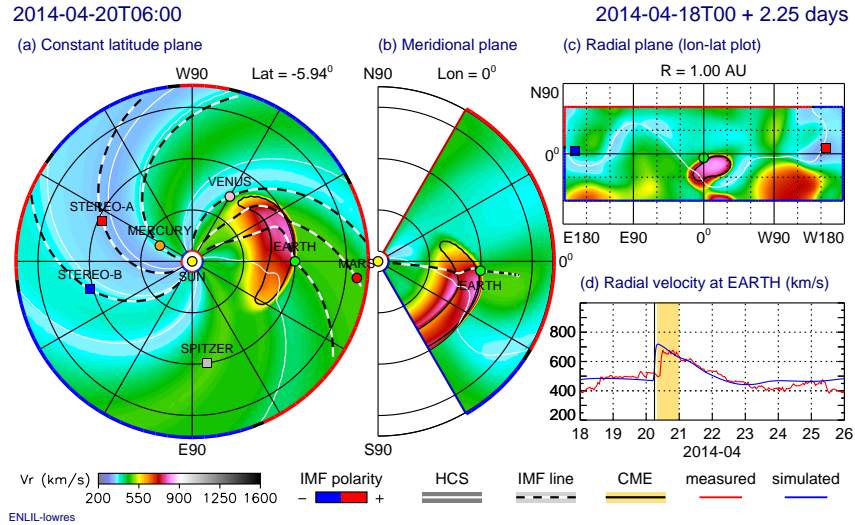}}
 \caption{Global view of the 18 April 2014 CME on 20 April at 06:00 UT: WSA-ENLIL+Cone scaled velocity contour plot for the (a) constant Earth latitude plane, (b) meridional plane of Earth, and (c) 1 AU sphere in cylindrical projection, for the ensemble member with median CME input parameters (speed of 1394 km/s, direction of 9$^{\circ}$ longitude, -35$^{\circ}$ latitude, and a half-width of 46$^{\circ}$). Panel (d) shows the measured (red) and simulated (blue) radial velocity profiles at Earth, with the simulated CME duration shown in yellow.}\label{fig:timvel1}
 \end{figure}

\begin{figure} 
\includegraphics[width=1.0\textwidth,angle=0,origin=c]{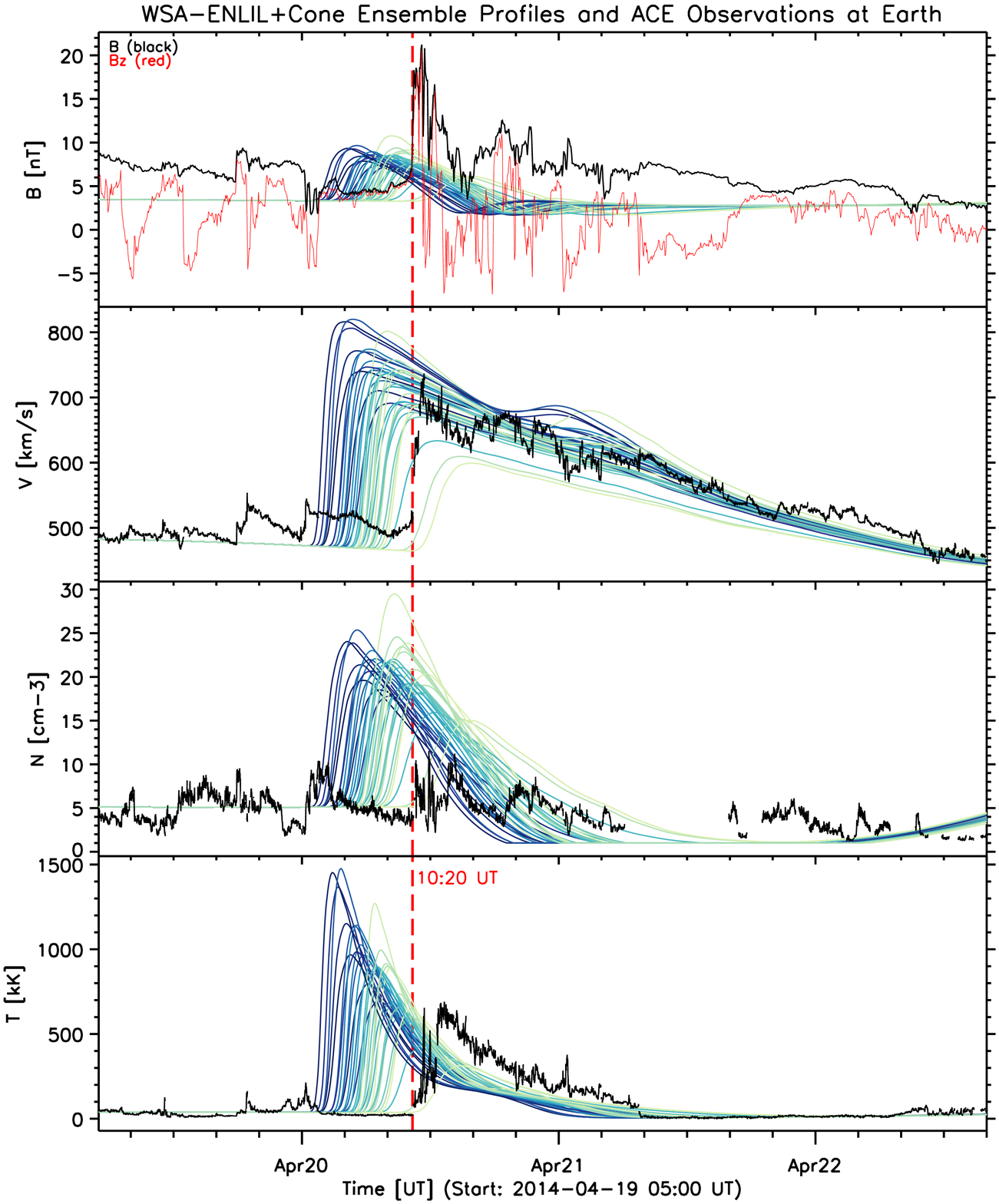}
\caption{18 April 2014 CME ensemble: Model calculated density, velocity, magnetic field, and temperature profiles at Earth for all 36  ensemble members plotted as color traces along with the observed in-situ L1 observations from ACE plotted in black (red for $B_z$).   The model traces are color coded by CME input speed such that slow to faster input speeds are colored from light green to dark blue.  The observations show clear signatures of the arrival of an ICME, including a leading shock (abrupt increase in all the solar wind parameters at around 10:20 UT) with enhanced post-shock temperatures, enhanced magnetic field with rotations in direction, and declining solar wind speed. The spread in the color traces show that most of the predictions are earlier than the observed arrival, with a mean predicted arrival at Earth of 20 April 2014 at 05:07 UT and a range from 20 April 2014 at 01:08 UT to 11:16 UT.}\label{fig:nvbt1}
 \end{figure}

\begin{figure} 
\centerline{\includegraphics[width=0.7\textwidth,angle=0,origin=c]{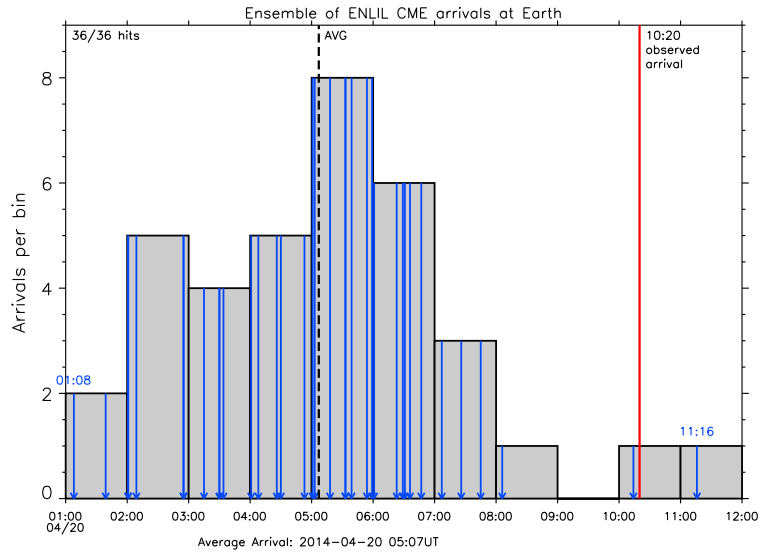}}
\caption{18 April 2014 CME: Histogram distribution of arrival time predictions at Earth (bin size of 1 hour) with individual arrivals marked by the blue arrows.  This figure shows a normal distribution with 50\% of the predicted arrivals are within one hour of the mean. The prediction error for the mean predicted CME arrival time is -5.2 hours and the observed arrival time was within the ensemble predicted spread.}\label{fig:e1}
 \end{figure}

Model results for the 36-member ensemble WSA-ENLIL+Cone run for this CME are shown in Figures \ref{fig:timvel1}-\ref{fig:nvbt1}.  For the ensemble member with median CME input parameters, Figure \ref{fig:timvel1} shows a scaled velocity contour plot for the (a) constant Earth latitude plane, (b) meridional plane of Earth, and (c) 1 AU sphere in cylindrical projection on 20 April at 06:00 UT.   Panel (d) shows the measured (red) and simulated (blue) radial velocity profiles at Earth, with the simulated CME duration shown in yellow. This simulation figure shows the northeastern portion of the CME impacting Earth. Figure \ref{fig:nvbt1} shows the modeled magnetic field, velocity, density, and temperature profiles at Earth plotted as color traces for all 36 ensemble members, along with the observed in-situ L1 observations from ACE, plotted in black.  The model traces are color coded by CME input speed such that slow to faster input speeds are colored from light green to dark blue.   The arrival of the CME-associated shock was observed by Wind and ACE on 20 April 2014 at around 10:20 UT, and energetic storm particles were observed by ACE.   The provisional SYM-H index ($\sim$1 minute $Dst$) shows a sudden storm commencement of +25 nT at 11:01 UT.  The observations in Figure \ref{fig:nvbt1} show clear signatures of the arrival of an Interplanetary Coronal Mass Ejection (ICME), including a leading shock (abrupt increase in all the solar wind parameters at around 10:20 UT) with enhanced post-shock temperatures, enhanced magnetic field with rotations in direction, and declining solar wind speed. This CME was predicted to arrive at Earth and also at Mars for all of the 36 runs.  The mean predicted arrival at Earth was on 20 April 2014 at 05:07 UT with arrival times from individual runs ranging from 20 April 2014 at 01:08 to 11:16 UT. A histogram showing the distribution of arrival times at Earth is shown in Figure \ref{fig:e1} with individual arrivals marked by the blue arrows.  This figure shows a normal distribution with 50\% of the predicted arrivals within one hour of the mean. The prediction error for the mean predicted CME arrival time was -5.2 hours and the observed arrival time was just within the ensemble predicted spread.  The spread in ensemble member predictions can also be seen in Figure \ref{fig:nvbt1} compared to the observations, showing that most of the predictions are earlier than the observed arrival with a few after.    From the CME input parameters plotted in Figure \ref{fig:catmeashist1} the ensemble members with arrival times closest to the observed time had CME input speeds in the range of 1200-1400 km/s, latitudes near -40$^{\circ}$ and half widths around 35$^{\circ}$-40$^{\circ}$.  This suggests that the early arrival time predictions for this event could be due to overestimations of the CME input speed and half width.

\begin{figure} 
\centerline{\includegraphics[width=0.65\textwidth,angle=0,origin=c]{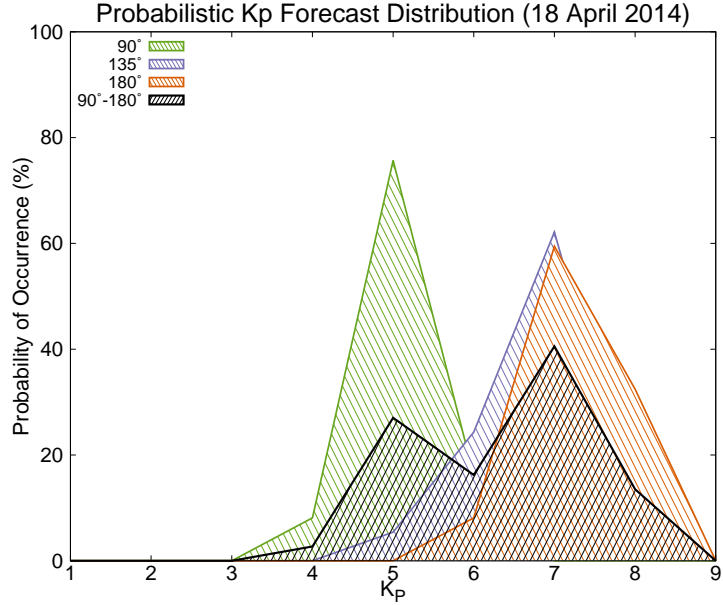}}
\caption{Distribution of $K_P$ probability forecast using ENLIL predicted solar wind quantities at Earth for three clock angle scenarios $\theta_{\rm C}=90^{\circ}$ (green), 135$^{\circ}$ (purple), 180$^{\circ}$ (orange), and all three angles combined 90$^{\circ}$-180$^{\circ}$ (black) (assuming equal likelihood).  The standard deviation of the overall $K_P$ forecast probability distribution is 1.1, with 84\% of the forecasts falling between $K_P$ = 5 to 7. The most likely forecast is for $K_P$=7 at 41\%, followed by $K_P$=5 at 27\% and $K_P$=6 at 16\% likelihood of occurrence. The NOAA real-time observed $K_P$ index (and the Potsdam final Kp) reached 5 during the synoptic period 12:00-15:00 UT on 20 April associated with the CME disturbance arrival.}\label{fig:kp1}
\end{figure}

The NOAA real-time observed $K_P$ index (and the Potsdam final Kp) reached 5 during the synoptic period 12:00-15:00 UT on 20 April associated with the CME shock arrival. The $Dst$ reached a minimum of -24 nT at 15:00 UT on 21 April and thus based on $Dst$, this CME only resulted in very weak geomagnetic activity. As discussed in Section \ref{method}, Eq. \ref{eq:kp} can be used to forecast the maximum $K_P$ index from maximum ENLIL predicted quantities at CME shock/sheath arrival at Earth (colored traces shown in Figure \ref{fig:nvbt1}).  Figure \ref{fig:kp1} shows the predicted probability distribution of $K_P$ for three clock angle scenarios $\theta_{\rm C}=90^{\circ}$ (green), 135$^{\circ}$ (purple), 180$^{\circ}$ (orange).  The figure also shows the overall $K_P$ forecast probability distribution calculated for all three angles combined 90$^{\circ}$-180$^{\circ}$, assuming each scenario is equally likely, in black.  The standard deviation of the overall $K_P$ forecast probability distribution is 1.1, with 84\% of the forecasts falling between $K_P$ = 5 to 7. The most likely forecast is for $K_P$=7 at 41\%, followed by $K_P$=5 at 27\% and $K_P$=6 at 16\% likelihood of occurrence.  Using the most likely forecast of $K_P$=7, the $K_P$ prediction error for this event is $\Delta {\rm K_P}_{\rm err}={\rm K_P}_{\rm predicted}-{\rm K_P}_{\rm observed}=2$ (overprediction).   The overprediction of $K_P$ may be related to the overestimation of the CME input speed. In Sections \ref{cmever}-\ref{kpver} and \ref{case} we discuss various factors which can contribute to early arrival time predictions and Kp overpredictions.

\section{Real-time Ensemble Modeling: First Results}\label{results}
For 35 Earth-directed CME events from January 2013 through June 2014, real-time ensemble modeling was carried out by the CCMC/SWRC team following the methods described in Sections \ref{method}-\ref{event}.  In Table \ref{table:stats} we list a summary of the ensemble simulation results for these 35 CME events. The first and second columns give the CME onset date and time based on the first appearance in C2 or COR2.  Generally, if two CMEs occur within a day of each other they will both be included in the same simulation as separate CMEs which may or may not merge during their propagation.  A few of the ensemble simulations listed in the table contain two CMEs as part of a single run.  In these cases, CMEs that were simulated together with the CME listed on the previous row are indicated by $^*$. The third column lists (for 2013) the 2nd order plane-of-the-sky (POS) speed at 20 $R_{\odot}$ reported in the SOHO LASCO CDAW CME catalog\footnote{http://cdaw.gsfc.nasa.gov/CME\_list} \cite{yashiro2004,gopalswamy2009}.  If measurements were not made to 20 $R_{\odot}$, the 2nd order POS speed at the time of last observation is used. The next four columns provide the median ensemble CME input parameters of $v$, latitude, longitude (HEEQ), and half-width ($w/2$) measured using StereoCAT.  In columns 8, 9 and 10 we list the mean predicted arrival time of all $n_{\rm tot}$ ensemble members, followed by the spread in arrival times in hours relative to the mean.  The next column (11) shows $n_{\rm predicted~ hits}$, the number of ensemble members out of $n_{\rm tot}$, the total number of ensemble members, that predict that the CME will arrive at Earth.  This ratio $p=n_{\rm predicted~ hits}/n_{\rm tot}$ gives a forecast probability and conveys the forecast uncertainty about the likelihood that the CME will arrive.  Columns 12, 13, and 14, list the actual arrival time of the CME-associated shock or disturbance observed in-situ at the Wind spacecraft, followed by the total in-situ observed CME transit time relative to the CME start time.  In the last column the prediction error $\Delta t_{\rm err}$ is calculated for predictions indicating hits. The prediction error is defined as $\Delta t_{\rm err}=t_{\rm predicted}-t_{\rm observed}$, which is negative when ENLIL predictions are earlier than the observed CME arrival time, and late predictions are positive.   When possible, ICME and magnetic cloud catalogs were used to help assess whether the CME did arrive at Earth. These included the \inlinecite{richardson2010} ICME catalog\footnote{\url{http://www.srl.caltech.edu/ACE/ASC/DATA/level3/icmetable2.htm}}, and the Wind ICME catalog\footnote{\url{http://wind.nasa.gov/index_WI_ICME_list.htm}} with circular flux rope model fitting (based on \inlinecite{hidalgo2000}).  Shocks identified by the SOHO CELIAS/MTOF/PM ``shockspotter'' program were also used in arrival time assessment.  Determining the measured in-situ arrival time of the CME-associated shock or disturbance can be subjective and therefore can be a source of error in the prediction error calculation.   Taking this into consideration, in-situ signatures which could not be unambiguously identified as the arrival of the CME-related disturbance are indicated by $\ddagger$ and these five ensembles are not included in the following forecast verification.  This reduces the sample size from 35 to 30 ensembles.

In the following subsections we discuss ensemble CME arrival and $K_P$ forecast verification inspired by methods used in ensemble weather forecasting and applied here for the first time.

\subsection{CME ARRIVAL FORECAST VERIFICATION}\label{cmever}
\begin{figure} 
\centerline{\includegraphics[width=0.5\textwidth,angle=0,origin=c]{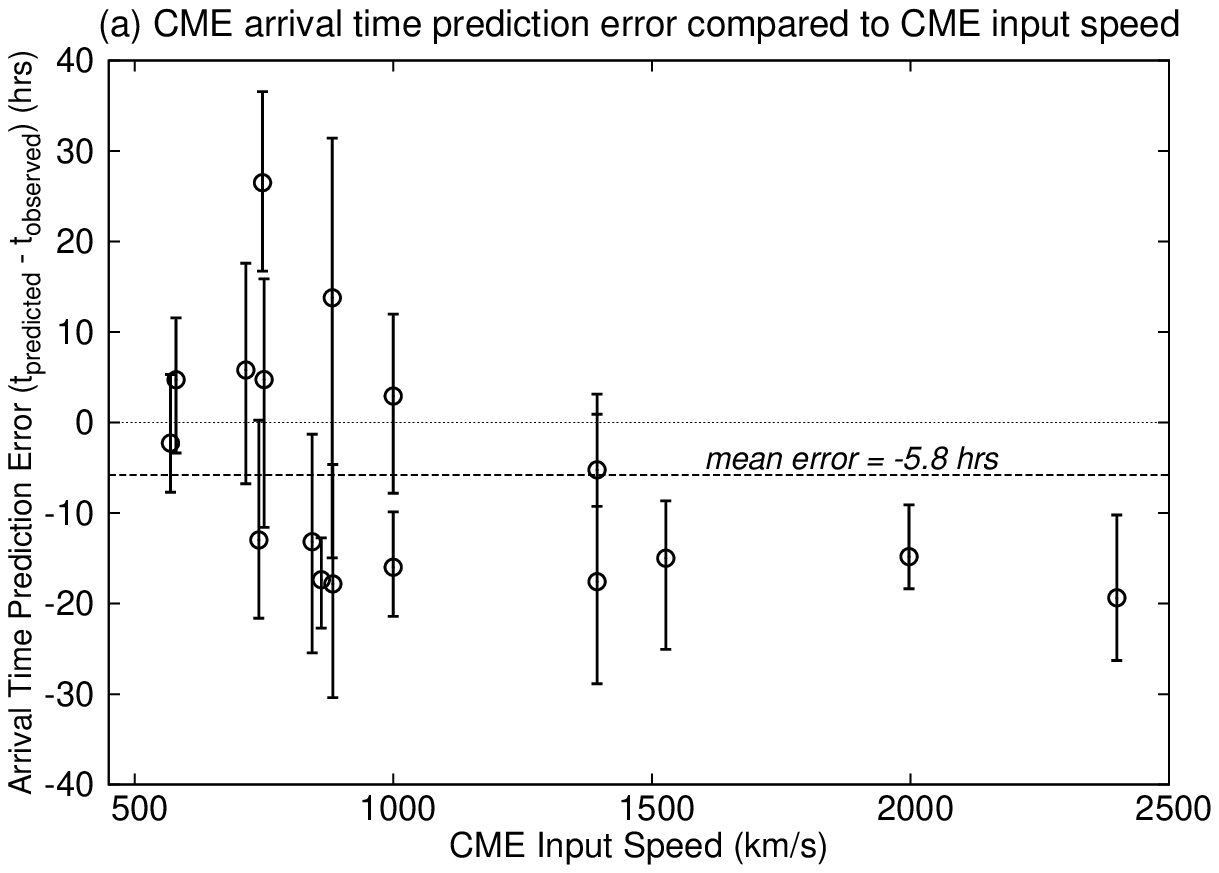}\includegraphics[width=0.5\textwidth,angle=0,origin=c]{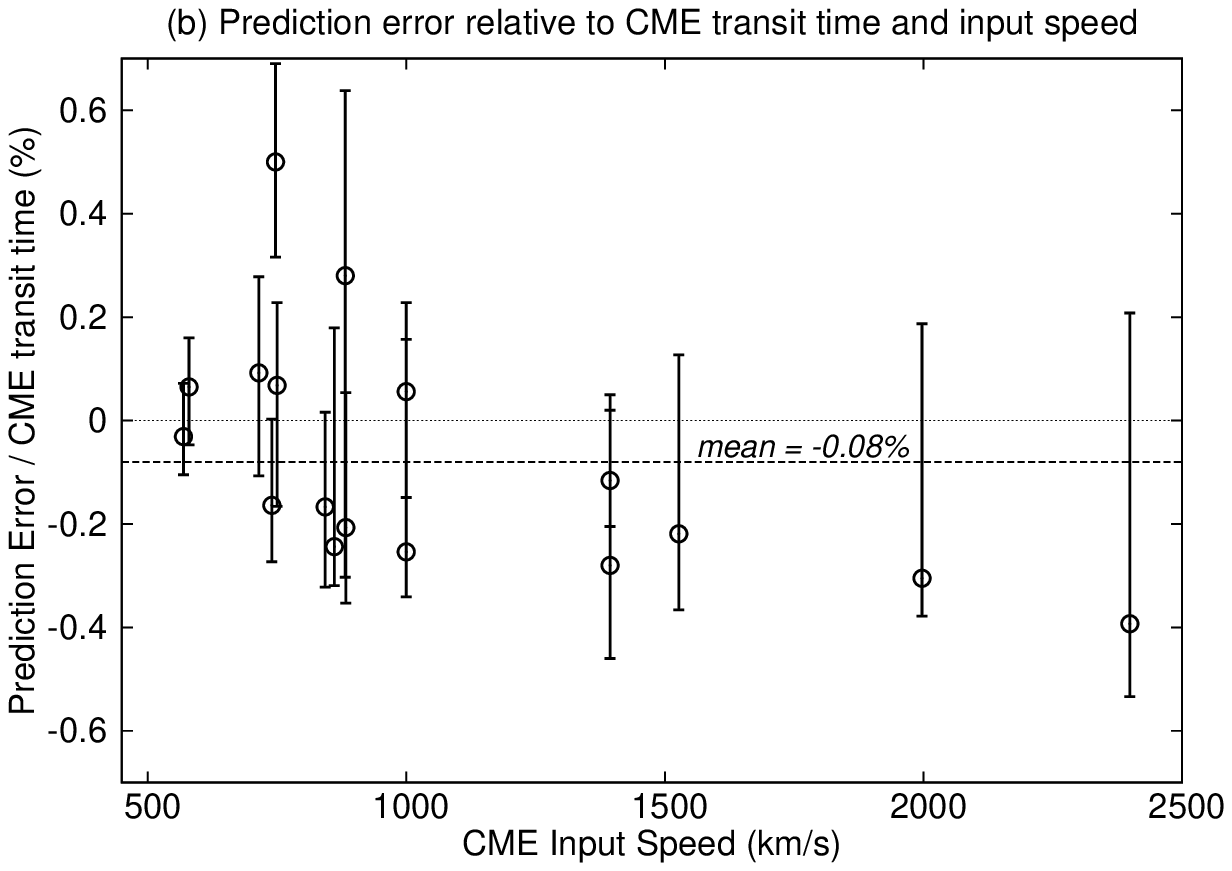}}
\caption{(a) CME arrival time prediction error plotted against the CME input speed.  (b) Prediction error as a percentage of the CME transit time, plotted against the CME input speed.  The error bars are computed using the predicted ensemble range as listed in Table \protect\ref{table:stats}.}\label{fig:relerr}
\end{figure}

To begin with a simple forecast evaluation of CME arrival time, the ensemble mean can be taken as a single forecast.  Using the prediction error  $\Delta t_{\rm err}=t_{\rm predicted}-t_{\rm observed}$ (last column of Table \ref{table:stats}, the mean absolute error (MAE) is 12.3 hours, the Root Mean Square Error (RMSE) is 13.9 hours, and the mean error (ME) is -5.8 hours (early) for all 17 ensembles containing hits. Considering the sample size in this study, these errors are comparable to CME arrival time prediction errors (a RMSE of $\sim$ 10 hours) reported by others \cite{millward2013,romano2013,vrsnak2014,mays2014b}. Similarly, \inlinecite{colaninno2013} used a variety of methods to evaluate CME arrival time predictions (non real-time) based on imaging data analysis only, and found an error $\pm$6 hours for 78\% of their sample, and $\pm$13 hours for their full sample of 9 CMEs. The CME arrival time prediction error is inevitably related to the CME propagation speed, thus it is useful to consider the input speed and in-situ observed transit time relative to the prediction error.  For this sample, the average in-situ observed transit time was 66 hours.  In Figure \ref{fig:relerr}a the CME arrival time prediction error is plotted against the CME input speed, and in Figure \ref{fig:relerr}b the prediction error as a percentage of the CME transit time is plotted against the CME input speed.  The error bars are computed using the predicted ensemble range as listed in column 10 of Table \ref{table:stats}.  The dashed horizontal line indicates the mean arrival time prediction error (a) and mean of the prediction error/transit time percentage (b).   These figures show a nearly consistent negative prediction error for fast CMEs  above $\sim$1000 km/s such that these fast CMEs are generally predicted to arrive earlier than they are observed.  This could be  a sign of the modeled CME having too much momentum as defined by a combination of the input speed and half width (which is related to the modeled CME mass). The overestimation of the modeled CME velocity compared to in-situ observed values is also due to the modeled CME having a lower magnetic pressure than is observed in typical magnetic clouds. 

\begin{landscape}
\scriptsize
\begin{longtable}{lcrrrrrcccr|ccr|r} 
\captionsetup{width=1.0\textwidth}
\caption{Summary of the ensemble simulation results for 35 CME events (January 2013 - June 2014). 
Columns 1-2: CME onset date and time. Column 3: SOHO LASCO CME Catalog plane-of-sky (POS) speed for 2013. Columns 4-7: median ensemble CME input parameters of $v$, latitude, longitude (HEEQ), and half-width ($w/2$). Columns 8-10: mean predicted arrival time of all $n_{\rm tot}$ ensemble members, and the spread in arrival times in hours relative to the mean.  Column 11: $n_{\rm predicted~ hits}$, the number of ensemble members predicting that the CME will arrive at Earth out of $n_{\rm tot}$, the total number of ensemble members. Columns 12-14: actual arrival time observed in-situ, and the observed CME transit time relative to the CME start time.  Column 15: prediction error $\Delta t_{\rm err}=t_{\rm predicted}-t_{\rm observed}$ for hits, or CR and FA for correct rejections and false alarms.}\label{table:stats}\\[-3pt]
 \hline
\multicolumn{2}{c}{CME Onset} & SOHO & \multicolumn{4}{r}{Median CME parameters} &  \multicolumn{3}{c}{Mean Predicted Arrival}  & $p_i=$~~ & \multicolumn{3}{r|}{In-situ Arrival~~~~~Transit}&\\
\multicolumn{2}{c}{Date  ~~~~~ Time} & $v_{\rm POS}$ & $v$ & Lat & Lon & $w/2$ &  \multicolumn{3}{c}{Date  ~~~~~~Time~~Spread} & $n_{\rm hits}/$ & \multicolumn{3}{r|}{Date ~~~~~ Time ~~~~~Time} & $\Delta t_{\rm err}$\\
\multicolumn{2}{c}{{\scriptsize(yyyy-mm-dd)}~~~~{\scriptsize(UT)}~~} & {\scriptsize(km/s)} & {\scriptsize(km/s)} & {\scriptsize($^{\circ}$)}  & {\scriptsize($^{\circ}$)} &  {\scriptsize($^{\circ}$)}  & \multicolumn{2}{c}{{\scriptsize(UT)}}& {\scriptsize(h)}& ~$n_{\rm tot}$ & \multicolumn{3}{r|}{{\scriptsize(UT) ~~~~~~~~~~~~~~ (h)~~}} & {\scriptsize(h)} \\ 
\hline
\endfirsthead
\multicolumn{15}{c}
{{\bfseries \tablename\ \thetable{}.}: Continued from previous page}\\
\hline
\multicolumn{2}{c}{CME Onset} & SOHO & \multicolumn{4}{r}{Median CME parameters} &  \multicolumn{3}{c}{Mean Predicted Arrival}  & $p_i=$~~ & \multicolumn{3}{r|}{In-situ Arrival~~~~~Transit}&\\
\multicolumn{2}{c}{Date  ~~~~~ Time} & $v_{\rm POS}$ & $v$ & Lat & Lon & $w/2$ &  \multicolumn{3}{c}{Date  ~~~~~~Time~~Spread} & $n_{\rm hits}/$ & \multicolumn{3}{r|}{Date ~~~~~ Time ~~~~~Time} & $\Delta t_{\rm err}$\\
\multicolumn{2}{c}{{\scriptsize(yyyy-mm-dd)}~~~~{\scriptsize(UT)}~~} & {\scriptsize(km/s)} & {\scriptsize(km/s)} & {\scriptsize($^{\circ}$)}  & {\scriptsize($^{\circ}$)} &  {\scriptsize($^{\circ}$)}  & \multicolumn{2}{c}{{\scriptsize(UT)}}& {\scriptsize(h)}& ~$n_{\rm tot}$ & \multicolumn{3}{r|}{{\scriptsize(UT) ~~~~~~~~~~~~~~ (h)~~}} & {\scriptsize(h)} \\ 
\hline
\endhead

\hline 
\multicolumn{15}{l}{{{\scriptsize $^*$  CMEs was simulated together with the CME listed on the previous row as part of a single ensemble.}}}\\ [-2pt]
\multicolumn{15}{l}{{{\scriptsize $\dagger$ 2nd-order plane-of-sky speed at last possible measured height.}}}\\ [-2pt]
\multicolumn{15}{l}{{{\scriptsize $\ddagger$  In-situ signature could not be unambiguously identified as arrival of CME-related disturbance and is not included in forecast verification.}}}\\ [-2pt]
\multicolumn{15}{c}{{Continued on next page}}\\ 
\endfoot

\hline
\multicolumn{15}{l}{{{\scriptsize $^*$  CMEs was simulated together with the CME listed on the previous row as part of a single ensemble.}}}\\ [-2pt]
\multicolumn{15}{l}{{{\scriptsize $\ddagger$ In-situ signature could not be unambiguously identified as arrival of CME-related disturbance and is not included in forecast verification.}}}\\ 
\endlastfoot
2013-01-13 & 07:24 & 229$^{\dagger}$ & 342 & 1 & 10 & 28 & 2013-01-17 & 06:30 & $^{+5.6}_{-6.2}$ & 20/48 & 2013-01-16 & 23:25$^{\ddagger}$ &\dotfill &\dotfill \\
2013-01-16 & 19:00 & 616~ & 750 & -26 & 52 & 42 & 2013-01-19 &  21:33 & $^{+11.1}_{-16.3}$ & 18/48 & 2013-01-19 & 16:48 & 69.8 & 4.8 \\
2013-02-06 & 00:24 & 1851~ & 1460 & 30 & -26 & 30 & 2013-02-08 & 05:37 & $^{+8.7}_{-6.8}$ & 19/48 & 2013-02-08 & 03:15$^{\ddagger}$ & \dotfill & \dotfill \\
2013-04-11 & 07:24 & 819~ & 1000 & 0 & -15 & 55 & 2013-04-13 & 06:14 & $^{+6.1}_{-5.4}$ & 36/36 & 2013-04-13 & 22:13 & 62.8 & -16.0 \\
2013-06-21 & 03:12 & 1903~ & 1997 & -15 & -48 & 60 & 2013-06-22 & 13:02 & $^{+5.7}_{-3.6}$ & 47/48 & 2013-06-23 & 03:51 & 48.7 & -14.8 \\
2013-06-30 & 02:24 & 410$^{\dagger}$ & 386 & 9 & 4 & 34 & 2013-07-02 & 20:56 & $^{+1.1}_{-0.6}$ & 4/36 & \dotfill & \dotfill & \dotfill & CR \\
2013-07-16 & 04:00 & 639~ & 795 & -19 & 9 & 19 & 2013-07-18 & 20:29 & $^{+6.9}_{-6.2}$ & 28/48 & 2013-07-18 & 12:55$^{\ddagger}$ & \dotfill &  \dotfill \\
2013-08-02 & 13:24 & 443~ & 596 & -16 & 28 & 13 & \dotfill & \dotfill & \dotfill & 0/24 & \dotfill & \dotfill & \dotfill & CR \\
2013-08-07 & 18:24 & 473~ & 570 & -25 & 11 & 44 & 2013-08-11 & 05:03 & $^{+6.8}_{-5.9}$ & 48/48 & \dotfill &  \dotfill &  \dotfill &  FA\\
2013-08-08 & 23:54 & 411$^{\dagger}$ & 454 & -17 & 14 & 18 & \dotfill & \dotfill & \dotfill & 0/48 & \dotfill & \dotfill & \dotfill &  CR\\
2013-08-30 & 02:48 & 884~ & 861 & 21 & -48 & 59 & 2013-09-01 & 08:34 & $^{+4.6}_{-5.4}$ & 46/48 & 2013-09-02 & 01:56 & 71.1 &  -17.4 \\
2013-09-19 & 03:36 & 449$^{\dagger}$ & 362 & -15 & -43 & 27 & \dotfill & \dotfill & \dotfill & 0/24 & \dotfill & \dotfill & \dotfill &  CR\\
2013-09-29 & 20:40 & 1164~ & 1000 & 26 & 30 & 66 & 2013-10-02 & 04:11 & $^{+9.1}_{-10.8}$ & 36/36 & 2013-10-02 & 01:15 & 52.6 & 2.9 \\
2013-10-06 & 14:39 & 710$^{\dagger}$ & 747 & 1 & 2 & 16 & 2013-10-09 & 22:10 & $^{+10.1}_{-9.8}$ & 22/24 & 2013-10-08 & 19:40 & 53.0 & 26.5 \\
2013-10-22 & 04:24 & 697~ & 764 & 51 & -10 & 49 & 2013-10-25 & 08:19 & $^{+10.2}_{-10.9}$ & 45/47 & \dotfill & \dotfill & \dotfill &  FA \\
2013-12-04 & 23:12 & 585~ & 697 & 41 & -9 & 46 & 2013-12-07 & 13:45 & $^{+4.1}_{-4.1}$ & 37/48 & \dotfill & \dotfill & \dotfill & FA \\[-3pt]
2013-12-05$^*$ & 00:00 & 623$^{\dagger}$ & 651 & 25 & 63 & 31 &  &  &   &  & & & &  \\
2013-12-12 & 03:36 & 943~ & 1067 & -32 & 51 & 50 & 2013-12-14 & 18:11 & $^{+16.1}_{-13.9}$ & 36/48 &  2013-12-15 & 16:30$^{\ddagger}$  &  \dotfill&   \dotfill\\[-3pt]
2013-12-12$^*$ & 06:24 & 695~ & 694 & -52 & 8 & 50 & & &  & & & & &  \\
2013-12-29 & 00:12 & 296$^{\dagger}$ & 682 & 32 & 8 & 47 & 2014-01-01 & 02:39 & $^{+13.2}_{-9.3}$ & 48/48 &  \dotfill &  \dotfill &  \dotfill &  FA \\[-3pt]
2013-12-29$^*$ & 05:48 & 477~ & 495 & -33 & -58 & 43 & & &  & 48/48 & & & &  \\
\pagebreak
2014-01-07 & 18:24 & 1714& 2399 & -28 & 38 & 64 & 2014-01-09 & 00:17 & $^{+9.2}_{-6.9}$ & 48/48 & 2014-01-09 & 19:39 & 49.3 & -19.4 \\
2014-01-30 & 16:24 &  940& 843 & -50 & -28 & 45 & 2014-02-02 & 10:10 & $^{+11.9}_{-12.3}$ & 13/24 & 2014-02-02 & 23:20 & 78.9 & -13.2 \\
2014-01-31 & 15:39 & 370& 718 & 12 & -29 & 40 & 2014-02-03 & 17:20 & $^{+9.1}_{-6.0}$ & 12/12 & \dotfill & \dotfill & \dotfill & FA\\
2014-02-04 & 01:09 & 501& 778 & -35 & 21 & 49 &  2014-02-06  & 20:39 & $^{+22.8}_{-18.03}$ & 23/24 & 2014-02-07 & 16:28$^{\ddagger}$ & \dotfill &  \dotfill\\[-3pt]
2014-02-04$^*$ & 16:24 & 323& 560 & -34 & 23 & 42 & & &  & & & & &  \\
2014-02-12 & 05:39 & 494& 740 & 8 & 5 & 59 & 2014-02-14 & 23:47 & $^{+13.2}_{-8.7}$ & 48/48 & 2014-02-15 & 12:46 & 79.1 &  -13.0\\
2014-02-18 & 01:25 & 712& 882 & -24 & -43 & 52 & 2014-02-20 & 16:29 & $^{+17.7}_{-28.7}$ & 29/36 & 2014-02-20 & 02:42 & 49.3 &  13.8 \\
2014-02-19 & 16:00 & 510& 883 & -32 & -10 & 29 & 2014-02-22 & 12:20 & $^{+13.2}_{-12.6}$ & 32/36 & 2014-02-23 & 06:09 & 86.2 &  -17.8\\
2014-02-25 & 01:09 & 2069& 1394 & -18 & -80 & 78 & 2014-02-26 & 22:15 & $^{+20.7}_{-11.3}$ & 40/48 & 2014-02-27 & 15:50 & 62.7 &  -17.6 \\
2014-03-23 & 03:36 & 834& 715 & -5 & -60 & 55 & 2014-03-26 & 00:58 & $^{+11.8}_{-12.6}$ & 38/48 & 2014-03-25 & 19:10 & 63.4 &  5.80\\
2014-03-23 & 06:12 & 536& 503 & 37 & -45 & 34 & \dotfill &\dotfill & \dotfill & 0/12 & \dotfill & \dotfill & \dotfill & CR \\
2014-03-29 & 18:12 & 505& 707 & 36 & 41 & 43 & 2014-04-01 & 21:30 & $^{+1.0}_{-1.0}$ & 2/36 & \dotfill & \dotfill & \dotfill &  CR \\
2014-04-02 & 13:36 & & 1527 & 19 & -55 & 51 & 2014-04-04 & 19:01 & $^{+6.3}_{-10.1}$ & 14/16 & 2014-04-05 & 10:00 & 68.4  &  -15.0\\
2014-04-18 & 13:09 & & 1394 & -35 & 9 & 46 & 2014-04-20 &  05:07 & $^{+6.1}_{-4.0}$ & 36/36 & 2014-04-20 & 10:20 & 45.2 &  -5.2\\
2014-06-04 & 15:48 & & 580 & -40 & -28 & 50 & 2014-06-07 & 20:56 & $^{+6.8}_{-8.1}$ & 22/36 & 2014-06-07 & 16:12 & 72.4 &  4.7\\
2014-06-10 & 13:09 & & 980 & -9 & -89 & 64 & 2014-06-12 & 20:28 & $^{+3.1 }_{-3.1}$ & 2/36 & \dotfill &\dotfill & \dotfill &  CR\\
2014-06-19 & 17:12 & & 569 & 3 & -20 & 44 & 2014-06-22 & 16:12 & $^{+7.6}_{-5.4}$ & 12/12 & 2014-06-22 & 18:28  & 73.3 & -2.3\\
2014-06-30 & 07:24 & & 751 & -12 & -63 & 29 & 2014-07-02 & 20:32 & $^{+7.4}_{-12.3}$ &  0/36 & \dotfill & \dotfill & \dotfill &  CR\\
 \hline
\end{longtable}
\normalsize
\end{landscape}


\begin{table}
\caption{Forecast performance contingency table for 30 ensembles.}\label{table:cont}
\begin{tabular}{r|cc}
\hline
 & \multicolumn{2}{|c}{CME arrival forecast}\\
Observation  & Will occur & Will not occur\\
\hline
 Occurs & Hit (17) & Miss (0)\\
 Does not occur & False alarm (5) & Correct rejection (8) \\
\hline
\end{tabular}
\end{table}

Ensemble modeling produces a probabilistic forecast $p_i$ of the likelihood of CME arrival for each ensemble $i$, but we begin with a more simple forecast evaluation by binning the probability $p_i$ into a categorical yes/no forecast.  Categorical forecasts only have two probabilities, zero and one. Therefore we start by binning the probability forecast $p$ into two categories: ``yes'' the CME will arrive, and ``no'' the CME will not arrive. In the signal detection theory model of weather forecasting, event forecasting performance can be evaluated in terms of a 2$\times$2 contingency table, as shown in Table \ref{table:cont} \cite{harvey1992,weigel2006,jolliffe2011}.  For CME arrival prediction, the ``event'' is taken as the ``CME arrival''.  Hits are then defined as CME arrivals which were both predicted and observed to occur.  Misses are defined as CME arrivals which were not predicted, but were observed to occur.  False alarms (FA) are defined as CME arrivals that were predicted to occur, but were observed not to occur.  And correct rejections (CR) are CME arrivals that were not predicted, and were observed not to occur.  To bin each ensemble's probabilistic forecast, correct rejections were identified when the criterion of the forecast probability $p_i=n_{\rm predicted ~ hits}/n_{\rm total~ members} < $ 15\% was met; i.e., that less than 15\% of the total predictions in the ensemble indicated CME arrival.  Similarly, the inverse criterion is used to identify hits.   Table \ref{table:cont} shows the contingency table definitions and values for this 30 event sample: 17 hits, 8 correct rejections, 5 false alarms, and 0 misses (see Table \ref{table:stats} for specific CR and FA events).  For this sample zero misses indicates that there were no ensemble simulations which did not predict CME arrivals which were observed to occur. There were 8 out of 30 correct rejections and 5 false alarms for events that were not observed in-situ, giving a correct rejection and false-alarm rate of 62\% (8/ 13) and 38\% (5/ 13) respectively. The correct alarm ratio, defined as the number of hits over the number of hits and false alarms, is 77\% and the false alarm ratio is 23\%.  

\begin{figure} 
\centerline{\includegraphics[width=0.45\textwidth,angle=0,origin=c]{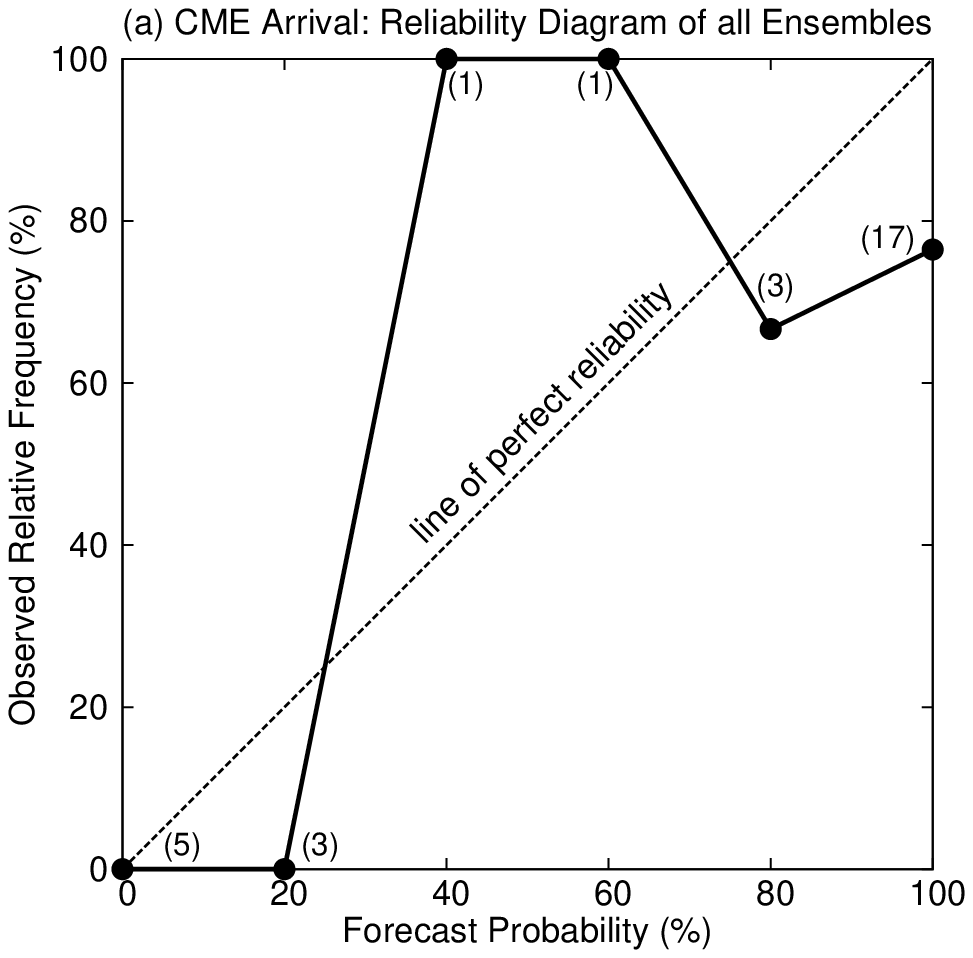}\includegraphics[width=0.45\textwidth,angle=0,origin=c]{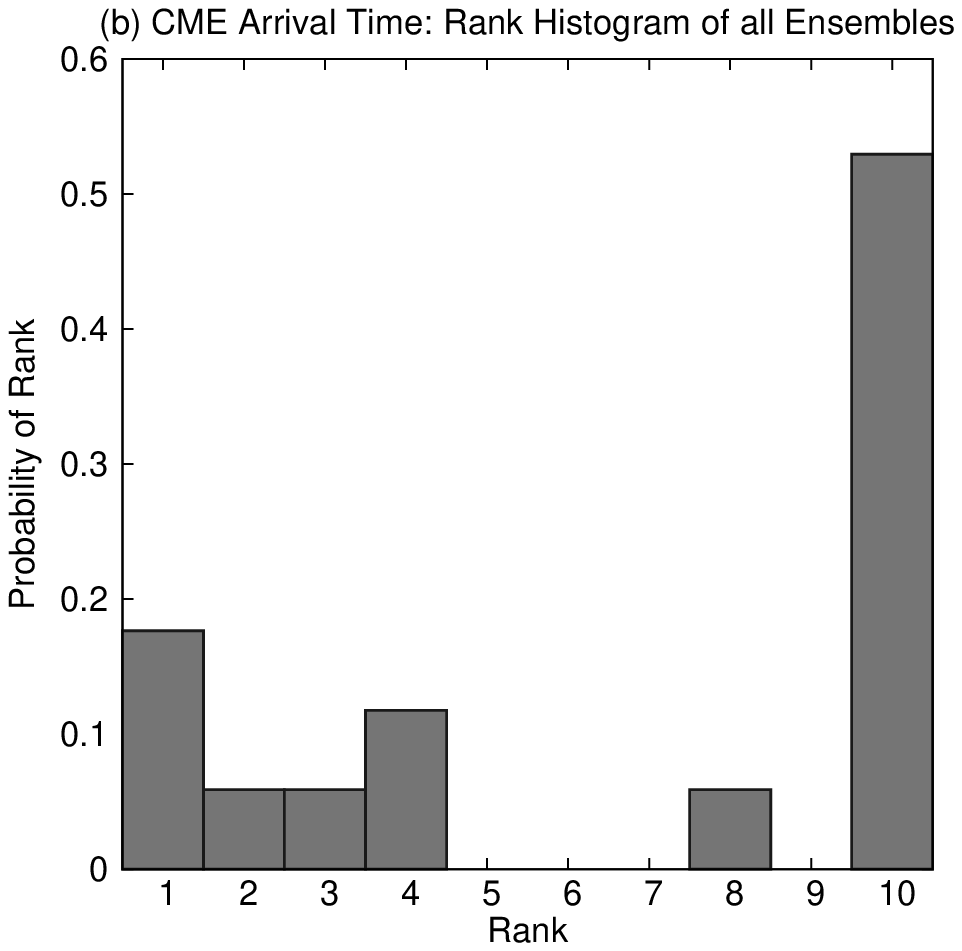}}
\caption{CME arrival time forecast verification: (a) Reliability diagram of the forecast probability of CME arrival for the 30 ensemble sample, with the ensemble results shown as the black line with points and the diagonal perfect reliability as a dotted black line.  The number of ensembles used in each calculation is shown next to each point.  The diagram indicates underforecasting in the forecast bins between 20-80\%, and slight overforecasting in the 1-20\% and 80-100\% forecast bins. Overforecasting is when the forecast probability of CME arrival is higher than observed; i.e. the CME is observed to arrive less often than is predicted.  Similarly, underforecasting is when the CME arrival forecast probability is lower than observed; i.e. the CME is observed to arrive more often than is predicted. (b) Rank histogram for the 17 ensembles containing hits indicates undervariability of initial conditions.}\label{fig:rank}
\end{figure}

Let us now consider a more nuanced technique to evaluate the probabilistic forecast without partitioning it into a categorical forecast with only two probabilities as described above.   A method defining the magnitude of probability forecast errors is the Brier Score ($BS$) \cite{brier1950,murphy1973,wilks1995}, defined as

\begin{equation}
BS={\frac{1} {N}{\sum\limits_{i = 1}^N {(p_{i} - o_{i} } })^{2} },\label{eq:bs}
\end{equation}
where $N$ is the number of events, $p_i$ is the forecast probability of occurrence for event $i$, and $o_i$ is 1 if the event was observed to occur and 0 if it did not occur.   For CME arrival prediction, the ``event'' here is taken as the ``CME arrival'' and $p_i$ is listed in column 11 of Table \ref{table:stats} for each ensemble.   This score is a probability  mean square error which weights larger errors more than small ones and ranges from 0 to 1, with 0 being a perfect forecast. The $BS$ computed from all $N$= 30 ensemble CME arrival probabilities (Table \ref{table:stats}, column 11) is 0.15, which indicates that in this sample, on average, the probability $p$ of the CME arriving is fairly accurate.  However, such verification scores reduce the problem to a single measure which can only consider one dimension, whereas there are many dimensions to the system.  For example, consider the aspect of forecast reliability.  Reliable forecasts are those where the observed frequencies of events are in agreement with the forecast probabilities.  To evaluate the reliability of probabilistic ensemble forecasts, a set of probabilistic forecasts $p_i$ must be evaluated using observations that demonstrate that those events either occurred or did not occur.  Multiple forecasts must be evaluated because a single probabilistic forecast cannot be simply assessed as ``right'' or ``wrong'' e.g. if a forecast suggests a 30\% chance of CME arrival, and the CME does arrive, the forecast is not clearly either ``right'' or ``wrong''. Therefore, to provide forecast verification for a $p$=30\% chance of CME arrival one would need to compile the statistics of observed CME arrivals for a set of forecasts that predicted a 30\% chance of arrival. In this way a reliability diagram can be constructed to determine how well the predicted probabilities of an event correspond to their observed frequencies \cite{wilks1995,jolliffe2011}. Figure \ref{fig:rank}a shows the reliability diagram of the likelihood of CME arrival forecast for the 30 event sample, with the reliability for this sample shown as the black line with points and the perfect reliability diagonal as a dotted line.  The line of perfect reliability is diagonal because, for example, when a 60\% probability forecast is made, it is considered perfectly reliable if the event is observed to occur 60\% of the time over multiple ensemble forecasts. The number of events used in each calculation is shown next to each point, and the sample size is smaller than needed for a robust diagram.  Nevertheless, the diagram shows that overall ensemble modeling is underforecasting in the forecast bins between 20-80\%, and slightly overforecasting in the 1-20\% and 80-100\% forecast bins. Overforecasting is when the forecast chance of CME arrival (forecast probability)  is higher than is actually observed; i.e., the CME is observed to arrive less often than is predicted.  Similarly, underforecasting is when the chance of CME arrival is lower than is actually observed; i.e., the CME is observed to arrive more often than is predicted.

Another aspect of forecast reliability is to assess how well the ensemble spread of the forecast represents the true variability of the observations.  For 8 out of 17 of the ensemble runs containing hits the observed CME arrival was within the spread of ensemble arrival time predictions.  This indicates that roughly half of the observations fall outside of the extremes of the predicted ensemble spread.  However, one aspect of a reliable forecast is that the set of ensemble member forecast values for a given event and observations should be considered as random samples from the same probability distribution. This reliability then implies that if an $n$ member ensemble and the observation are sorted from earliest to latest arrival times, the observation is equally likely to occur in each of the $n+1$ possible ``ranks''. Therefore a histogram of the rank of the observation, ``rank histogram'', tallied over many events should show be uniform (flat) \cite{anderson1996,hamill1997,talagrand1997}. While more samples would be desirable, it is still instructive to examine the rank histogram for the CME arrival time predictions from the 17 ensembles containing hits in this sample, shown in Figure \ref{fig:rank}b.  Since each ensemble run in our sample does not have the same number of members, the rank has been normalized to 10 (9 member ensemble). To construct this rank histogram the CME arrival time predictions of each ensemble are sorted from earliest to latest and the rank of where the observed arrival falls among the predicted times is noted. For example, an ensemble with a rank of 8 has the meaning that 7 arrival time predictions fall before the observed arrival, a rank of 10 would mean that all 9 predictions occur before the observation, and a rank of 1 means that the observation occurs before all of the predictions. The non-uniform U-shape of this histogram partly illustrates that roughly half of the observed arrivals are outside the spread of predictions (ranks 1 and 10), with a tendency for an overall early spread of predictions (rank=10) compared to observations (also quantified by mean arrival time error of -7.0 hours).   U-shaped rank histograms can indicate lack of variability in the ensemble, but can also be a sign of a combination of conditional biases in the model \cite{hamill2001}.  However, when evaluating the WSA-ENLIL+Cone model in this sample of ensembles, and $>$70 regular runs containing hits performed by SWRC \cite{romano2013}, an overall negative bias (early predictions) was found, with less bias for CME input speeds below $\sim$1000 km/s.  Therefore, it is unlikely that a combination of positive and negative model biases within the ensembles contributed to the U-shaped rank histogram for our sample.  Most likely, the U-shape suggests undervariability, indicating that these ensembles to not sample a  wide enough spread in CME input parameters.

\subsection{$K_P$ FORECAST VERIFICATION}\label{kpver}
For each event for which a hit is predicted in Table \ref{table:stats}, ensemble modeling provides a probabilistic $K_P$ forecast (see Section \ref{method}) for three magnetic field clock angles scenarios of 90$^{\circ}$ (westward), 135$^{\circ}$ (south-westward), 180$^{\circ}$ (southward).  An overall probabilistic $K_P$ forecast can then be obtained by making the simple assumption that each clock angle is equally likely to occur. Table \ref{table:kp} lists the overall probabilistic $K_P$ forecast $p({{\rm K_P}=b})$ for each $K_P$ bin $b$ (e.g. the distribution shown in Figure \ref{fig:kp1} in black) for these 17 events. The observed $K_P$, sudden storm commencement (SSC) and minimum $Dst$ indices are also shown.  The mean predicted $K_P$ is listed in column 12, along with the overall predicted $K_P$ spread (using plus or minus notation). Underlined $K_P$ probabilities indicate that the NOAA real-time observation falls within this bin, and the final definitive $K_P$ values are listed in column 13. The $Dst$ values are from the real-time (quicklook) $Dst$ index provided by the World Data Center for Geomagnetism in Kyoto, Japan. In order to estimate the reliability of the probabilistic $K_P$ forecast, the Brier Score is calculated for each $K_P$ bin and listed on the last line of the table.  

To evaluate forecast performance, a single categorical predicted $K_P$ forecast can be derived from the probabilistic $K_P$ forecast $p(K_P=b)$ distribution. For example, the single categorical forecast ${K_P}_{\rm predicted}$ can be taken as the mean predicted $K_P$, or the most probable $K_P$ value. This allows a $K_P$ prediction error to be computed as $\Delta {K_P}_{\rm err}={K_P}_{\rm predicted}-{K_P}_{\rm observed}$ for each ensemble, where positive values of $\Delta {K_P}_{\rm err}$ indicate an over prediction of the $K_P$ index and negative values indicate that $K_P$ has been under predicted.  If the categorical ${K_P}_{\rm predicted}$ is taken as the $K_P$ bin $b$ which has the highest likelihood in the probabilistic $K_P$ forecast $p(K_P=b)$ for each ensemble, the prediction errors are calculated to give a mean absolute error (MAE) of 1.9, Root Mean Square Error (RMSE) of 2.5, and mean error (ME) of +1.4.   However, if the categorical ${K_P}_{\rm predicted}$ is taken as the mean predicted $K_P$ in each ensemble (last column of  Table \ref{table:kp}) these errors are reduced to MAE=1.5, RMSE=2.0, and ME=+0.6.  Consequently, utilizing the ensemble mean $K_P$ yields a more accurate forecast in this sample, however both forecast choices show an overall tendency for the overprediction of $K_P$. Given that the modeled CMEs do not have an internal magnetic field structure, the \inlinecite{newell2007} $K_P$ coupling function using ENLIL results as input performs surprisingly well.  For comparison, using ACE solar wind data as input to the coupling function for this sample gives $K_P$ prediction errors of MAE=0.67, RMSE=0.77, and ME=+0.22. 

\begin{figure} 
\centerline{\includegraphics[width=0.6\textwidth,angle=0,origin=c]{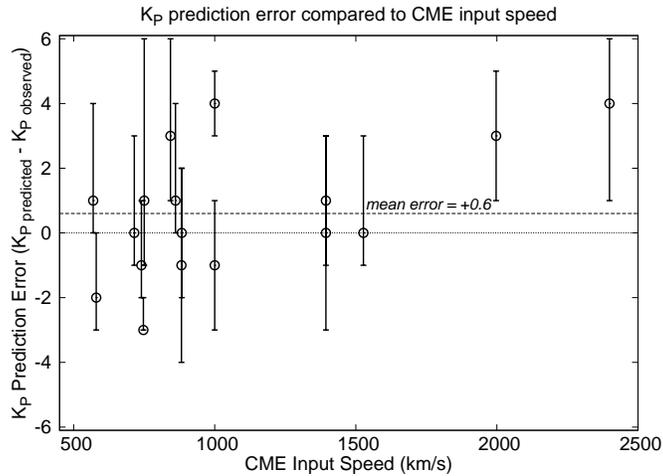}}
\caption{$K_P$ prediction error (computed from the ensemble mean $K_P$) compared to the CME input speed shows an overprediction of the $K_P$ value for  CME input speeds above $\sim$1000 km/s. Error bars indicate the ensemble $K_P$ prediction spread listed in \protect\ref{table:kp}.}\label{fig:kpspeed}
\end{figure}

In Figure \ref{fig:kpspeed} the $K_P$ prediction error (from the ensemble mean $K_P$) is compared to the CME input speed; the error bars show the ensemble $K_P$ prediction spread.  This figure shows that $K_P$ is usually overpredicted when  CME input speeds are above $\sim$1000 km/s.   This bias is also apparent in the $K_P$ predictions made from a sample of $>$70 regular WSA-ENLIL+Cone runs reported by \inlinecite{romano2013}.  The $K_P$ overprediction is most likely due to an overestimation of the CME dynamic pressure at Earth by the WSA-ENLIL+Cone model, due to the CME having a lower magnetic pressure than is observed in typical magnetic clouds. Also since the CME dynamic pressure is linearly related to the density and the square of the velocity, this quantity will be in particular more sensitive to higher CME input speeds, and produce higher in-situ speeds than those measured. Another factor in the higher CME dynamic pressure can arise from the approximation of the CME as a cloud with homogeneous density in the model.   

\setlength{\tabcolsep}{.43em}
\begin{table}
 \caption{Summary of $K_P$ prediction results for 17 ensemble runs containing hits.  Columns 1-2: CME start date and time. Columns 3-11: overall probabilistic $K_P$ forecast for each $K_P$ bin assuming equal likelihood of three clock angle scenarios.  Underlined $K_P$ probabilities indicate that the NOAA real-time $K_P$ observation falls in this bin and the observed definitive $K_P$ is in column 13. The mean predicted $K_P$ is listed in column 12, along with the overall predicted $K_P$ spread (using plus or minus notation). The Brier Score $BS$ is calculated for each $K_P$ bin and listed on the last line of the table. The $Dst$ sudden storm commencement and minimum values are listed in the last two columns. }\label{table:kp}
 \begin{tabular}{llrrrrrrrrrc|lrr} 
 \hline
\multicolumn{2}{c}{CME Onset}& \multicolumn{9}{c}{\scriptsize Binned Probabilistic $K_P$ Forecast (\%)} & \multicolumn{1}{c|}{\scriptsize Mean $K_P$}& \multicolumn{1}{c}{Obs.} & \multicolumn{2}{c}{$Dst$ (nT)}\\
\multicolumn{2}{c}{Date ~~ Time {\scriptsize(UT)}} & 1 & 2 & 3 & 4 & 5 & 6 & 7 & 8 & 9 & \multicolumn{1}{c|}{\scriptsize \& spread} & \multicolumn{1}{c}{$K_P$} & \multicolumn{1}{c}{SSC} &\multicolumn{1}{c}{min.}\\ 
\hline
2013-01-16 & 19:00&  0&  13&  \uline{26}&  28&  6&  11&  9&  4&  4&    4 $^{+5}_{-2}$ &   4- &   +6 & -34  \\
2013-04-11 & 07:24&  0 &  0&  \uline{0}&  0&  0&  33&  5&  62&  0&    7 $^{+1}_{-1}$ &  3+  & +21     & -7 \\
2013-06-21 & 03:12&  0 &  0&  0& \uline{0}&  4&  16&  23&  43&  15&    7 $^{+2}_{-2}$ &  5+  & \dotfill   & -49\\
2013-08-30 & 02:48&  0 &  0&  \uline{6}&  31&  28&  33&  2&  0&  0&    4 $^{+3}_{-1}$ &  3+  &  \dotfill    & -31\\
2013-09-29 & 20:40&  0 &  0&  6&  26&  24&  \uline{39}&  5&  0&  0&    5 $^{+2}_{-2}$ &  8-  & +30     & -67\\
2013-10-06 & 14:39&  0 & 67 & 33 & 0 &  \uline{0} & 0 & 0 & 0 & 0  &  2 $^{+1}_{-0}$  & 6-   & +21     & -65\\
2014-01-07 & 18:24&  0 &  0&  \uline{0}&  6&  8&  19& 25 &  26&  16&    7 $^{+2}_{-3}$ & 3-   &  +2    & -14\\
2014-01-30 & 16:24&  0 &  \uline{0}&  15& 13&  33&  13&  18&  8&  0&    5 $^{+3}_{-2}$ & 2+   & +15     & -7 \\
2014-02-12 & 05:39&  0 &  0&  12&  25&  \uline{40}&  24&  0&  0&  0&    4 $^{+2}_{-1}$ & 5o   & +52     & -16 \\
2014-02-18 & 01:25&  0 &  1&  10&  21&  29&  \uline{26}&  10&  2&  0&    5 $^{+3}_{-3}$ &  6o  &  \dotfill    & -86\\
2014-02-19 & 16:00&  0 &  2&  30& \uline{34}&  28&  5&  0&  0&  0&    4 $^{+2}_{-2}$ &  4+  &  +4    & -56\\
2014-02-25 & 01:09&  0 &  0&  1&  11&  16&  \uline{21}&  22&  21&  9&    6 $^{+3}_{-3}$ &  5+   & \dotfill     & -99 \\
2014-03-23 & 03:48&  0 &  0&  16&  \uline{28}&  28&  24&  4&  0&  0&    4 $^{+3}_{-1}$ & 4-   & +20     & -18 \\
2014-04-02 & 13:36&  0 &  0&  21&  \uline{19}&  40&  12&  7&  0&  0&    4 $^{+3}_{-1}$ & 4o   & +16     & -16 \\
2014-04-18 & 13:09&  0 &  0&  0&  3&  \uline{27}&  17& 40 & 14 &  0&    6 $^{+2}_{-2}$ & 5o   & +25     & -24\\
2014-06-04 & 15:48&  0 & 0 & 18 & 29 & 36 & \uline{17} & 0 & 0 & 0 &    4 $^{+2}_{-1}$ & 6+   & +31     & -38\\
2014-06-19 & 17:12&  0 & 0 & \uline{8} & 31 & 33 & 25 & 3 & 0 & 0 &    4 $^{+3}_{-1}$ &  3o  &  +14    & -9\\
 \hline
\multicolumn{2}{c}{{\it Brier Score}} & 0.00 & 0.09 & 0.27 & 0.19 & 0.17 & 0.17 & 0.02 & 0.04 & 0.00 &  &     \\
\hline
 \end{tabular}
 \end{table}

\begin{figure} 
\centerline{\includegraphics[width=0.5\textwidth,angle=0,origin=c]{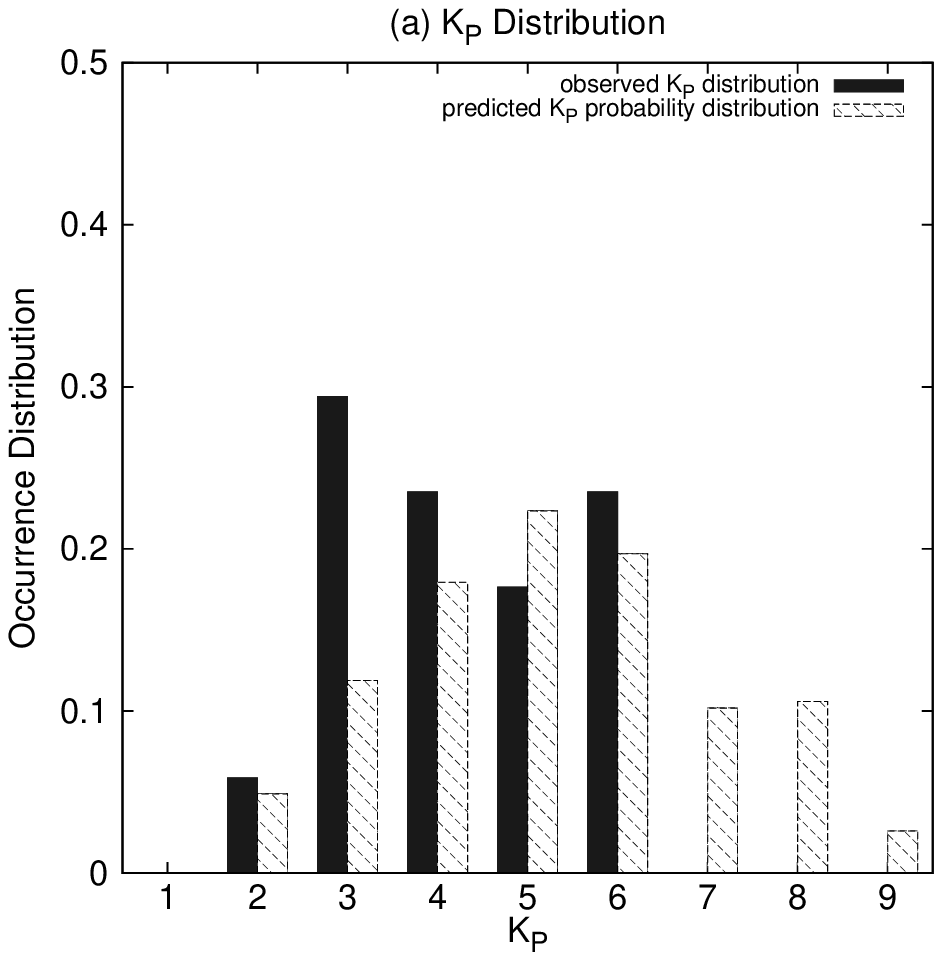}\includegraphics[width=0.5\textwidth,angle=0,origin=c]{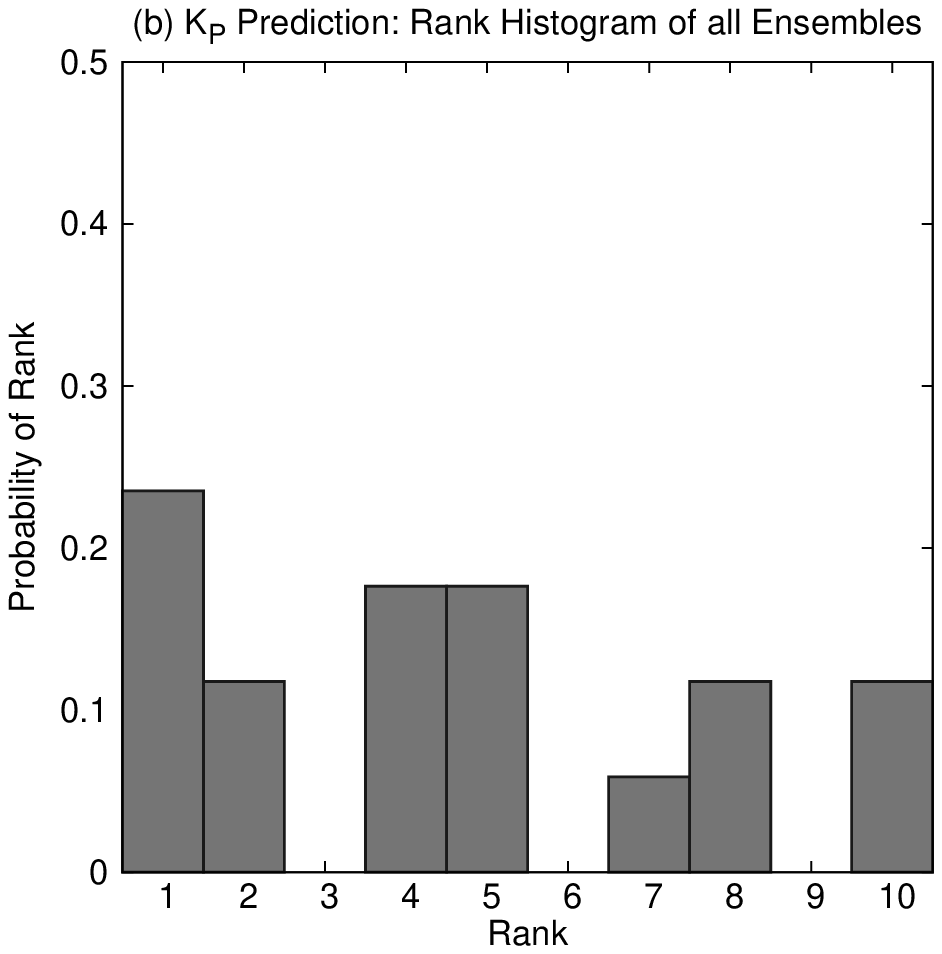}}
\caption{$K_P$ forecast verification: (a) Histogram of the observed $K_P$ values (black) and the forecast $K_P$ probability distribution (hashed) for this sample (see Table \protect\ref{table:kp}). (b) Rank histogram of $K_P$ predictions for all ensembles.}\label{fig:kpbs}
\end{figure}

Other factors contributing to $K_P$ overprediction may include the magnetic field direction - two out of the three field configurations assumed produce persistent southward fields (135$^{\circ}$ and 180$^{\circ}$), so there is a bias towards geoeffective field configurations. Examining the distribution of north-south magnetic fields associated with the ICMEs of \inlinecite{richardson2010} and the associated sheaths, in only 2\% of cases are southward fields completely absent, so the bias towards geoeffective field configurations is consistent with observations.  However, both small and large maximum southward fields are observed relatively infrequently (e.g., maximum southward fields are $<$4 nT in 17\% of events, and $>$15 nT in 16\%), suggesting that the weighting of 90$^{\circ}$ and 180$^{\circ}$ clock angles should be reduced.  In particular, reducing the 180$^{\circ}$ clock angle weight would be expected to reduce the Kp overprediction.    

The last line of Table \ref{table:kp} lists Brier Score calculated for each $K_P$ bin.  Here, the $BS$ is a measure of the magnitude of error in the $K_P$ probability forecast (how likely a given $K_P$ bin will occur) in each bin.  The $BS$ values indicate that in this sample, the $K_P$ probability forecast is reliable for the $K_P$=5 and 6 bins ($BS$=0.17 for both), and less so for the $K_P$=3 and 4 bins ($BS$=0.27 and 0.19). Although the scores also indicate that the forecast is most reliable for the smallest and largest $K_P$ bins, most of the observations in this sample did not fall in these extreme bins, hence a larger sample is needed to verify forecast reliability for these bins.  Figure \ref{fig:kpbs}a shows the overall observed $K_P$ distribution and the forecast $K_P$ probability distribution for the events in Table \ref{table:kp} used to calculate the $BS$.    

To further evaluate $K_P$ probability forecast reliability we compare the observed $K_P$ to the spread in ensemble predictions. For most (12 out of 17) of the ensembles, the observed $K_P$ was within the overall predicted $K_P$ spread (column 11). The observed $K_P$ was also within the predicted mean $K_P \pm$1 for 11 out of 17 of the ensembles.  A rank histogram was also constructed for the $K_P$ predictions for all ensembles and is shown in Figure \ref{fig:kpbs}b, again normalized to an ensemble size of 9.  To construct this rank histogram the $K_P$ predictions are sorted from smallest to largest and the rank of where the observed $K_P$ value falls among the predicted $K_P$ values is noted. For example, an ensemble with a rank of 6 has the meaning that 5 $K_P$ predictions are less than the observed $K_P$ value, a rank of 10 would mean that all 9 of the $K_P$ predictions are less than the observed $K_P$ (underprediction), and a rank of 1 means that the observed $K_P$ value is less than all of the $K_P$ predictions (overprediction).   The histogram has an overall flat shape, with more occurrence at rank 1 (the observed $K_P$ was less than the predicted range) and less occurrence in the higher ranks which shows the bias for $K_P$ overprediction (mean error=+0.6). Note, that the rank histogram does not indicate how ``good'' forecasts are but only measures whether the observed probability distribution is well represented by the ensemble. Therefore, a uniform, flat rank histogram is a necessary but not sufficient condition for determining the reliability of ensembles \cite{hamill2001}. 

\section{ENLIL Parameter Sensitivity: 11 April 2013 Event Case Study}\label{case}
In the current configuration, other than the measured CME speed, direction, and size, the real-time WSA-ENLIL+Cone ensemble simulations use the default values for the model CME free parameters.   In this section, we present a case study which examines the effect of changing these model free parameters on the ensemble modeling. The CME starting on 11 April 2013 at 07:24 UT was chosen for this study due to the large early arrival time prediction error obtained for all members of the model ensemble. 
\inlinecite{taktak2010} studied the dependence of arrival time predictions on the uncertainty in CME input parameters (speed, width, density ratio) for three Earth directed CME events of varying speeds.  A similar procedure was adopted for this case study, and by employing the ensemble modeling technique, the parameter space can be sampled more systematically.

\begin{figure}  
\centerline{\includegraphics[width=0.335\textwidth,angle=0,origin=c]{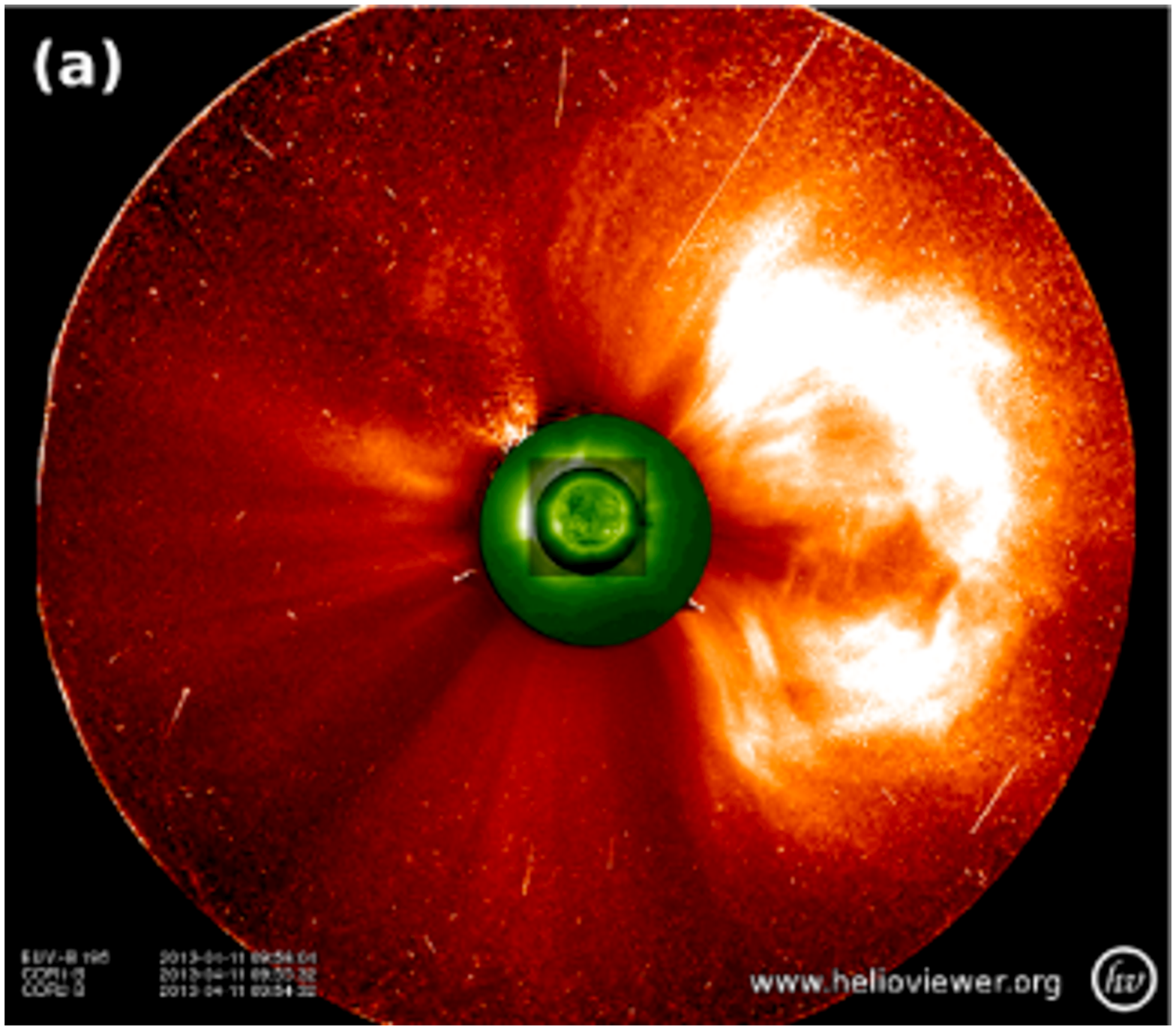}\includegraphics[width=0.335\textwidth,angle=0,origin=c]{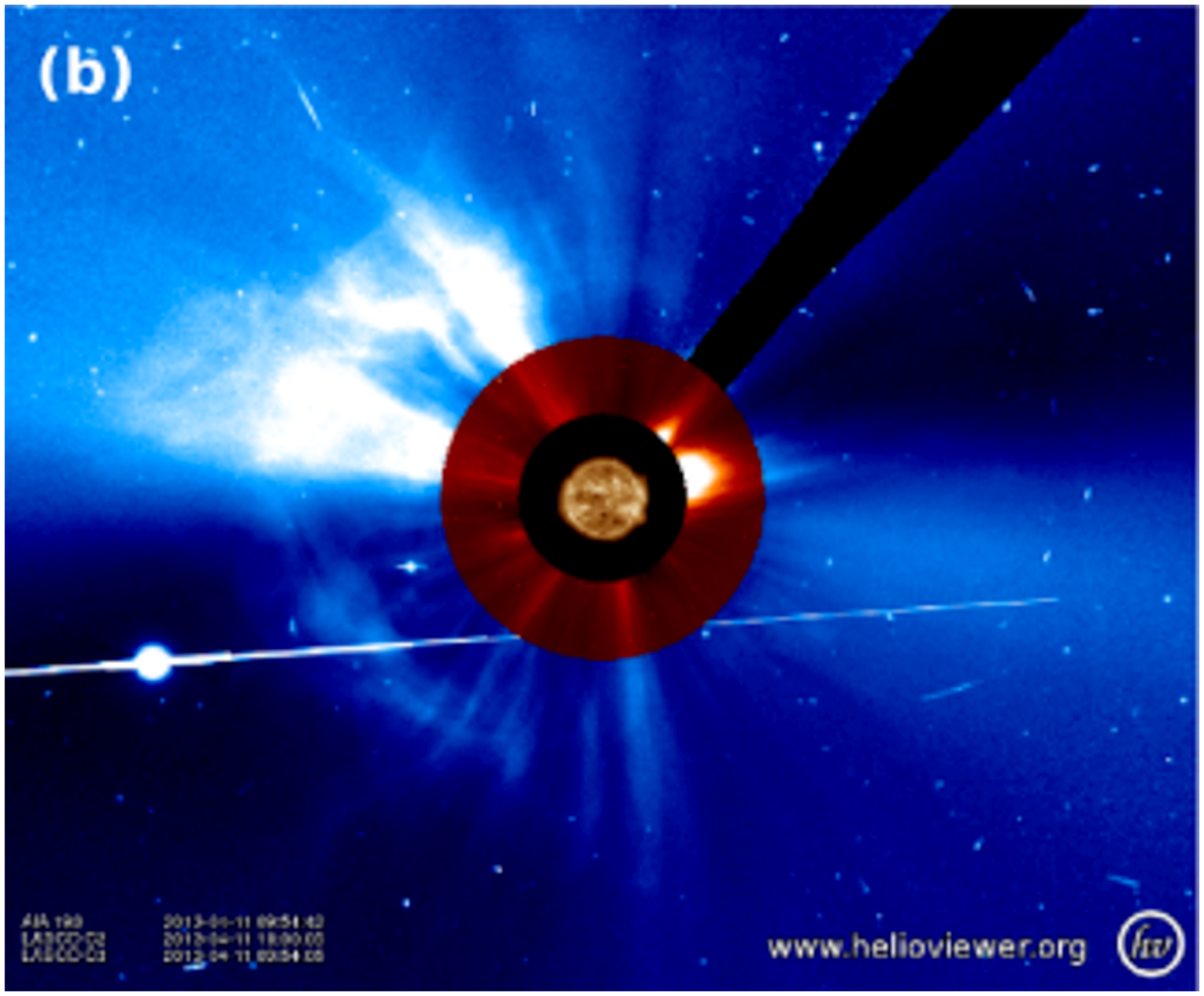}\includegraphics[width=0.335\textwidth,angle=0,origin=c]{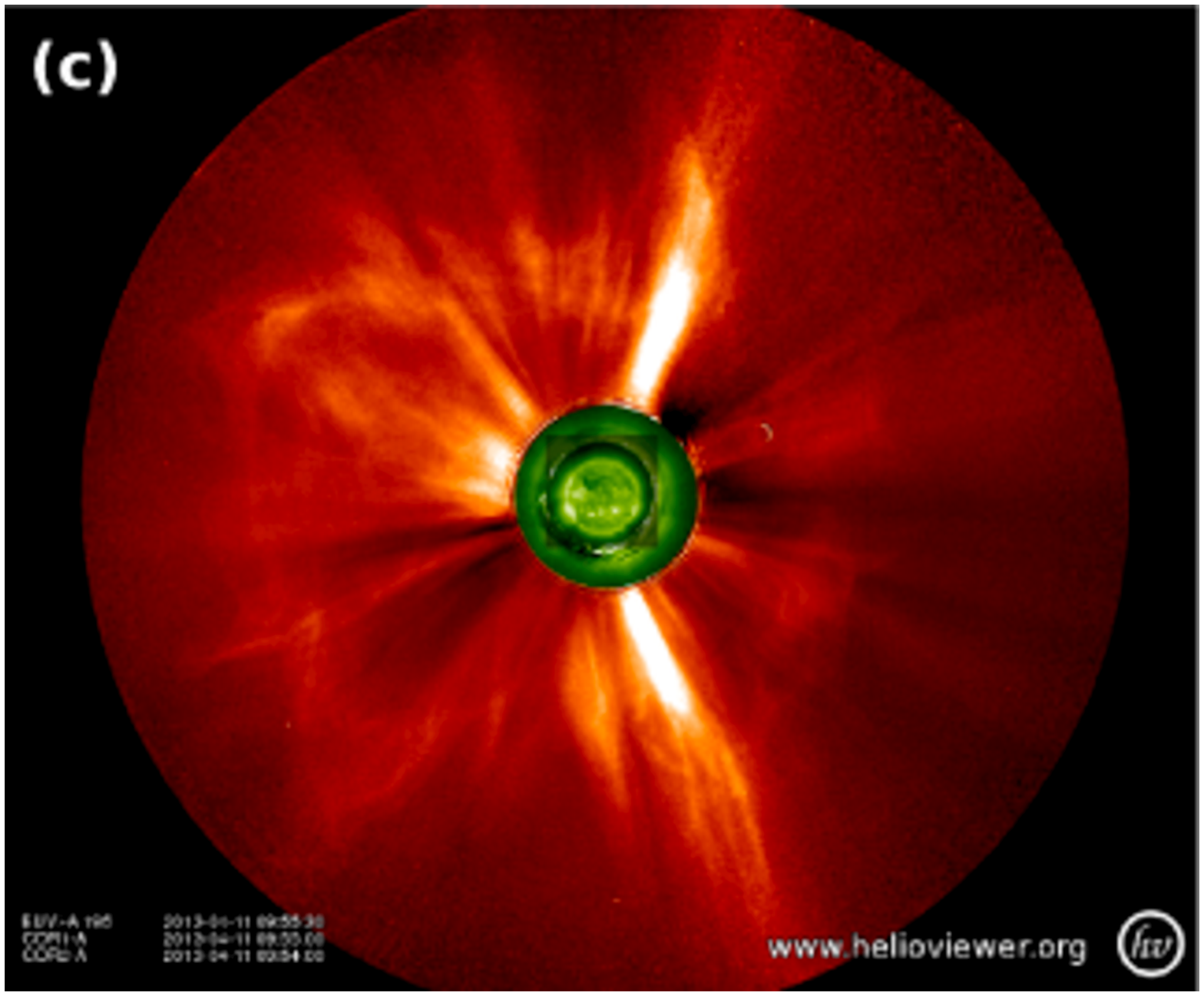}}
\caption{Coronagraph observations of the 11 April 2013 07:24 UT CME as viewed from (b) SOHO LASCO, (a) STEREO SECCHI/COR2 B and (c) A near the time of 09:55 UT.}\label{fig:apr2013cme}
 \end{figure}

\begin{figure}  
\centerline{\includegraphics[width=0.7\textwidth,angle=0,origin=c]{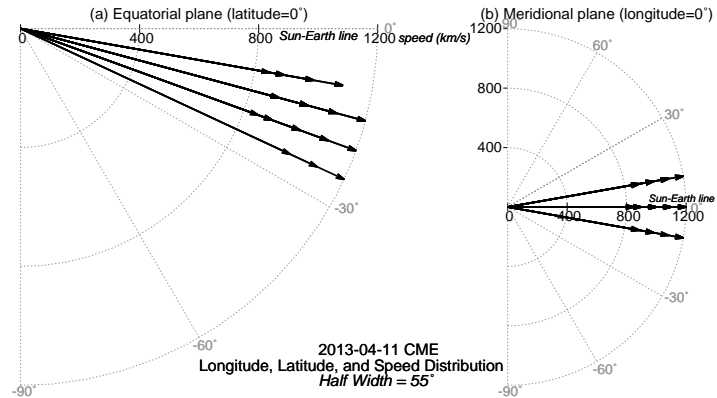}}
\caption{Distribution of the 11 April 2013 CME input parameters shown as speed vectors (all CME half widths are 55$^{\circ}$), in the same format as Figure \protect\ref{fig:catmeashist1}.  Median CME parameters are: speed of 1000 km/s, direction of -15$^{\circ}$ longitude, 0$^{\circ}$ latitude, and a half-width of 55$^{\circ}$.   This figure shows that custom ensemble members were chosen with speeds of 850, 900, 1000, 1100, and 1200 km/s, between $\pm$-10$^{\circ}$ latitude, -10$^{\circ}$ to -25$^{\circ}$ longitude with a half width of 55$^{\circ}$.}
\label{fig:catmeashist2}
\end{figure}

The original set of simulations performed in real-time were chosen as the ``base ensemble'' (ensemble I). Subsequently, ten ensemble runs (ensembles II-XI), each containing 36 members for 360 total simulations, were performed to assess the sensitivity of the CME arrival time prediction to changes in the model free parameters and ambient solar wind model, while keeping the CME speed and direction input parameters fixed.  The ENLIL model free parameters considered in this study include the CME half width, CME density ratio, CME cavity ratio, and ambient solar wind reduction factor.  The CME density ratio (\url{dcld}) is a free parameter which is by default is a set factor of 4 times larger than typical mean values in the ambient fast wind providing a pressure of four times larger than that of the ambient fast wind.  The cavity ratio \url{radcav} is defined as the ratio of the radial CME cavity width to the CME width, with the default being no cavity \url{radcav}=0.   The ambient speed reduction factor \url{vred} reduces the solar wind speed provided by the WSA coronal map in order to account for expansion of the solar wind from the WSA boundary to 1 AU since WSA is calibrated against 1 AU in-situ observations.

Figure \ref{fig:apr2013cme} shows the CME starting on 11 April 2013 at 07:24 UT as viewed from SOHO LASCO C3, STEREO A and B SECCHI/COR2 near the time of 09:55 UT. On this date the STEREO B spacecraft was located at -142$^{\circ}$ and STEREO A was at 133$^{\circ}$ in HEEQ coordinates.   This CME was associated with an M6.5 class flare from AR 11719 located at N07E13 with peak intensity at 07:16 UT.  The eruption, coronal dimming and wave were visible mostly southeast of the active region in SDO/AIA 193{\AA}. Additionally, an increase in solar energetic particle proton flux was observed starting at around 07:40 UT by the SOHO COSTEP (reaching 1 pfu/MeV, in the 16-40 MeV energy range), ACE EPAM (100 pfu/MeV, 1.22-4.94 MeV), and  GOES-13 EPEAD (5 pfu/MeV, 15-40 MeV energy range) instruments starting at 08:00 UT, and by the IMPACT HET instruments on STEREO B (5 pfu/MeV, 24-41 MeV energy range) and A (0.001 pfu/MeV, 24-41 MeV energy range).  This solar energetic particle event and its longitudinal extent is studied in detail by \inlinecite{cohen2014} and \inlinecite{lario2014}.

Due to the lack of availability of real-time concurrent coronagraph images, triangulation of CME parameters with the StereoCAT ``ensemble'' mode method was not possible for this CME.  Therefore, the ensemble was composed of ``custom'' members.  The CME parameters for each member were derived from plane-of-sky CME speed measurements combined with the source location at the Sun. The distribution CME input parameters for 36 ensemble members are shown in Figure \ref{fig:catmeashist2}. Median CME parameters are: speed of 1000 km/s, direction of -15$^{\circ}$ longitude, 0$^{\circ}$ latitude, and a half-width of 55$^{\circ}$.   This figure shows that custom ensemble members were chosen with speeds of 850, 900, 1000, 1100, and 1200 km/s, between $\pm$10$^{\circ}$ latitude, -10$^{\circ}$ to -25$^{\circ}$ longitude with a half width of 55$^{\circ}$.    Subsequent re-analysis of the CME height-time evolution gives average plane-of-sky speeds of $\sim$800 km/s and $\sim$700 km/s for SECCHI COR2B and LASCO C3 respectively, yielding a triangulated speed of 850$\pm$200 km/s, -5$^{\circ}$$\pm$5$^{\circ}$ latitude,  -15$^{\circ}$$\pm$10$^{\circ}$ longitude, 50$^{\circ}$$\pm$5$^{\circ}$ half width, which is represented within the ensemble members derived in real-time.   

\begin{figure} 
\centerline{\includegraphics[width=0.75\textwidth,angle=0,origin=c]{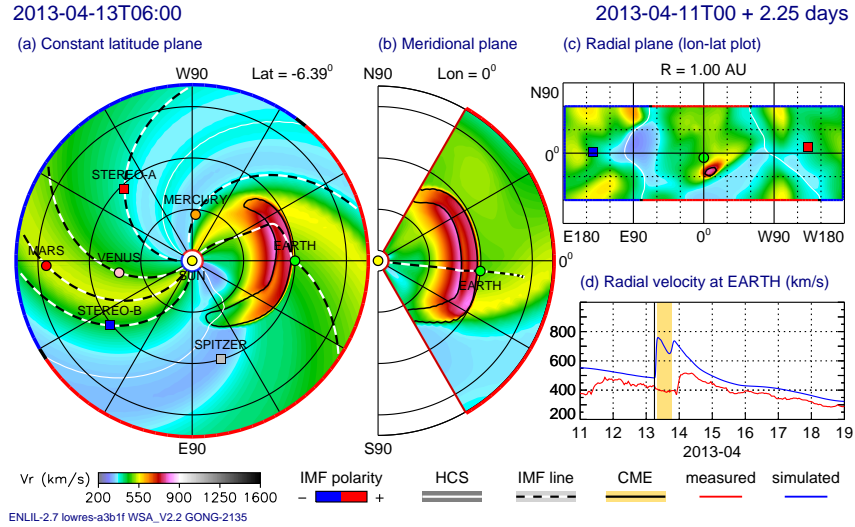}}
\caption{Global view of 11 April 2013 CME on 13 April at 06:00 UT: WSA-ENLIL+Cone scaled velocity contour plot in the same format as \protect\ref{fig:timvel1} for the ensemble member with median CME input parameters (speed of 1394 km/s, direction of 9$^{\circ}$ longitude, -35$^{\circ}$ latitude, and a half-width of 46$^{\circ}$). }\label{fig:timvel2}
\end{figure}

\begin{figure} 
\includegraphics[width=1.0\textwidth,angle=0,origin=c]{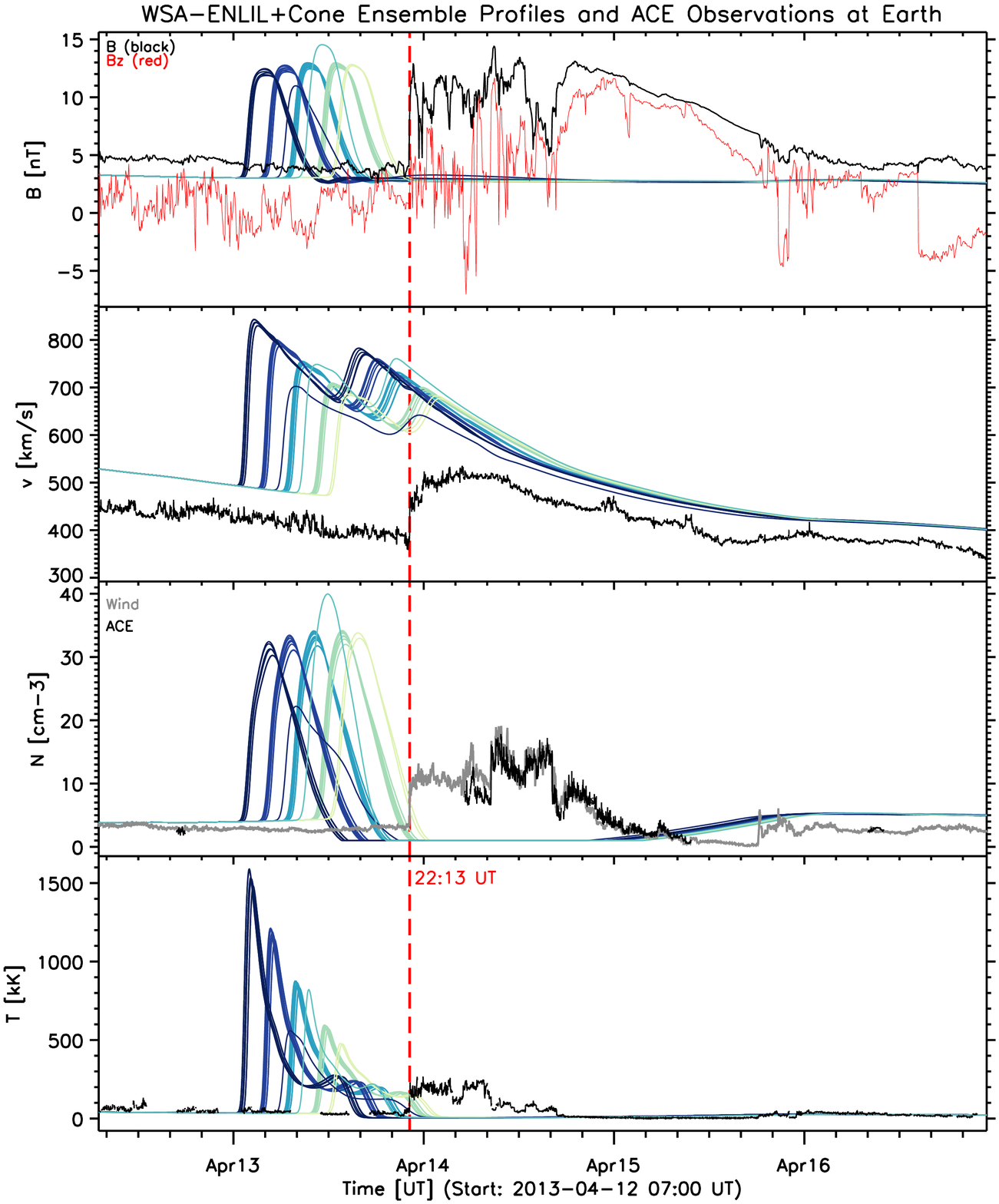}
\caption{11 April 2013 CME base ensemble: Model calculated magnetic field, velocity, density, and temperature profiles at Earth for each ensemble member along with the observed in-situ L1 observations from ACE in black (red for $B_z$). Wind density observations are plotted in grey due to missing ACE values.  The model traces are color coded by CME input speed such that slow to faster input speeds are colored from light green to dark blue.  The CME-associated shock was observed to arrive at ACE and Wind on 13 April at around 22:13 UT with clear ICME signatures starting around  16:45 UT on 14 April through about 18:30 UT on 15 April.
All of the arrival times indicated by the model results are earlier than the observed shock arrival, and are clustered by CME input speed. The mean predicted arrival time at Earth is 13 April 06:14 UT, with a range from 13 April 00:47 to 12:20 UT.}\label{fig:nvbt2}
 \end{figure}

\begin{figure} 
   \centerline{\includegraphics[width=0.8\textwidth,angle=0,origin=c]{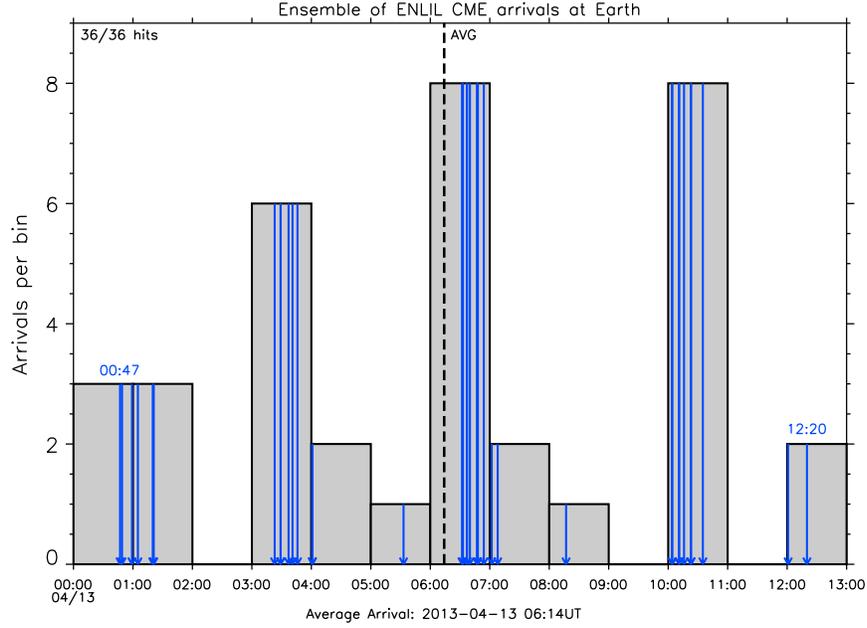}}
\caption{11 April 2013 base ensemble: Histogram of distribution of arrival time predictions at Earth (one hour bin size). The actual arrival was observed on 13 April at around 22:13 UT by Wind.  The clustering of predicted arrival times reflects the limited number of different CME input speeds represented in the ensemble (see Figure \ref{fig:catmeashist2}), with faster CMEs arriving first.}\label{fig:e2}
 \end{figure}

The WSA-ENLIL+Cone model scaled velocity contour plot is shown in Figure \ref{fig:timvel2} on 13 April at 06:00 UT for the ensemble member with median CME input parameters. This simulation figure shows a nearly direct CME impact at Earth, slightly eastward.  Figure \ref{fig:nvbt2} shows the base ensemble WSA-ENLIL+Cone modeled quantities for all 36 ensemble members (color traces) at Earth along with in-situ ACE (black) and Wind (grey) observations (when there are ACE datagaps).  The model traces are color coded by CME input speed such that slow to faster input speeds are colored from light green to dark blue.   All 36 of the ensemble members predicted that the CME would arrive (100\%) and the mean predicted arrival at Earth was 13 April 06:14 UT (range from 13 April 00:47 to 12:20 UT). The histogram of the distribution of arrival times is shown in Figure \ref{fig:e2}.  The clustering of predicted arrival times in this histogram (and also in Figure \ref{fig:nvbt2}) reflects the limited number of discrete CME input speeds represented in the ensemble (see Figure \ref{fig:catmeashist2}), with faster CMEs arriving first. 
The CME-associated shock was observed to arrive at ACE and Wind on 13 April at 22:13 UT, giving an average prediction error of -16 hours. Clear ICME signatures including enhanced low variability magnetic field, declining solar wind speed, and low proton temperatures, start at around 16:45 UT on 14 April through about 18:30 UT on 15 April. The overall spread in arrival time predictions of all of the members in the base ensemble (including the clustering by CME input speed) can also be seen in Figure \ref{fig:nvbt2} as the color traces increase ahead of the observed arrival.  The traces also show that the velocity, density, and temperature are overpredicted, while the maximum magnetic field strength is similar to that actually observed. The passage of this CME did not produce a geomagnetic storm due to an almost persistently northward magnetic field, shown in red in the top panel of Figure \ref{fig:nvbt2}.  The NOAA real-time observed $K_P$ index reached 3 during the synoptic period 21-24:00 UT on 13 April, while the Potsdam final $K_P$ was 3+. The $Dst$ index shows a sudden storm commencement of +21 nT at 23:00 UT on 13 April, and reached a minimum of only -7 nT at 11:00 UT on 15 April.

 \begin{figure} 
   \centerline{\includegraphics[width=0.75\textwidth,angle=0,origin=c]{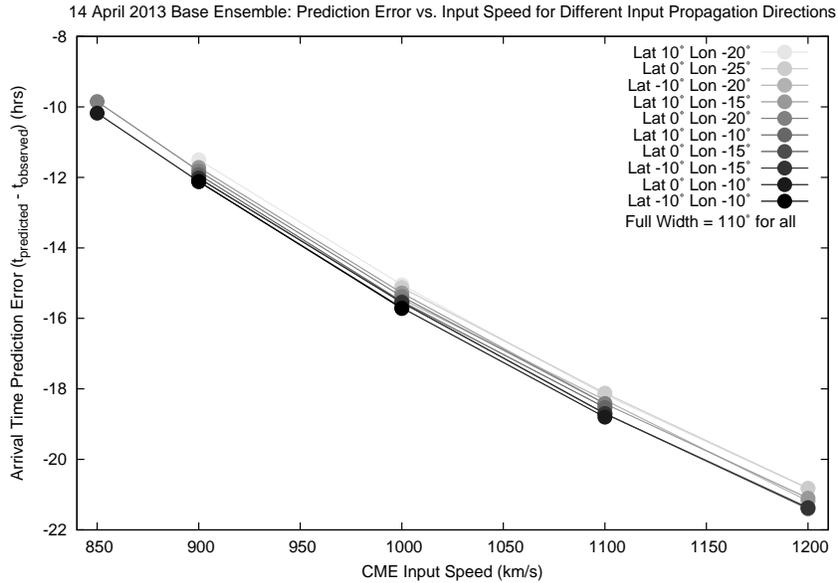}}
 \caption{11 April 2013 base ensemble: CME arrival time prediction error ($\Delta t_{\rm err}=t_{\rm predicted}-t_{\rm observed}$) for the ensemble members plotted against the CME input speed for different CME input propagation directions (gray scale coded).}\label{fig:dtv}
 \end{figure}

In Figure \ref{fig:dtv} the arrival time prediction error ($\Delta t_{\rm err}=t_{\rm predicted}-t_{\rm observed}$) for the members in the base ensemble is plotted against the CME input speed for different CME input propagation directions (gray scale coded) and a fixed half width of 55$^{\circ}$ (full angular width of 110$^{\circ}$).  On 11 April 2013 Earth was located at -5.9$^{\circ}$ latitude and 0$^{\circ}$ longitude in HEEQ coordinates, thus the input propagation direction of -10$^{\circ}$ latitude and -10$^{\circ}$ longitude (black, and dark blue in subsequent figures) represent the members with the most direct impact. This Figure shows that the arrival time prediction error ranges from -9.9 hours to -21.4 hours and increases with initial CME speed.  

\begin{figure} 
 \centerline{\includegraphics[width=0.75\textwidth,angle=0]{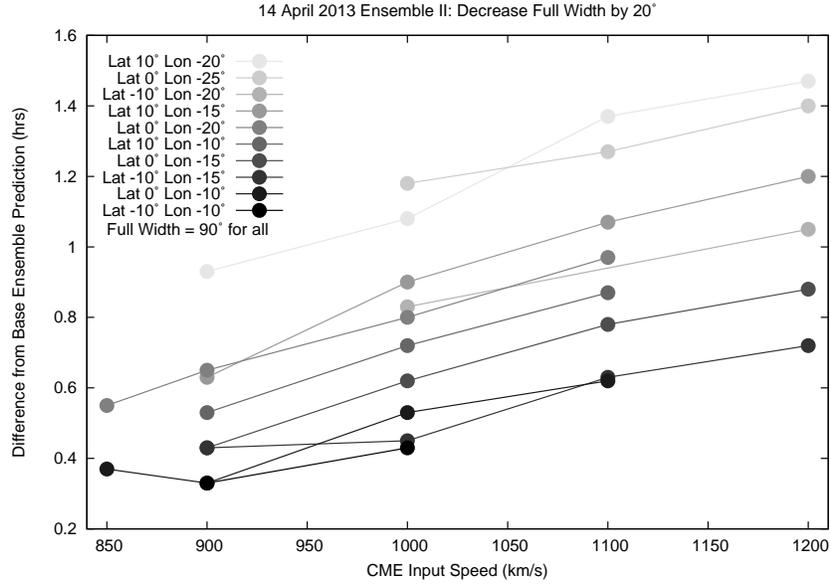}}
 \caption{11 April 2013 ensemble II: Difference from the base ensemble (as shown in Figure \ref{fig:dtv}) in hours when the CME input half width is decreased by 10$^{\circ}$ against the CME input speed for different propagation directions.}\label{fig:w10}
 \end{figure}

Considering that a source of the prediction error could be due to uncertainty in the CME width, an identical ensemble (II) simulation was performed with the same input conditions but decreasing the half width by 10$^{\circ}$ (full angular width decreased from 110$^{\circ}$ to 90$^{\circ}$).   Figure \ref{fig:w10} shows the {\it difference} from the original predicted arrival times from the base ensemble against the CME input speed for different propagation directions (as shown in Figure \ref{fig:dtv}) when the full angular width decreased from 110$^{\circ}$ to 90$^{\circ}$.  In this figure (and those subsequent), the new CME arrival time prediction error is shown in hours relative to the original ``base ensemble'' prediction error.   Compared to the original arrival time estimates, the overall prediction error decreases by 0.2 to 1.8 hours with increasing initial CME speed.  Since all of the predictions in the base ensemble were too early (negative prediction error), a decrease in prediction error means that the new predictions are shifted to later times, closer to the observed arrival time. Nevertheless, the improvement is small compared to the prediction error.

\begin{figure} 
 \centerline{\includegraphics[width=0.75\textwidth,angle=0]{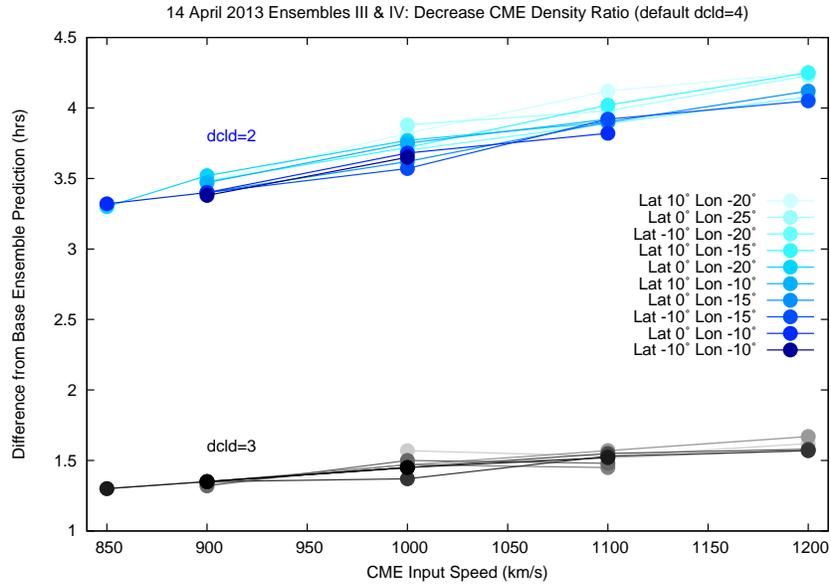}}
 \caption{11 April 2013 ensembles III and IV: Difference from the base ensemble (as shown in Figure \ref{fig:dtv}) in hours when the CME density ratio \protect\url{dcld} is decreased to \protect\url{dcld=3} and \protect\url{dcld=2} (default \protect\url{dcld}=4) against the CME input speed for different propagation directions}\label{fig:dcld}
 \end{figure}

Next, the dependence of prediction on the input CME density ratio \url{dcld} was considered. Two ensembles were performed (III and IV) for which all parameters of the base ensemble were held fixed but the CME density ratio was adjusted from four (default), to two, and three.  The results are shown in Figure \ref{fig:dcld} which shows the difference from the predicted arrival time for the base ensemble as a function of CME speed for the two different density ratios. The prediction error decreases by 3.3 to 4.3 hrs for a CME density ratio of two and by 1.3 to 1.7 hrs for a density ratio of three, as a function of increasing initial CME speed. Hence, reducing the density ratio from four to two or three improves the arrival time prediction by around 3.5 or 1.5 hours, respectively.

\begin{figure} 
 \centerline{\includegraphics[width=0.7\textwidth,angle=0]{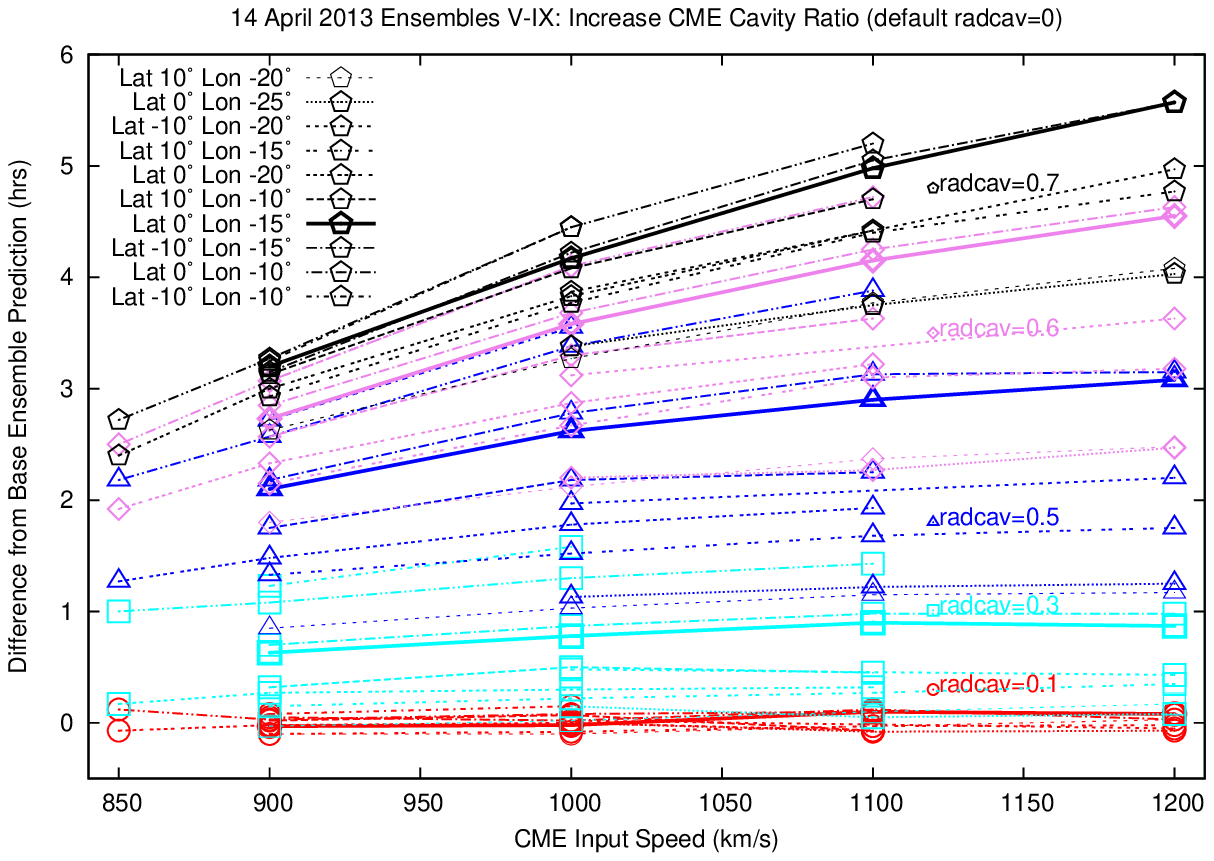}}
 \caption{11 April 2013 ensembles V-IX: Difference from the base ensemble (as shown in Figure \ref{fig:dtv}) in hours when the CME cavity ratio is increased (\protect\url{radcav}=0.1, 0.3, 0.5. 0.6, and 0.7) against the CME input speed for different propagation directions.  Different CME input directions are indicated by the symbol/line type and each ensemble is indicated by a different line color. The cavity ratio \protect\url{radcav} is defined as the ratio of the radial CME cavity width to the CME width, and the default is no cavity \protect\url{radcav}=0.}\label{fig:radcav}
 \end{figure}

Another ENLIL CME parameter is the cavity ratio which allows the CME to be represented by spherical shell of plasma, based on coronagraph observations of CME cavities.  The cavity ratio \url{radcav} is defined as the ratio of the radial CME cavity width to the CME width, with the default being no cavity \url{radcav}=0.  Figure \ref{fig:radcav} shows the results of five ensembles (V-IX) with the CME cavity ratio adjusted to 0.1, 0.3, 0.5. 0.6, and 0.7, i.e., the CME is modeled as a progressively thinner shell as the ratio increases, using the base ensemble for all other parameters fixed. Specifically, the differences from the arrival times obtained for the base ensemble are plotted as a function of CME speed and direction (indicated by the symbol/line type) for each of these ensembles (indicated by the line color). For a cavity ratio of 0.1 the prediction remains largely unchanged compared to the base ensemble (with 0.15 hours).   For the other cavity ratios, increasing differences in Figure \ref{fig:radcav} as the cavity ratio increases correspond to the predicted arrival moving to later times, reducing the prediction error.  Furthermore, for each cavity ratio there is a spread (1 to 3 hours) in prediction time difference (compared to the base ensemble) for the different CME input directions with the more Earth directed inputs showing the largest difference from the base ensemble.  Overall the prediction error decreases by between 0-1.6 hrs, 0.9-3.9 hrs, 1.8-4.7 hrs, 2.4-5.6 hrs for cavity ratios of 0.3, 0.5, 0.5, 0.6 respectively.

\begin{figure} 
 \centerline{\includegraphics[width=0.75\textwidth,angle=0]{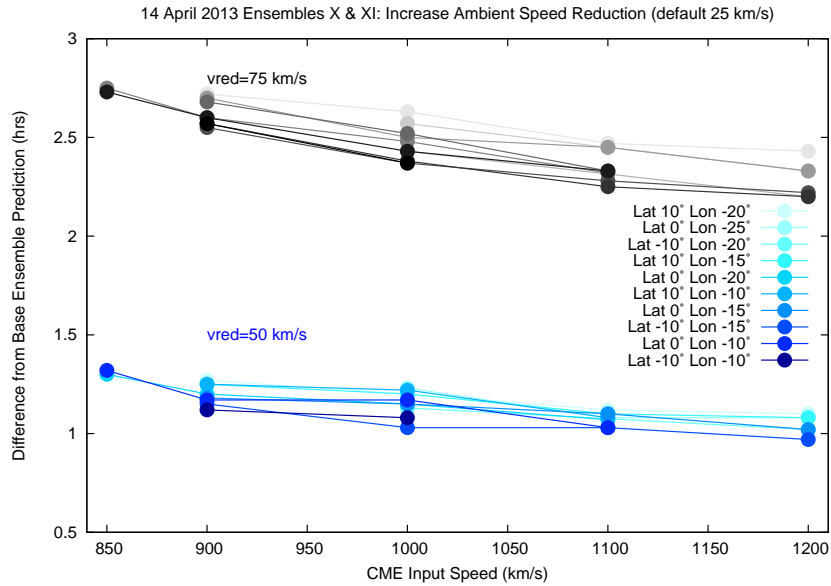}}
 \caption{11 April 2013 ensembles X and XI: Difference from the base ensemble (as shown in Figure \ref{fig:dtv}) in hours when the ENLIL ambient speed reduction factor \protect\url{vred} is increased to 50 km/s and 75 km/s (from the default value of 25 km/s) against the CME input speed for different propagation directions.}\label{fig:vred}
 \end{figure}

\begin{figure} 
 \centerline{\includegraphics[width=0.7\textwidth,angle=0,origin=c]{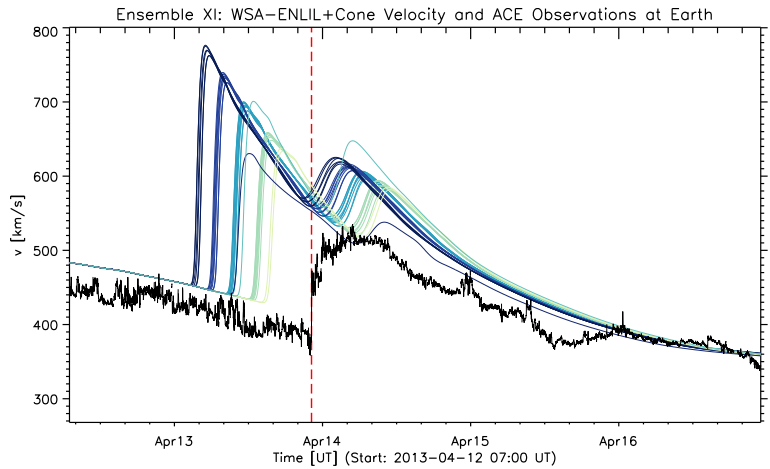}}
\caption{11 April 2013 ensemble XI: The predicted velocities (color traces) better match the observed in-situ values at ACE (black) when \protect\url{vred} is increased to 75 km/s compared to the default \protect\url{vred}=25 km/s results shown in Figure \ref{fig:nvbt2}. The second peaks in the predicted solar wind speed are artifacts of the ENLIL modeled CME as a spherical cloud.} \label{fig:vred75}
\end{figure}

Considering now the influence of the ENLIL ambient solar wind solution, the in-situ data-model comparison (Figure \ref{fig:nvbt2}) for the base ensemble indicates that the modeled background solar wind speed is $\sim$125 km/s higher than the observed in-situ values, whereas the default value of ambient solar wind reduction factor \url{vred} is 25 km/s . To examine the role of the speed reduction factor in the prediction error, \url{vred} was increased to 50 km/s and 75 km/s for two ensembles (X and XI).  This factor reduces the speed provided by the WSA coronal map in order to account for expansion of the solar wind from the WSA boundary to 1AU since WSA is calibrated against 1 AU in-situ observations.  Figure \ref{fig:vred} shows the prediction time difference from the base ensemble for these two ensembles which show differences of 1-1.3 hrs and 2.2-2.8 hrs for \url{vred} of 50 km/s and 75 km/s respectively. Since the differences are positive, this indicates that the predicted arrival times are moved later, reducing the error relative to the observed arrival time.  As might be expected, the modeled CME propagates more slowly when the ambient solar wind is slower.  Figure \ref{fig:vred75} illustrates how the modeled background solar wind speed better matches the observed speed prior to CME arrival when the ambient speed reduction factor \url{vred} is increased to 75 km/s in ensemble XI.

Overall, this parametric case study shows that after the CME input speed, the cavity ratio and density ratio assumed in ENLIL have the greatest effects on the predicted CME arrival time, each changing this time by about 3 hours on average.  Their effect is also more noticeable with higher CME input speeds.  The CME input speed, cavity and density ratios define the CME momentum which defines the CME deceleration.  In the addition to the using the default values, new ensemble runs could be performed with changes to the CME cavity ratio and density ratio as informed by coronagraph measurements of the CME.  Here, we have only examined the effect of changing the ad hoc ambient speed reduction factor in ENLIL, but we could also produce an ensemble of ambient solar wind WSA-ENLIL simulations using different ambient speed reduction factors which can be compared to in-situ measurements to determine ``best'' factor to use in subsequent CME simulations.   An ensemble forecast reflecting uncertainties in the background solar wind could also be produced by using a variety of magnetograms (from different observatories or processed using different techniques) as input to the WSA or WSA-ADAPT models.

\section{Summary and Discussion}\label{disc}
This study evaluates the first ensemble CME prediction system of its kind employed in a real-time environment, providing unique space weather information for NASA users.  The ensemble prediction approach provides a probabilistic forecast which includes an estimation of arrival time uncertainty from the spread in predictions and a forecast confidence in the likelihood of CME arrival.  The current implementation explores the sensitivity of CME arrival time predictions from the WSA-ENLIL+Cone model to initial CME parameters. First results give a mean absolute arrival time error of 12.3 hours, RMSE of 13.9 hours, and mean error of -5.8 hours (early bias), based on a sample of 30 CME events for which ensemble simulations were performed.  The arrival time is generally based on the arrival of the CME-generated shock at the Earth. The ensemble mean absolute error and RMSE are both comparable with other CME arrival time prediction errors reported in the literature.   

When considering the overall performance of CME arrival prediction, it was found that the correct rejection rate is 62\%, the false-alarm rate is 38\%, correct alarm ratio is 77\%, and false alarm ratio is 23\%.  Each ensemble CME arrival time forecast includes a forecast probability $p=n_{\rm predicted~ hits}/n_{\rm tot}$, which conveys a forecast uncertainty about the likelihood that the CME will arrive, which can be compared with observations to determine forecast reliability. The Brier Score ($BS$) of 0.15 for all 30 ensemble CME arrival probabilities indicates that in this sample, on average, the predicted probability of the CME arriving is fairly accurate.  (A $BS$ of 0 on a range of 0 to 1 is a perfect forecast.)  However, the reliability diagram (Figure \ref{fig:rank}a) shows that the ensemble simulations are underforecasting the likelihood that the CME will arrive in the forecast bins between 20-80\%, and slightly overforecasting in the 1-20\% and 80-100\% forecast bins.   Overforecasting is when the forecast chance of CME arrival is higher than is actually observed; i.e., the CME is observed to arrive less often than is predicted.  More ensemble simulations are needed for a more robust forecast verification of these probabilistic CME arrival time forecasts.

For 8 out of 17 of the ensemble runs containing hits, the observed CME arrival was within the spread of ensemble arrival time predictions. The initial distribution of CME input parameters was shown to be an important influence on the accuracy of CME arrival time predictions.  Particularly, the median and spread of the input distribution should accurately represent the range of CME parameters derived from observations.  This is evidenced by the rank histogram (Figure \ref{fig:rank}b) which illustrates that roughly half of the observed arrivals are outside the spread of predictions, and also suggests undervariability in initial conditions; i.e., these ensembles do not sample a wide enough spread in CME input parameters.

Each set of ensemble simulations also provides a probabilistic $K_P$ forecast $p({{\rm K_P}=b})$ for each $K_P$ bin $b$ which can be compared with observations to determine forecast reliability.  The Brier Score ($BS$) for the probabilistic $K_P$ forecast bins show reliability for the $K_P$=5 and 6 bins ($BS$=0.17 for both), and less so for the $K_P$=3 and 4 bins ($BS$=0.27 and 0.19).  If choosing a single categorical $K_P$ forecast value, the mean predicted $K_P$ was found to have smaller prediction errors compared to using the $K_P$ bin with the highest likelihood from the probabilistic $K_P$ forecast.  The observed $K_P$ was within $\pm$1  of the predicted mean $K_P$ for 11 out of 17 of the ensembles.   The $K_P$ prediction errors computed from the mean predicted $K_P$ show a mean absolute error of 1.4, RMSE of 1.8, and mean error +0.4. There is a known overall tendency for the overprediction of $K_P$, generally found for CME input speeds above 800-1000 km/s.   Again, more ensemble simulations are needed to provide better forecast verification and to calibrate the $K_P$ forecast.

This paper focuses on the forecast verification of the ensemble modeling aspect of CME arrival and $K_P$ predictions.   More events, as well as comparison of results using different CME propagation models, are needed for more comprehensive forecast verification. These aspects are being investigated in a separate verification study which evaluates $>$400 single WSA-ENLIL+Cone simulations (of which there are $>$70 simulations containing CME arrivals) performed at the CCMC/SWRC.

The parameter sensitivity studied discussed in Section \ref{case} suggests future directions for this ensemble system. In the addition to the using the default model values, new ensemble runs could be performed with changes to the CME cavity ratio and density ratio as informed by coronagraph measurements of the CME. As discussed in Section \ref{model} an accurate representation of the background solar wind is necessary for simulating transients, and prediction errors arising from background characterization and other model limitations should be considered. An ensemble forecast reflecting uncertainties in the background solar wind could be produced by using a variety of magnetograms (from different observatories or processed using different techniques) as input to the WSA or WSA-ADAPT models.  From these results one can produce an ensemble of ambient solar wind WSA-ENLIL model outputs which can be compared to in-situ measurements to determine ``best'' coronal maps/model instance.   These sub-selected WSA or WSA-ADAPT maps could then be used for a series of ensemble WSA-ENLIL+Cone CME simulations.  Such an improved ensemble forecast would produce predictions which also reflect uncertainties in the WSA-ENLIL modeled background solar wind in addition to the uncertainties in CME input parameters (as considered in this work).

Another improvement could be the use of real-time interplanetary scintillation (IPS) observations by the Ooty Radio Telescope \cite{manoharan2006}. These data can provide crucial information about the CME propagation and interaction with the surrounding solar wind which could be used to provide updated information on CME parameters as the CME moves out from the Sun.  This information could then be used to refine model predictions of the propagation of the CME. The STEREO Heliospheric Imagers also provide CME propagation information out to 1AU. However, it is not always possible to extract this information from real-time data, and the imagers do not always have an optimal viewing angle for Earth-directed CMEs.  
Comparisons of CME propagation from WSA-ENLIL with near real-time observations of the CME location inferred from IPS, the STEREO heliospheric imagers, or some other source, can be used to select ensemble members with the best agreement using quantitative and visual inspection employing advanced visualization techniques such as ``3D volumetric rendering'' \cite{bock2014}.  

Finally, the forecasting of CME arrival would benefit from the use of other propagation models, in addition to WSA-ENLIL, each with its own set of independently assessed input parameters, leading to a community-wide ensemble prediction capability.  A first step to such a capability is provided by the {\it CME Scoreboard}, described in Section \ref{intro}, where anyone is invited to post their estimate of the arrival time of a recently observed CME in real-time.

%

%

%

%
\begin{acks}
The work was carried out as a part of NASA's Game Changing Development Program Advanced Radiation Protection Integrated Solar Energetic Proton (ISEP) project. LKJ acknowledges the support of NSF grants AGS 1242798 and 1321493. MLM thanks T. Nieves-Chinchilla and B.J. Thompson for useful discussions. We gratefully acknowledge the participants of the CME Arrival Time Scoreboard \url{http://kauai.ccmc.gsfc.nasa.gov/CMEscoreboard}. The ACE and Wind solar wind plasma and magnetic field data were obtained at NASA's CDAWeb (\url{http://cdaweb.gsfc.nasa.gov}). OMNI data was obtained from NASA's COHOWeb (\url{http://omniweb.gsfc.nasa.gov/coho}). The $Dst$ geomagnetic index was obtained from the World Data Center for Geomagnetism in Kyoto, Japan.  Estimated real-time planetary $K_P$ indices are from NOAA and the NGDC, and final definitive $K_P$ indices are from the Helmholtz Center Potsdam GFZ German Research Centre for Geosciences.  The SOHO LASCO CME catalog is generated and maintained at the CDAW Data Center by NASA and the Catholic University of America in cooperation with the Naval Research Laboratory.  SOHO is a mission of international cooperation between the European Space Agency and NASA.  The STEREO/SECCHI data are produced by an international consortium of the NRL, LMSAL and NASA GSFC (USA), RAL and University of Birmingham (UK), MPS (Germany), CSL (Belgium), IOTA and IAS (France). Some figure colors based on \url{ColorBrewer.org}.
\end{acks}

%
%

\begin{thebibliography}{70}
\ifx \bisbn   \undefined \def \bisbn  #1{ISBN #1}\fi
\ifx \binits  \undefined \def \binits#1{#1}\fi
\ifx \bauthor  \undefined \def \bauthor#1{#1}\fi
\ifx \batitle  \undefined \def \batitle#1{#1}\fi
\ifx \bjtitle  \undefined \def \bjtitle#1{\textit{#1}}\fi
\ifx \bvolume  \undefined \def \bvolume#1{\textbf{#1}}\fi
\ifx \byear  \undefined \def \byear#1{#1}\fi
\ifx \bissue  \undefined \def \bissue#1{#1}\fi
\ifx \bfpage  \undefined \def \bfpage#1{#1}\fi
\ifx \blpage  \undefined \def \blpage #1{#1}\fi
\ifx \burl  \undefined \def \burl#1{\textsf{#1}}\fi
\ifx \href  \undefined \def \href#1#2{\textsf{#2}}\fi
\ifx \doiurl  \undefined \def
  \doiurl#1{\href{http://dx.doi.org/#1}{\textsf{#1}}}\fi
\ifx \betal  \undefined \def \betal{\textit{et al.}}\fi
\ifx \binstitute  \undefined \def \binstitute#1{#1}\fi
\ifx \bctitle  \undefined \def \bctitle#1{#1}\fi
\ifx \beditor  \undefined \def \beditor#1{#1}\fi
\ifx \bpublisher  \undefined \def \bpublisher#1{#1}\fi
\ifx \bbtitle  \undefined \def \bbtitle#1{\textit{#1}}\fi
\ifx \bedition  \undefined \def \bedition#1{#1}\fi
\ifx \bseriesno  \undefined \def \bseriesno#1{\textbf{#1}}\fi
\ifx \blocation  \undefined \def \blocation#1{#1}\fi
\ifx \bsertitle  \undefined \def \bsertitle#1{\textit{#1}}\fi
\ifx \bsnm \undefined \def \bsnm#1{#1}\fi
\ifx \bsuffix \undefined \def \bsuffix#1{#1}\fi
\ifx \bparticle \undefined \def \bparticle#1{#1}\fi
\ifx \barticle \undefined \def \barticle#1{}\fi
\ifx \botherref \undefined \def \botherref#1{}\fi
\ifx \url \undefined \def \url#1{\textsf{#1}}\fi
\ifx \bchapter \undefined \def \bchapter#1{}\fi
\ifx \bbook \undefined \def \bbook#1{}\fi
\ifx \bcomment \undefined \def \bcomment#1{#1}\fi
\ifx \oauthor \undefined \def \oauthor#1{#1}\fi
\ifx \citeauthoryear \undefined \def \citeauthoryear#1{#1}\fi
\def \endbibitem {}
\ifx \bconflocation  \undefined \def \bconflocation#1{#1} \fi

\bibitem[\protect\citeauthoryear{{Anderson}}{1996}]{anderson1996}
\begin{barticle}
\bauthor{\bsnm{{Anderson}}, \binits{J.L.}}:
\byear{1996},
\batitle{{A Method for Producing and Evaluating Probabilistic Forecasts from
  Ensemble Model Integrations.}}
\bjtitle{Journal of Climate}
\bvolume{9},
\bfpage{1518}\,--\,\blpage{1530}.
\end{barticle}
\endbibitem

\bibitem[\protect\citeauthoryear{{Arge} and {Pizzo}}{2000}]{arge2000}
\begin{barticle}
\bauthor{\bsnm{{Arge}}, \binits{C.N.}},
\bauthor{\bsnm{{Pizzo}}, \binits{V.J.}}:
\byear{2000},
\batitle{{Improvement in the prediction of solar wind conditions using
  near-real time solar magnetic field updates}}.
\bjtitle{\jgr}
\bvolume{105},
\bfpage{10465}\,--\,\blpage{10480}.
doi:\doiurl{10.1029/1999JA000262}.
\end{barticle}
\endbibitem

\bibitem[\protect\citeauthoryear{{Arge} \textit{et~al.}}{2004}]{arge2004}
\begin{barticle}
\bauthor{\bsnm{{Arge}}, \binits{C.N.}},
\bauthor{\bsnm{{Luhmann}}, \binits{J.G.}},
\bauthor{\bsnm{{Odstr{\v c}il}}, \binits{D.}},
\bauthor{\bsnm{{Schrijver}}, \binits{C.J.}},
\bauthor{\bsnm{{Li}}, \binits{Y.}}:
\byear{2004},
\batitle{{Stream structure and coronal sources of the solar wind during the May
  12th, 1997 CME}}.
\bjtitle{Journal of Atmospheric and Solar-Terrestrial Physics}
\bvolume{66},
\bfpage{1295}\,--\,\blpage{1309}.
doi:\doiurl{10.1016/j.jastp.2004.03.018}.
\end{barticle}
\endbibitem

\bibitem[\protect\citeauthoryear{{Arge} \textit{et~al.}}{2010}]{arge2010}
\begin{barticle}
\bauthor{\bsnm{{Arge}}, \binits{C.N.}},
\bauthor{\bsnm{{Henney}}, \binits{C.J.}},
\bauthor{\bsnm{{Koller}}, \binits{J.}},
\bauthor{\bsnm{{Compeau}}, \binits{C.R.}},
\bauthor{\bsnm{{Young}}, \binits{S.}},
\bauthor{\bsnm{{MacKenzie}}, \binits{D.}},
\bauthor{\bsnm{{Fay}}, \binits{A.}},
\bauthor{\bsnm{{Harvey}}, \binits{J.W.}}:
\byear{2010},
\batitle{{Air Force Data Assimilative Photospheric Flux Transport (ADAPT)
  Model}}.
\bjtitle{Twelfth International Solar Wind Conference}
\bvolume{1216},
\bfpage{343}\,--\,\blpage{346}.
doi:\doiurl{10.1063/1.3395870}.
\end{barticle}
\endbibitem

\bibitem[\protect\citeauthoryear{{Bartels}, {Heck}, and
  {Johnston}}{1939}]{bartels1939}
\begin{barticle}
\bauthor{\bsnm{{Bartels}}, \binits{J.}},
\bauthor{\bsnm{{Heck}}, \binits{N.H.}},
\bauthor{\bsnm{{Johnston}}, \binits{H.F.}}:
\byear{1939},
\batitle{{The three-hour-range index measuring geomagnetic activity}}.
\bjtitle{Terrestrial Magnetism and Atmospheric Electricity (Journal of
  Geophysical Research)}
\bvolume{44},
\bfpage{411}.
doi:\doiurl{10.1029/TE044i004p00411}.
\end{barticle}
\endbibitem

\bibitem[\protect\citeauthoryear{{Bock} \textit{et~al.}}{2014}]{bock2014}
\begin{botherref}
\oauthor{\bsnm{{Bock}}, \binits{A.}},
\oauthor{\bsnm{{Mays}}, \binits{M.L.}},
\oauthor{\bsnm{{Rastaetter}}, \binits{L.}},
\oauthor{\bsnm{{Ynnerman}}, \binits{A.}},
\oauthor{\bsnm{{Ropinski}}, \binits{T.}}:
2014,
{VCMass: A Framework for Verification of Coronal Mass Ejection Ensemble
  Simulations}.
\textit{IEEE Scientific Visualization Conference Abstracts}.
\end{botherref}
\endbibitem

\bibitem[\protect\citeauthoryear{{Brier}}{1950}]{brier1950}
\begin{barticle}
\bauthor{\bsnm{{Brier}}, \binits{G.W.}}:
\byear{1950},
\batitle{{Verification of Forecasts Expressed in Terms of Probability}}.
\bjtitle{Monthly Weather Review}
\bvolume{78},
\bfpage{1}.
\end{barticle}
\endbibitem

\bibitem[\protect\citeauthoryear{{Brueckner} \textit{et~al.}}{1995}]{lasco}
\begin{barticle}
\bauthor{\bsnm{{Brueckner}}, \binits{G.E.}},
\bauthor{\bsnm{{Howard}}, \binits{R.A.}},
\bauthor{\bsnm{{Koomen}}, \binits{M.J.}},
\bauthor{\bsnm{{Korendyke}}, \binits{C.M.}},
\bauthor{\bsnm{{Michels}}, \binits{D.J.}},
\bauthor{\bsnm{{Moses}}, \binits{J.D.}},
\bauthor{\bsnm{{\it et al.}}}:
\byear{1995},.
\bjtitle{Solar Phys.}
\bvolume{162},
\bfpage{357}\,--\,\blpage{402}.
\end{barticle}
\endbibitem

\bibitem[\protect\citeauthoryear{{Cohen} \textit{et~al.}}{2014}]{cohen2014}
\begin{barticle}
\bauthor{\bsnm{{Cohen}}, \binits{C.M.S.}},
\bauthor{\bsnm{{Mason}}, \binits{G.M.}},
\bauthor{\bsnm{{Mewaldt}}, \binits{R.A.}},
\bauthor{\bsnm{{Wiedenbeck}}, \binits{M.E.}}:
\byear{2014},
\batitle{{The Longitudinal Dependence of Heavy-ion Composition in the 2013
  April 11 Solar Energetic Particle Event}}.
\bjtitle{\apj}
\bvolume{793},
\bfpage{35}.
doi:\doiurl{10.1088/0004-637X/793/1/35}.
\end{barticle}
\endbibitem

\bibitem[\protect\citeauthoryear{{Colaninno}, {Vourlidas}, and
  {Wu}}{2013}]{colaninno2013}
\begin{barticle}
\bauthor{\bsnm{{Colaninno}}, \binits{R.C.}},
\bauthor{\bsnm{{Vourlidas}}, \binits{A.}},
\bauthor{\bsnm{{Wu}}, \binits{C.C.}}:
\byear{2013},
\batitle{{Quantitative comparison of methods for predicting the arrival of
  coronal mass ejections at Earth based on multiview imaging}}.
\bjtitle{Journal of Geophysical Research (Space Physics)}
\bvolume{118},
\bfpage{6866}\,--\,\blpage{6879}.
doi:\doiurl{10.1002/2013JA019205}.
\end{barticle}
\endbibitem

\bibitem[\protect\citeauthoryear{{Davies} \textit{et~al.}}{2013}]{davies2013}
\begin{barticle}
\bauthor{\bsnm{{Davies}}, \binits{J.A.}},
\bauthor{\bsnm{{Perry}}, \binits{C.H.}},
\bauthor{\bsnm{{Trines}}, \binits{R.M.G.M.}},
\bauthor{\bsnm{{Harrison}}, \binits{R.A.}},
\bauthor{\bsnm{{Lugaz}}, \binits{N.}},
\bauthor{\bsnm{{M{\"o}stl}}, \binits{C.}},
\bauthor{\bsnm{{Liu}}, \binits{Y.D.}},
\bauthor{\bsnm{{Steed}}, \binits{K.}}:
\byear{2013},
\batitle{{Establishing a Stereoscopic Technique for Determining the Kinematic
  Properties of Solar Wind Transients based on a Generalized Self-similarly
  Expanding Circular Geometry}}.
\bjtitle{\apj}
\bvolume{777},
\bfpage{167}.
doi:\doiurl{10.1088/0004-637X/777/2/167}.
\end{barticle}
\endbibitem

\bibitem[\protect\citeauthoryear{{Domingo}, {Fleck}, and {Poland}}{1995}]{soho}
\begin{barticle}
\bauthor{\bsnm{{Domingo}}, \binits{V.}},
\bauthor{\bsnm{{Fleck}}, \binits{B.}},
\bauthor{\bsnm{{Poland}}, \binits{A.I.}}:
\byear{1995},
\batitle{{The SOHO Mission: an Overview}}.
\bjtitle{Solar Phys.}
\bvolume{162},
\bfpage{1}\,--\,\blpage{37}.
\end{barticle}
\endbibitem

\bibitem[\protect\citeauthoryear{{Dryer}}{1974}]{dryer1974}
\begin{barticle}
\bauthor{\bsnm{{Dryer}}, \binits{M.}}:
\byear{1974},
\batitle{{Interplanetary Shock Waves Generated by Solar Flares}}.
\bjtitle{\ssr}
\bvolume{51},
\bfpage{403}\,--\,\blpage{468}.
\end{barticle}
\endbibitem

\bibitem[\protect\citeauthoryear{{Dryer} \textit{et~al.}}{2001}]{dryer2001}
\begin{barticle}
\bauthor{\bsnm{{Dryer}}, \binits{M.}},
\bauthor{\bsnm{{Fry}}, \binits{C.D.}},
\bauthor{\bsnm{{Sun}}, \binits{W.}},
\bauthor{\bsnm{{Deehr}}, \binits{C.}},
\bauthor{\bsnm{{Smith}}, \binits{Z.}},
\bauthor{\bsnm{{Akasofu}}, \binits{S.-I.}},
\bauthor{\bsnm{{Andrews}}, \binits{M.D.}}:
\byear{2001},
\batitle{{Prediction in Real Time of the 2000 July 14 Heliospheric Shock Wave
  and its Companions During the `Bastille' Epoch$^{*}$}}.
\bjtitle{\solphys}
\bvolume{204},
\bfpage{265}\,--\,\blpage{284}.
doi:\doiurl{10.1023/A:1014200719867}.
\end{barticle}
\endbibitem

\bibitem[\protect\citeauthoryear{{Emmons} \textit{et~al.}}{2013}]{emmons2013}
\begin{barticle}
\bauthor{\bsnm{{Emmons}}, \binits{D.}},
\bauthor{\bsnm{{Acebal}}, \binits{A.}},
\bauthor{\bsnm{{Pulkkinen}}, \binits{A.}},
\bauthor{\bsnm{{Taktakishvili}}, \binits{A.}},
\bauthor{\bsnm{{MacNeice}}, \binits{P.}},
\bauthor{\bsnm{{Odstr{\v c}il}}, \binits{D.}}:
\byear{2013},
\batitle{{Ensemble forecasting of coronal mass ejections using the WSA-ENLIL
  with CONED Model}}.
\bjtitle{Space Weather}
\bvolume{11},
\bfpage{95}\,--\,\blpage{106}.
doi:\doiurl{10.1002/swe.20019}.
\end{barticle}
\endbibitem

\bibitem[\protect\citeauthoryear{{Fry} \textit{et~al.}}{2003}]{fry2003}
\begin{barticle}
\bauthor{\bsnm{{Fry}}, \binits{C.D.}},
\bauthor{\bsnm{{Dryer}}, \binits{M.}},
\bauthor{\bsnm{{Smith}}, \binits{Z.}},
\bauthor{\bsnm{{Sun}}, \binits{W.}},
\bauthor{\bsnm{{Deehr}}, \binits{C.S.}},
\bauthor{\bsnm{{Akasofu}}, \binits{S.-I.}}:
\byear{2003},
\batitle{{Forecasting solar wind structures and shock arrival times using an
  ensemble of models}}.
\bjtitle{Journal of Geophysical Research (Space Physics)}
\bvolume{108},
\bfpage{1070}.
doi:\doiurl{10.1029/2002JA009474}.
\end{barticle}
\endbibitem

\bibitem[\protect\citeauthoryear{{Gopalswamy}
  \textit{et~al.}}{2009}]{gopalswamy2009}
\begin{barticle}
\bauthor{\bsnm{{Gopalswamy}}, \binits{N.}},
\bauthor{\bsnm{{Yashiro}}, \binits{S.}},
\bauthor{\bsnm{{Michalek}}, \binits{G.}},
\bauthor{\bsnm{{Stenborg}}, \binits{G.}},
\bauthor{\bsnm{{Vourlidas}}, \binits{A.}},
\bauthor{\bsnm{{Freeland}}, \binits{S.}},
\bauthor{\bsnm{{Howard}}, \binits{R.}}:
\byear{2009},
\batitle{{The SOHO/LASCO CME Catalog}}.
\bjtitle{Earth Moon and Planets}
\bvolume{104},
\bfpage{295}\,--\,\blpage{313}.
doi:\doiurl{10.1007/s11038-008-9282-7}.
\end{barticle}
\endbibitem

\bibitem[\protect\citeauthoryear{{Hamill}}{2001}]{hamill2001}
\begin{barticle}
\bauthor{\bsnm{{Hamill}}, \binits{T.M.}}:
\byear{2001},
\batitle{{Interpretation of Rank Histograms for Verifying Ensemble Forecasts}}.
\bjtitle{Monthly Weather Review}
\bvolume{129},
\bfpage{550}.
\end{barticle}
\endbibitem

\bibitem[\protect\citeauthoryear{{Hamill} and {Colucci}}{1997}]{hamill1997}
\begin{barticle}
\bauthor{\bsnm{{Hamill}}, \binits{T.M.}},
\bauthor{\bsnm{{Colucci}}, \binits{S.J.}}:
\byear{1997},
\batitle{{Verification of Eta RSM Short-Range Ensemble Forecasts}}.
\bjtitle{Monthly Weather Review}
\bvolume{125},
\bfpage{1312}.
\end{barticle}
\endbibitem

\bibitem[\protect\citeauthoryear{{Harvey} \textit{et~al.}}{1996}]{harvey1996}
\begin{barticle}
\bauthor{\bsnm{{Harvey}}, \binits{J.W.}},
\bauthor{\bsnm{{Hill}}, \binits{F.}},
\bauthor{\bsnm{{Hubbard}}, \binits{R.P.}},
\bauthor{\bsnm{{Kennedy}}, \binits{J.R.}},
\bauthor{\bsnm{{Leibacher}}, \binits{J.W.}},
\bauthor{\bsnm{{Pintar}}, \binits{J.A.}},
\bauthor{\bsnm{{Gilman}}, \binits{P.A.}},
\bauthor{\bsnm{{Noyes}}, \binits{R.W.}},
\bauthor{\bsnm{{Title}}, \binits{A.M.}},
\bauthor{\bsnm{{Toomre}}, \binits{J.}},
\bauthor{\bsnm{{Ulrich}}, \binits{R.K.}},
\bauthor{\bsnm{{Bhatnagar}}, \binits{A.}},
\bauthor{\bsnm{{Kennewell}}, \binits{J.A.}},
\bauthor{\bsnm{{Marquette}}, \binits{W.}},
\bauthor{\bsnm{{Patron}}, \binits{J.}},
\bauthor{\bsnm{{Saa}}, \binits{O.}},
\bauthor{\bsnm{{Yasukawa}}, \binits{E.}}:
\byear{1996},
\batitle{{The Global Oscillation Network Group (GONG) Project}}.
\bjtitle{Science}
\bvolume{272},
\bfpage{1284}\,--\,\blpage{1286}.
doi:\doiurl{10.1126/science.272.5266.1284}.
\end{barticle}
\endbibitem

\bibitem[\protect\citeauthoryear{{Harvey} \textit{et~al.}}{1992}]{harvey1992}
\begin{barticle}
\bauthor{\bsnm{{Harvey}}, \binits{L.O.}},
\bauthor{\bsnm{{Hammond}}, \binits{K.R.}},
\bauthor{\bsnm{{Lusk}}, \binits{C.M.}},
\bauthor{\bsnm{{Mross}}, \binits{E.F.}}:
\byear{1992},
\batitle{{The Application of Signal Detection Theory to Weather Forecasting
  Behavior}}.
\bjtitle{Monthly Weather Review}
\bvolume{120},
\bfpage{863}.
\end{barticle}
\endbibitem

\bibitem[\protect\citeauthoryear{{Henney} \textit{et~al.}}{2012}]{henney2012}
\begin{barticle}
\bauthor{\bsnm{{Henney}}, \binits{C.J.}},
\bauthor{\bsnm{{Toussaint}}, \binits{W.A.}},
\bauthor{\bsnm{{White}}, \binits{S.M.}},
\bauthor{\bsnm{{Arge}}, \binits{C.N.}}:
\byear{2012},
\batitle{{Forecasting F$_{10.7}$ with solar magnetic flux transport modeling}}.
\bjtitle{Space Weather}
\bvolume{10},
\bfpage{2011}.
doi:\doiurl{10.1029/2011SW000748}.
\end{barticle}
\endbibitem

\bibitem[\protect\citeauthoryear{{Hidalgo} \textit{et~al.}}{2000}]{hidalgo2000}
\begin{barticle}
\bauthor{\bsnm{{Hidalgo}}, \binits{M.A.}},
\bauthor{\bsnm{{Cid}}, \binits{C.}},
\bauthor{\bsnm{{Medina}}, \binits{J.}},
\bauthor{\bsnm{{Vi{\~n}as}}, \binits{A.F.}}:
\byear{2000},
\batitle{{A new model for the topology of magnetic clouds in the solar wind}}.
\bjtitle{\solphys}
\bvolume{194},
\bfpage{165}\,--\,\blpage{174}.
doi:\doiurl{10.1023/A:1005206107017}.
\end{barticle}
\endbibitem

\bibitem[\protect\citeauthoryear{{Howard} \textit{et~al.}}{2008}]{secchi}
\begin{barticle}
\bauthor{\bsnm{{Howard}}, \binits{R.A.}},
\bauthor{\bsnm{{Moses}}, \binits{J.D.}},
\bauthor{\bsnm{{Vourlidas}}, \binits{A.}},
\bauthor{\bsnm{{Newmark}}, \binits{J.S.}},
\bauthor{\bsnm{{Socker}}, \binits{D.G.}},
\bauthor{\bsnm{{Plunkett}}, \binits{S.P.}},
\bauthor{\bsnm{{\it et al.}}}:
\byear{2008},
\batitle{{Sun Earth Connection Coronal and Heliospheric Investigation
  (SECCHI)}}.
\bjtitle{Space Science Reviews}
\bvolume{136},
\bfpage{67}\,--\,\blpage{115}.
doi:\doiurl{10.1007/s11214-008-9341-4}.
\end{barticle}
\endbibitem

\bibitem[\protect\citeauthoryear{{Howard} and {DeForest}}{2012}]{howard2012}
\begin{barticle}
\bauthor{\bsnm{{Howard}}, \binits{T.A.}},
\bauthor{\bsnm{{DeForest}}, \binits{C.E.}}:
\byear{2012},
\batitle{{The Thomson Surface. I. Reality and Myth}}.
\bjtitle{\apj}
\bvolume{752},
\bfpage{130}.
doi:\doiurl{10.1088/0004-637X/752/2/130}.
\end{barticle}
\endbibitem

\bibitem[\protect\citeauthoryear{{Jackson} \textit{et~al.}}{2011}]{jackson2011}
\begin{barticle}
\bauthor{\bsnm{{Jackson}}, \binits{B.V.}},
\bauthor{\bsnm{{Hick}}, \binits{P.P.}},
\bauthor{\bsnm{{Buffington}}, \binits{A.}},
\bauthor{\bsnm{{Bisi}}, \binits{M.M.}},
\bauthor{\bsnm{{Clover}}, \binits{J.M.}},
\bauthor{\bsnm{{Tokumaru}}, \binits{M.}},
\bauthor{\bsnm{{Kojima}}, \binits{M.}},
\bauthor{\bsnm{{Fujiki}}, \binits{K.}}:
\byear{2011},
\batitle{{Three-dimensional reconstruction of heliospheric structure using
  iterative tomography: A review}}.
\bjtitle{Journal of Atmospheric and Solar-Terrestrial Physics}
\bvolume{73},
\bfpage{1214}\,--\,\blpage{1227}.
doi:\doiurl{10.1016/j.jastp.2010.10.007}.
\end{barticle}
\endbibitem

\bibitem[\protect\citeauthoryear{{Jian} \textit{et~al.}}{2011}]{jian2011}
\begin{barticle}
\bauthor{\bsnm{{Jian}}, \binits{L.K.}},
\bauthor{\bsnm{{Russell}}, \binits{C.T.}},
\bauthor{\bsnm{{Luhmann}}, \binits{J.G.}},
\bauthor{\bsnm{{MacNeice}}, \binits{P.J.}},
\bauthor{\bsnm{{Odstr{\v c}il}}, \binits{D.}},
\bauthor{\bsnm{{Riley}}, \binits{P.}},
\bauthor{\bsnm{{Linker}}, \binits{J.A.}},
\bauthor{\bsnm{{Skoug}}, \binits{R.M.}},
\bauthor{\bsnm{{Steinberg}}, \binits{J.T.}}:
\byear{2011},
\batitle{{Comparison of Observations at ACE and Ulysses with Enlil Model
  Results: Stream Interaction Regions During Carrington Rotations 2016 -
  2018}}.
\bjtitle{\solphys}
\bvolume{273},
\bfpage{179}\,--\,\blpage{203}.
doi:\doiurl{10.1007/s11207-011-9858-7}.
\end{barticle}
\endbibitem

\bibitem[\protect\citeauthoryear{{Jolliffe} and
  {Stephenson}}{2011}]{jolliffe2011}
\begin{bbook}
\beditor{\bsnm{{Jolliffe}}, \binits{I.T.}},
\beditor{\bsnm{{Stephenson}}, \binits{D.B.}} (eds.):
\byear{2011},
\bbtitle{{Forecast Verification: A Practioner's Guide in Atmospheric Science}},
\bedition{2nd} edn.
\bpublisher{Wiley},
\blocation{New Jersey, USA}.
\end{bbook}
\endbibitem

\bibitem[\protect\citeauthoryear{{Kaiser} \textit{et~al.}}{2008}]{stereo}
\begin{barticle}
\bauthor{\bsnm{{Kaiser}}, \binits{M.L.}},
\bauthor{\bsnm{{Kucera}}, \binits{T.A.}},
\bauthor{\bsnm{{Davila}}, \binits{J.M.}},
\bauthor{\bsnm{{St.~Cyr}}, \binits{O.C.}},
\bauthor{\bsnm{{Guhathakurta}}, \binits{M.}},
\bauthor{\bsnm{{Christian}}, \binits{E.}}:
\byear{2008},
\batitle{{The STEREO Mission: An Introduction}}.
\bjtitle{Space Science Reviews}
\bvolume{136},
\bfpage{5}\,--\,\blpage{16}.
doi:\doiurl{10.1007/s11214-007-9277-0}.
\end{barticle}
\endbibitem

\bibitem[\protect\citeauthoryear{{Lario} \textit{et~al.}}{2014}]{lario2014}
\begin{barticle}
\bauthor{\bsnm{{Lario}}, \binits{D.}},
\bauthor{\bsnm{{Raouafi}}, \binits{N.E.}},
\bauthor{\bsnm{{Kwon}}, \binits{R.-Y.}},
\bauthor{\bsnm{{Zhang}}, \binits{J.}},
\bauthor{\bsnm{{G{\'o}mez-Herrero}}, \binits{R.}},
\bauthor{\bsnm{{Dresing}}, \binits{N.}},
\bauthor{\bsnm{{Riley}}, \binits{P.}}:
\byear{2014},
\batitle{{The Solar Energetic Particle Event on 2013 April 11: An Investigation
  of its Solar Origin and Longitudinal Spread}}.
\bjtitle{\apj}
\bvolume{797},
\bfpage{8}.
doi:\doiurl{10.1088/0004-637X/797/1/8}.
\end{barticle}
\endbibitem

\bibitem[\protect\citeauthoryear{{LaSota}}{2013}]{lasota2013}
\begin{botherref}
\oauthor{\bsnm{{LaSota}}, \binits{J.A.}}:
2013,
{STEREO Analysis}.
{Undergraduate Honors Thesis},
{University of Alaska Fairbanks}.
\end{botherref}
\endbibitem

\bibitem[\protect\citeauthoryear{{Lee} \textit{et~al.}}{2013}]{lee2013}
\begin{barticle}
\bauthor{\bsnm{{Lee}}, \binits{C.O.}},
\bauthor{\bsnm{{Arge}}, \binits{C.N.}},
\bauthor{\bsnm{{Odstr{\v c}il}}, \binits{D.}},
\bauthor{\bsnm{{Millward}}, \binits{G.}},
\bauthor{\bsnm{{Pizzo}}, \binits{V.}},
\bauthor{\bsnm{{Quinn}}, \binits{J.M.}},
\bauthor{\bsnm{{Henney}}, \binits{C.J.}}:
\byear{2013},
\batitle{{Ensemble Modeling of CME Propagation}}.
\bjtitle{\solphys}
\bvolume{285},
\bfpage{349}\,--\,\blpage{368}.
doi:\doiurl{10.1007/s11207-012-9980-1}.
\end{barticle}
\endbibitem

\bibitem[\protect\citeauthoryear{{Liu} \textit{et~al.}}{2010}]{liu2010a}
\begin{barticle}
\bauthor{\bsnm{{Liu}}, \binits{Y.}},
\bauthor{\bsnm{{Davies}}, \binits{J.A.}},
\bauthor{\bsnm{{Luhmann}}, \binits{J.G.}},
\bauthor{\bsnm{{Vourlidas}}, \binits{A.}},
\bauthor{\bsnm{{Bale}}, \binits{S.D.}},
\bauthor{\bsnm{{Lin}}, \binits{R.P.}}:
\byear{2010},
\batitle{{Geometric Triangulation of Imaging Observations to Track Coronal Mass
  Ejections Continuously Out to 1 AU}}.
\bjtitle{\apjl}
\bvolume{710},
\bfpage{L82}\,--\,\blpage{L87}.
doi:\doiurl{10.1088/2041-8205/710/1/L82}.
\end{barticle}
\endbibitem

\bibitem[\protect\citeauthoryear{{Lugaz} \textit{et~al.}}{2010}]{lugaz2010}
\begin{barticle}
\bauthor{\bsnm{{Lugaz}}, \binits{N.}},
\bauthor{\bsnm{{Hernandez-Charpak}}, \binits{J.N.}},
\bauthor{\bsnm{{Roussev}}, \binits{I.I.}},
\bauthor{\bsnm{{Davis}}, \binits{C.J.}},
\bauthor{\bsnm{{Vourlidas}}, \binits{A.}},
\bauthor{\bsnm{{Davies}}, \binits{J.A.}}:
\byear{2010},
\batitle{{Determining the Azimuthal Properties of Coronal Mass Ejections from
  Multi-Spacecraft Remote-Sensing Observations with STEREO SECCHI}}.
\bjtitle{\apj}
\bvolume{715},
\bfpage{493}\,--\,\blpage{499}.
doi:\doiurl{10.1088/0004-637X/715/1/493}.
\end{barticle}
\endbibitem

\bibitem[\protect\citeauthoryear{{MacNeice}}{2009}]{macneice2009}
\begin{barticle}
\bauthor{\bsnm{{MacNeice}}, \binits{P.}}:
\byear{2009},
\batitle{{Validation of community models: 2. Development of a baseline using
  the Wang-Sheeley-Arge model}}.
\bjtitle{Space Weather}
\bvolume{7},
\bfpage{12002}.
doi:\doiurl{10.1029/2009SW000489}.
\end{barticle}
\endbibitem

\bibitem[\protect\citeauthoryear{{Manoharan}}{2006}]{manoharan2006}
\begin{barticle}
\bauthor{\bsnm{{Manoharan}}, \binits{P.K.}}:
\byear{2006},
\batitle{{Evolution of Coronal Mass Ejections in the Inner Heliosphere: A Study
  Using White-Light and Scintillation Images}}.
\bjtitle{\solphys}
\bvolume{235},
\bfpage{345}\,--\,\blpage{368}.
doi:\doiurl{10.1007/s11207-006-0100-y}.
\end{barticle}
\endbibitem

\bibitem[\protect\citeauthoryear{{Mays} \textit{et~al.}}{{2014}}]{mays2014b}
\begin{botherref}
\oauthor{\bsnm{{Mays}}, \binits{M.L.}},
\oauthor{\bsnm{{Taktakishvili}}, \binits{A.}},
\oauthor{\bsnm{{Romano}}, \binits{M.}},
\oauthor{\bsnm{{MacNeice}}, \binits{P.J.}},
\oauthor{\bsnm{{Zheng}}, \binits{Y.}},
\oauthor{\bsnm{{Pulkkinen}}, \binits{A.A.}},
\oauthor{\bsnm{{Kuznetsova}}, \binits{M.M.}},
\oauthor{\bsnm{{Odstr{\v c}il}}, \binits{D.}}:
{2014},
{Validation of Real-time Modeling of Coronal Mass Ejections Using the
  WSA-ENLIL+Cone Heliospheric Model}.
\textit{Space Weather}.
{In preparation}.
\end{botherref}
\endbibitem

\bibitem[\protect\citeauthoryear{{McKenna-Lawlor}
  \textit{et~al.}}{2006}]{mckenna2006}
\begin{barticle}
\bauthor{\bsnm{{McKenna-Lawlor}}, \binits{S.M.P.}},
\bauthor{\bsnm{{Dryer}}, \binits{M.}},
\bauthor{\bsnm{{Kartalev}}, \binits{M.D.}},
\bauthor{\bsnm{{Smith}}, \binits{Z.}},
\bauthor{\bsnm{{Fry}}, \binits{C.D.}},
\bauthor{\bsnm{{Sun}}, \binits{W.}},
\bauthor{\bsnm{{Deehr}}, \binits{C.S.}},
\bauthor{\bsnm{{Kecskemety}}, \binits{K.}},
\bauthor{\bsnm{{Kudela}}, \binits{K.}}:
\byear{2006},
\batitle{{Near real-time predictions of the arrival at Earth of flare-related
  shocks during Solar Cycle 23}}.
\bjtitle{Journal of Geophysical Research (Space Physics)}
\bvolume{111},
\bfpage{11103}.
doi:\doiurl{10.1029/2005JA011162}.
\end{barticle}
\endbibitem

\bibitem[\protect\citeauthoryear{{Menvielle} and
  {Berthelier}}{1991}]{menvielle1991}
\begin{barticle}
\bauthor{\bsnm{{Menvielle}}, \binits{M.}},
\bauthor{\bsnm{{Berthelier}}, \binits{A.}}:
\byear{1991},
\batitle{{The K-derived planetary indices - Description and availability}}.
\bjtitle{Reviews of Geophysics}
\bvolume{29},
\bfpage{415}\,--\,\blpage{432}.
doi:\doiurl{10.1029/91RG00994}.
\end{barticle}
\endbibitem

\bibitem[\protect\citeauthoryear{{Millward}
  \textit{et~al.}}{2013}]{millward2013}
\begin{barticle}
\bauthor{\bsnm{{Millward}}, \binits{G.}},
\bauthor{\bsnm{{Biesecker}}, \binits{D.}},
\bauthor{\bsnm{{Pizzo}}, \binits{V.}},
\bauthor{\bsnm{{Koning}}, \binits{C.A.}}:
\byear{2013},
\batitle{{An operational software tool for the analysis of coronagraph images:
  Determining CME parameters for input into the WSA-Enlil heliospheric model}}.
\bjtitle{Space Weather}
\bvolume{11},
\bfpage{57}\,--\,\blpage{68}.
doi:\doiurl{10.1002/swe.20024}.
\end{barticle}
\endbibitem

\bibitem[\protect\citeauthoryear{{M\"{u}ller}
  \textit{et~al.}}{2009}]{muller2009}
\begin{botherref}
\oauthor{\bsnm{{M\"{u}ller}}, \binits{D.}},
\oauthor{\bsnm{{Dimitoglou}}, \binits{G.}},
\oauthor{\bsnm{{Caplins}}, \binits{B.}},
\oauthor{\bsnm{{Ireland}}, \binits{J.}},
\oauthor{\bsnm{{Wamsler}}, \binits{B.}},
\oauthor{\bsnm{{Hughitt}}, \binits{K.}},
\oauthor{\bsnm{{Agheksanterian}}, \binits{D.} \bsuffix{A.~{Amadigwe}}}:
2009,
{Jhelioviewer - Visualizing large sets of solar data using JPEG 2000}.
\textit{Comput. Sci. Eng. 11}
\textbf{38}.
\end{botherref}
\endbibitem

\bibitem[\protect\citeauthoryear{{Murphy}}{1973}]{murphy1973}
\begin{barticle}
\bauthor{\bsnm{{Murphy}}, \binits{A.H.}}:
\byear{1973},
\batitle{{A New Vector Partition of the Probability Score.}}
\bjtitle{Journal of Applied Meteorology}
\bvolume{12},
\bfpage{595}\,--\,\blpage{600}.
\end{barticle}
\endbibitem

\bibitem[\protect\citeauthoryear{{Newell} \textit{et~al.}}{2007}]{newell2007}
\begin{barticle}
\bauthor{\bsnm{{Newell}}, \binits{P.T.}},
\bauthor{\bsnm{{Sotirelis}}, \binits{T.}},
\bauthor{\bsnm{{Liou}}, \binits{K.}},
\bauthor{\bsnm{{Meng}}, \binits{C.-I.}},
\bauthor{\bsnm{{Rich}}, \binits{F.J.}}:
\byear{2007},
\batitle{{A nearly universal solar wind-magnetosphere coupling function
  inferred from 10 magnetospheric state variables}}.
\bjtitle{Journal of Geophysical Research (Space Physics)}
\bvolume{112},
\bfpage{1206}.
doi:\doiurl{10.1029/2006JA012015}.
\end{barticle}
\endbibitem

\bibitem[\protect\citeauthoryear{{Odstr{\v c}il}}{2003}]{odstrcil2003}
\begin{barticle}
\bauthor{\bsnm{{Odstr{\v c}il}}, \binits{D.}}:
\byear{2003},
\batitle{{Modeling 3-D solar wind structure}}.
\bjtitle{Advances in Space Research}
\bvolume{32},
\bfpage{497}\,--\,\blpage{506}.
doi:\doiurl{10.1016/S0273-1177(03)00332-6}.
\end{barticle}
\endbibitem

\bibitem[\protect\citeauthoryear{{Odstr{\v c}il} and
  {Pizzo}}{1999a}]{odstrcil1999_1}
\begin{barticle}
\bauthor{\bsnm{{Odstr{\v c}il}}, \binits{D.}},
\bauthor{\bsnm{{Pizzo}}, \binits{V.J.}}:
\byear{1999}a,
\batitle{{Three-dimensional propagation of CMEs in a structured solar wind
  flow: 1. CME launched within the streamer belt}}.
\bjtitle{\jgr}
\bvolume{104},
\bfpage{483}\,--\,\blpage{492}.
doi:\doiurl{10.1029/1998JA900019}.
\end{barticle}
\endbibitem

\bibitem[\protect\citeauthoryear{{Odstr{\v c}il} and
  {Pizzo}}{1999b}]{odstrcil1999_2}
\begin{barticle}
\bauthor{\bsnm{{Odstr{\v c}il}}, \binits{D.}},
\bauthor{\bsnm{{Pizzo}}, \binits{V.J.}}:
\byear{1999}b,
\batitle{{Three-dimensional propagation of coronal mass ejections in a
  structured solar wind flow 2. CME launched adjacent to the streamer belt}}.
\bjtitle{\jgr}
\bvolume{104},
\bfpage{493}\,--\,\blpage{504}.
doi:\doiurl{10.1029/1998JA900038}.
\end{barticle}
\endbibitem

\bibitem[\protect\citeauthoryear{{Odstr{\v c}il}, {Riley}, and
  {Zhao}}{2004}]{odstrcil2004}
\begin{barticle}
\bauthor{\bsnm{{Odstr{\v c}il}}, \binits{D.}},
\bauthor{\bsnm{{Riley}}, \binits{P.}},
\bauthor{\bsnm{{Zhao}}, \binits{X.P.}}:
\byear{2004},
\batitle{{Numerical simulation of the 12 May 1997 interplanetary CME event}}.
\bjtitle{\jgr}
\bvolume{109},
\bfpage{2116}.
doi:\doiurl{10.1029/2003JA010135}.
\end{barticle}
\endbibitem

\bibitem[\protect\citeauthoryear{{Odstr{\v c}il}, {Smith}, and
  {Dryer}}{1996}]{ods1996}
\begin{barticle}
\bauthor{\bsnm{{Odstr{\v c}il}}, \binits{D.}},
\bauthor{\bsnm{{Smith}}, \binits{Z.}},
\bauthor{\bsnm{{Dryer}}, \binits{M.}}:
\byear{1996},
\batitle{{Distortion of the heliospheric plasma sheet by interplanetary
  shocks}}.
\bjtitle{\grl}
\bvolume{23},
\bfpage{2521}\,--\,\blpage{2524}.
doi:\doiurl{10.1029/96GL00159}.
\end{barticle}
\endbibitem

\bibitem[\protect\citeauthoryear{{Pizzo} and {Biesecker}}{2004}]{pizzo2004}
\begin{barticle}
\bauthor{\bsnm{{Pizzo}}, \binits{V.J.}},
\bauthor{\bsnm{{Biesecker}}, \binits{D.A.}}:
\byear{2004},
\batitle{{Geometric localization of STEREO CMEs}}.
\bjtitle{\grl}
\bvolume{31},
\bfpage{21802}.
doi:\doiurl{10.1029/2004GL021141}.
\end{barticle}
\endbibitem

\bibitem[\protect\citeauthoryear{{Pulkkinen}, {Oates}, and
  {Taktakishvili}}{2010}]{pulkkinen2009}
\begin{barticle}
\bauthor{\bsnm{{Pulkkinen}}, \binits{A.}},
\bauthor{\bsnm{{Oates}}, \binits{T.}},
\bauthor{\bsnm{{Taktakishvili}}, \binits{A.}}:
\byear{2010},
\batitle{{Automatic Determination of the Conic Coronal Mass Ejection Model
  Parameters}}.
\bjtitle{\solphys}
\bvolume{261},
\bfpage{115}\,--\,\blpage{126}.
doi:\doiurl{10.1007/s11207-009-9473-z}.
\end{barticle}
\endbibitem

\bibitem[\protect\citeauthoryear{{Pulkkinen}
  \textit{et~al.}}{2011}]{pulkkinen2011}
\begin{botherref}
\oauthor{\bsnm{{Pulkkinen}}, \binits{A.A.}},
\oauthor{\bsnm{{Taktakishvili}}, \binits{A.}},
\oauthor{\bsnm{{Odstr{\v c}il}}, \binits{D.}},
\oauthor{\bsnm{{MacNeice}}, \binits{P.J.}}:
2011,
{Ensemble forecasting of coronal mass ejection propagation in the
  interplanetary medium}.
\textit{{NOAA Space Weather Workshop Abstracts}}.
\end{botherref}
\endbibitem

\bibitem[\protect\citeauthoryear{{Richardson} and
  {Cane}}{2010}]{richardson2010}
\begin{barticle}
\bauthor{\bsnm{{Richardson}}, \binits{I.G.}},
\bauthor{\bsnm{{Cane}}, \binits{H.V.}}:
\byear{2010},
\batitle{{Near-Earth Interplanetary Coronal Mass Ejections During Solar Cycle
  23 (1996 - 2009): Catalog and Summary of Properties}}.
\bjtitle{\solphys}
\bvolume{264},
\bfpage{189}\,--\,\blpage{237}.
doi:\doiurl{10.1007/s11207-010-9568-6}.
\end{barticle}
\endbibitem

\bibitem[\protect\citeauthoryear{{Riley}, {Linker}, and
  {Miki{\'c}}}{2001}]{riley2001}
\begin{barticle}
\bauthor{\bsnm{{Riley}}, \binits{P.}},
\bauthor{\bsnm{{Linker}}, \binits{J.A.}},
\bauthor{\bsnm{{Miki{\'c}}}, \binits{Z.}}:
\byear{2001},
\batitle{{An empirically-driven global MHD model of the solar corona and inner
  heliosphere}}.
\bjtitle{\jgr}
\bvolume{106},
\bfpage{15889}\,--\,\blpage{15902}.
doi:\doiurl{10.1029/2000JA000121}.
\end{barticle}
\endbibitem

\bibitem[\protect\citeauthoryear{{Romano} \textit{et~al.}}{2013}]{romano2013}
\begin{botherref}
\oauthor{\bsnm{{Romano}}, \binits{M.}},
\oauthor{\bsnm{{Mays}}, \binits{M.L.}},
\oauthor{\bsnm{{Taktakishvili}}, \binits{A.}},
\oauthor{\bsnm{{MacNeice}}, \binits{P.J.}},
\oauthor{\bsnm{{Zheng}}, \binits{Y.}},
\oauthor{\bsnm{{Pulkkinen}}, \binits{A.A.}},
\oauthor{\bsnm{{Kuznetsova}}, \binits{M.M.}},
\oauthor{\bsnm{{Odstr{\v c}il}}, \binits{D.}}:
2013,
{Validation of Real-time Modeling of Coronal Mass Ejections Using the
  WSA-ENLIL+Cone Heliospheric Model}.
\textit{AGU Fall Meeting Abstracts},
A2156.
\end{botherref}
\endbibitem

\bibitem[\protect\citeauthoryear{{Rostoker}}{1972}]{rostoker1972}
\begin{barticle}
\bauthor{\bsnm{{Rostoker}}, \binits{G.}}:
\byear{1972},
\batitle{{Geomagnetic indices.}}
\bjtitle{Reviews of Geophysics and Space Physics}
\bvolume{10},
\bfpage{935}\,--\,\blpage{950}.
doi:\doiurl{10.1029/RG010i004p00935}.
\end{barticle}
\endbibitem

\bibitem[\protect\citeauthoryear{{Sivillo}, {Ahlquist}, and
  {Toth}}{1997}]{sivillo1997}
\begin{barticle}
\bauthor{\bsnm{{Sivillo}}, \binits{J.K.}},
\bauthor{\bsnm{{Ahlquist}}, \binits{J.E.}},
\bauthor{\bsnm{{Toth}}, \binits{Z.}}:
\byear{1997},
\batitle{{An Ensemble Forecasting Primer}}.
\bjtitle{Weather and Forecasting}
\bvolume{12},
\bfpage{809}\,--\,\blpage{818}.
\end{barticle}
\endbibitem

\bibitem[\protect\citeauthoryear{Smith and Dryer}{1990}]{smith1990}
\begin{barticle}
\bauthor{\bsnm{Smith}, \binits{Z.}},
\bauthor{\bsnm{Dryer}, \binits{M.}}:
\byear{1990},
\batitle{Mhd study of temporal and spatial evolution of simulated
  interplanetary shocks in the ecliptic plane within 1 au}.
\bjtitle{Solar Physics}
\bvolume{129}(\bissue{2}),
\bfpage{387}\,--\,\blpage{405}.
doi:\doiurl{10.1007/BF00159049}.
\burl{http://dx.doi.org/10.1007/BF00159049}.
\end{barticle}
\endbibitem

\bibitem[\protect\citeauthoryear{{Smith} \textit{et~al.}}{2009}]{smith2009}
\begin{barticle}
\bauthor{\bsnm{{Smith}}, \binits{Z.K.}},
\bauthor{\bsnm{{Dryer}}, \binits{M.}},
\bauthor{\bsnm{{McKenna-Lawlor}}, \binits{S.M.P.}},
\bauthor{\bsnm{{Fry}}, \binits{C.D.}},
\bauthor{\bsnm{{Deehr}}, \binits{C.S.}},
\bauthor{\bsnm{{Sun}}, \binits{W.}}:
\byear{2009},
\batitle{{Operational validation of HAFv2's predictions of interplanetary shock
  arrivals at Earth: Declining phase of Solar Cycle 23}}.
\bjtitle{Journal of Geophysical Research (Space Physics)}
\bvolume{114},
\bfpage{5106}.
doi:\doiurl{10.1029/2008JA013836}.
\end{barticle}
\endbibitem

\bibitem[\protect\citeauthoryear{{Sugiura}}{1964}]{sugiuraDst}
\begin{barticle}
\bauthor{\bsnm{{Sugiura}}, \binits{M.}}:
\byear{1964},
\batitle{{Hourly values of equatorial Dst for the IGY}}.
\bjtitle{Ann. Int. Geophys. Year}
\bvolume{35}(\bissue{9}),
\bfpage{945}.
\end{barticle}
\endbibitem

\bibitem[\protect\citeauthoryear{{Taktakishvili}, {MacNeice}, and {Odstr{\v
  c}il}}{2010}]{taktak2010}
\begin{barticle}
\bauthor{\bsnm{{Taktakishvili}}, \binits{A.}},
\bauthor{\bsnm{{MacNeice}}, \binits{P.}},
\bauthor{\bsnm{{Odstr{\v c}il}}, \binits{D.}}:
\byear{2010},
\batitle{{Model uncertainties in predictions of arrival of coronal mass
  ejections at Earth orbit}}.
\bjtitle{Space Weather}
\bvolume{8},
\bfpage{6007}.
doi:\doiurl{10.1029/2009SW000543}.
\end{barticle}
\endbibitem

\bibitem[\protect\citeauthoryear{{Talagrand}, {Vautard}, and
  {Strauss}}{1997}]{talagrand1997}
\begin{bchapter}
\bauthor{\bsnm{{Talagrand}}, \binits{O.}},
\bauthor{\bsnm{{Vautard}}, \binits{R.}},
\bauthor{\bsnm{{Strauss}}, \binits{B.}}:
\byear{1997},
\bctitle{{Evaluation of probabilistic prediction systems}}.
In: \bbtitle{Proceedings, ECMWF Workshop on Predictability},
\bfpage{1}\,--\,\blpage{25}.
\end{bchapter}
\endbibitem

\bibitem[\protect\citeauthoryear{{Thernisien}, {Howard}, and
  {Vourlidas}}{2006}]{thernisien2006}
\begin{barticle}
\bauthor{\bsnm{{Thernisien}}, \binits{A.F.R.}},
\bauthor{\bsnm{{Howard}}, \binits{R.A.}},
\bauthor{\bsnm{{Vourlidas}}, \binits{A.}}:
\byear{2006},
\batitle{{Modeling of Flux Rope Coronal Mass Ejections}}.
\bjtitle{\apj}
\bvolume{652},
\bfpage{763}\,--\,\blpage{773}.
doi:\doiurl{10.1086/508254}.
\end{barticle}
\endbibitem

\bibitem[\protect\citeauthoryear{{Vr\v{s}nak}
  \textit{et~al.}}{2014}]{vrsnak2014}
\begin{barticle}
\bauthor{\bsnm{{Vr\v{s}nak}}, \binits{B.}},
\bauthor{\bsnm{{Temmer}}, \binits{M.}},
\bauthor{\bsnm{{{\v Z}ic}}, \binits{T.}},
\bauthor{\bsnm{{Taktakishvili}}, \binits{A.}},
\bauthor{\bsnm{{Dumbovi\'{c}}}, \binits{M.}},
\bauthor{\bsnm{{M\"{o}stl}}, \binits{C.}},
\bauthor{\bsnm{{Veronig}}, \binits{A.M.}},
\bauthor{\bsnm{{Mays}}, \binits{M.L.}},
\bauthor{\bsnm{{Odstr{\v c}il}}, \binits{D.}}:
\byear{2014},
\batitle{{Heliospheric Propagation of Coronal Mass Ejections: Comparison of
  Numerical WSA-ENLIL+Cone Model and Analytical Drag-based Model}}.
\bjtitle{\apjs}
\bvolume{213},
\bfpage{21}.
doi:\doiurl{10.1088/0067-0049/213/2/21}.
\end{barticle}
\endbibitem

\bibitem[\protect\citeauthoryear{{Weigel} \textit{et~al.}}{2006}]{weigel2006}
\begin{barticle}
\bauthor{\bsnm{{Weigel}}, \binits{R.S.}},
\bauthor{\bsnm{{Detman}}, \binits{T.}},
\bauthor{\bsnm{{Rigler}}, \binits{E.J.}},
\bauthor{\bsnm{{Baker}}, \binits{D.N.}}:
\byear{2006},
\batitle{{Decision theory and the analysis of rare event space weather
  forecasts}}.
\bjtitle{Space Weather}
\bvolume{4},
\bfpage{5002}.
doi:\doiurl{10.1029/2005SW000157}.
\end{barticle}
\endbibitem

\bibitem[\protect\citeauthoryear{Wilks}{1995}]{wilks1995}
\begin{bbook}
\bauthor{\bsnm{Wilks}, \binits{D.S.}}:
\byear{1995},
\bbtitle{{Statistical Methods in Atmospheric Sciences: An Introduction}},
\bpublisher{Academic Press},
\blocation{Massachusetts, USA}.
\end{bbook}
\endbibitem

\bibitem[\protect\citeauthoryear{{Xie}, {Ofman}, and
  {Lawrence}}{2004}]{xie2004}
\begin{barticle}
\bauthor{\bsnm{{Xie}}, \binits{H.}},
\bauthor{\bsnm{{Ofman}}, \binits{L.}},
\bauthor{\bsnm{{Lawrence}}, \binits{G.}}:
\byear{2004},
\batitle{{Cone model for halo CMEs: Application to space weather forecasting}}.
\bjtitle{\jgr}
\bvolume{109},
\bfpage{3109}.
doi:\doiurl{10.1029/2003JA010226}.
\end{barticle}
\endbibitem

\bibitem[\protect\citeauthoryear{{Yashiro} \textit{et~al.}}{2004}]{yashiro2004}
\begin{barticle}
\bauthor{\bsnm{{Yashiro}}, \binits{S.}},
\bauthor{\bsnm{{Gopalswamy}}, \binits{N.}},
\bauthor{\bsnm{{Michalek}}, \binits{G.}},
\bauthor{\bsnm{{St.~Cyr}}, \binits{O.C.}},
\bauthor{\bsnm{{Plunkett}}, \binits{S.P.}},
\bauthor{\bsnm{{Rich}}, \binits{N.B.}},
\bauthor{\bsnm{{Howard}}, \binits{R.A.}}:
\byear{2004},
\batitle{{A catalog of white light coronal mass ejections observed by the SOHO
  spacecraft}}.
\bjtitle{Journal of Geophysical Research (Space Physics)}
\bvolume{109},
\bfpage{7105}.
doi:\doiurl{10.1029/2003JA010282}.
\end{barticle}
\endbibitem

\bibitem[\protect\citeauthoryear{{Zhao} and {Dryer}}{2014}]{zhao2014}
\begin{barticle}
\bauthor{\bsnm{{Zhao}}, \binits{X.}},
\bauthor{\bsnm{{Dryer}}, \binits{M.}}:
\byear{2014},
\batitle{{Current status of CME/shock arrival time prediction}}.
\bjtitle{Space Weather}
\bvolume{12},
\bfpage{448}\,--\,\blpage{469}.
doi:\doiurl{10.1002/2014SW001060}.
\end{barticle}
\endbibitem

\bibitem[\protect\citeauthoryear{{Zhao}, {Plunkett}, and
  {Liu}}{2002}]{zhao2002}
\begin{barticle}
\bauthor{\bsnm{{Zhao}}, \binits{X.P.}},
\bauthor{\bsnm{{Plunkett}}, \binits{S.P.}},
\bauthor{\bsnm{{Liu}}, \binits{W.}}:
\byear{2002},
\batitle{{Determination of geometrical and kinematical properties of halo
  coronal mass ejections using the cone model}}.
\bjtitle{\jgr}
\bvolume{107},
\bfpage{1223}.
doi:\doiurl{10.1029/2001JA009143}.
\end{barticle}
\endbibitem

\bibitem[\protect\citeauthoryear{{Zheng} \textit{et~al.}}{2013}]{zheng2013}
\begin{barticle}
\bauthor{\bsnm{{Zheng}}, \binits{Y.}},
\bauthor{\bsnm{{Macneice}}, \binits{P.}},
\bauthor{\bsnm{{Odstr{\v c}il}}, \binits{D.}},
\bauthor{\bsnm{{Mays}}, \binits{M.L.}},
\bauthor{\bsnm{{Rastaetter}}, \binits{L.}},
\bauthor{\bsnm{{Pulkkinen}}, \binits{A.}},
\bauthor{\bsnm{{Taktakishvili}}, \binits{A.}},
\bauthor{\bsnm{{Hesse}}, \binits{M.}},
\bauthor{\bsnm{{Masha Kuznetsova}}, \binits{M.}},
\bauthor{\bsnm{{Lee}}, \binits{H.}},
\bauthor{\bsnm{{Chulaki}}, \binits{A.}}:
\byear{2013},
\batitle{{Forecasting propagation and evolution of CMEs in an operational
  setting: What has been learned}}.
\bjtitle{Space Weather}
\bvolume{11},
\bfpage{557}\,--\,\blpage{574}.
doi:\doiurl{10.1002/swe.20096}.
\end{barticle}
\endbibitem

\end{thebibliography}

\end{article} 
\end{document}